\documentclass[%
 reprint,
 amsmath,amssymb,
 aps,
]{revtex4-2}

\usepackage{graphicx}
\usepackage{dcolumn}
\usepackage{bm}
\usepackage{xcolor}

\begin{document}

\preprint{APS/123-QED}

\title{Glassy dynamics in deep neural networks: A structural comparison}

\author{Max Kerr Winter}
\author{Liesbeth M. C. Janssen}
\affiliation{Eindhoven University of Technology}
\date{\today}

\begin{abstract}
Deep Neural Networks (DNNs) share important similarities with structural glasses. Both have many degrees of freedom, and their dynamics are governed by a high-dimensional, non-convex landscape representing either the loss or energy, respectively. Furthermore, both experience gradient descent dynamics subject to noise. In this work we investigate, by performing quantitative measurements on realistic networks trained on the MNIST and CIFAR-10 datasets, the extent to which this qualitative similarity gives rise to glass-like dynamics in neural networks. We demonstrate the existence of a Topology Trivialisation Transition as well as the previously studied under-to-overparameterised transition analogous to jamming. By training DNNs with overdamped Langevin dynamics in the resulting disordered phases, we do not observe diverging relaxation times at non-zero temperature, nor do we observe any caging effects, in contrast to glass phenomenology. However, the weight overlap function follows a power law in time, with exponent $\approx -0.5$, in agreement with the Mode-Coupling Theory of structural glasses. In addition, the DNN dynamics obey a form of time-temperature superposition. Finally, dynamic heterogeneity and ageing are observed at low temperatures. These results highlight important and surprising points of both difference and agreement between the behaviour of DNNs and structural glasses.
\end{abstract}

\maketitle



\section{\label{sec:Intro}Introduction}
Deep neural networks (DNNs) have experienced an explosion of popularity in the last decade as they have proven to be highly effective at tackling a wide range of problems across the natural sciences and engineering. Consequently, there is significant interest from the statistical physics community in understanding the fundamental mechanisms by which neural networks train and generalise \cite{Bahri2020,Li2020,Goldt2020,Geiger2020,Zhang2018}. An important strand of this work is the similarity between DNNs and glasses. The link between spin glasses and neural networks has received considerable attention since the 1980s, originally with an emphasis on the mapping between the Hopfield network and Ising-like systems, with attention turning to feed-forward networks in the last decade or so \cite{Hopfield1982,Little1974,Amit1985,Amit1985a,Mezard1986,Krauth1987,Gardner1988,Sompolinsky1988,Barra2012,Ghio2023,Biroli2024,Mignacco2022, Becker2020, Choromanska2015}. Several studies have focused on the nature of the loss landscape, in particular demonstrating the surprising result that in many cases the landscape, while non-convex, has no poorly performing local minima \cite{Kawaguchi2016,Soudry2016,Kawaguchi2020}. More recently, training dynamics of networks trained with Stochastic Gradient Descent (SGD) have been investigated in the context of anomalous diffusion \cite{Chen2022}, and in comparison with spin glasses where both similarities and differences were found \cite{Baity-Jesi2019}. There is also some significant overlap between spin and structural glass phenomenology \cite{Baity-Jesi2019b}, particularly in the limit of infinite dimensions \cite{Kirkpatrick1989,Charbonneau2014a,Folena2020,Laudicina2024}. However, the possible analogy between neural networks and structural glasses, which in general exhibit a richer phenomenology than spin glasses, is still largely unexplored. Moreover, since modern deep learning comprises a wide variety of (typically feed forward, as opposed to Hopfield) architectures, one may also wonder whether this analogy might be extended to feed forward DNNs with more complex interactions between parameters. 

Structural glasses exhibit a wide variety of typical phenomena (growing and ultimately diverging relaxation times, caging, dynamic heterogeneity, violation of the Stokes-Einstein relation, ageing) that have been the subject of intense research for many decades, resulting in a rich body of theory of the kind that is currently lacking in deep learning \cite{Janssen2018,Das2004,Micoulaut2016,Berthier2011}. Structural glasses are a potentially valuable analogy for DNNs as they involve complicated interaction potentials between the constituent particles. This is reminiscent of the situation in DNNs where parameters also interact via highly non-linear, indirect couplings. Furthermore, recent studies have made a quantitative comparison between the jamming transition (a phenomenon closely related to the structural glass transition \cite{Parisi2010}) and an under-to-overparameterised transition (UOT) in DNNs \cite{Geiger2019, Spigler2019, Franz2019,Franz2016}. Large networks are able to fit their training data exactly, resulting in zero loss (equivalent to an unjammed state with zero interaction energy, and sometimes referred to as a canyon landscape \cite{Urbani2023,Urbani2023a}), whereas small networks have a non-zero loss similar to the positive interaction energy of jammed particles. It is highly counter-intuitive that modern deep learning works so successfully in this overparameterised phase while avoiding the problems of overfitting. Much like the energy landscape of an unjammed system, the loss landscape in the long-time limit of overparameterised networks has been shown to consist of a large number of flat directions \cite{Baity-Jesi2019}. These recent results raise the intriguing possibility, explored in this work, of whether similar connections can be drawn between the dynamics of glass forming particle systems and DNNs. 

As well as the analogy with jamming, results derived from high-dimensional random energy landscapes \cite{Fyodorov2004} are also relevant to the statistical physics of deep learning. Random energy landscapes, which are in some respects highly idealised models of glasses, have been widely applied in physics \cite{Ros2019,Subag2021,Naumis2005} and are now increasingly being used to study DNNs and related machine learning systems \cite{Wainrib2013,SaraoMannelli2020,Mannelli2019}. Of particular relevance to the current work is the Topology Trivialisation Transition (TTT), where the energy landscape of a system switches from a single minimum to many minima on decreasing a quadratic energy penalty. The existence of TTT-like transitions has been demonstrated in high-dimensional elastic models subject to random fields \cite{Cugliandolo1996,Cugliandolo1996a}, in large ecosystems \cite{Altieri2021}, in systems of coupled differential equations \cite{Arous2021}, and in random neural networks \cite{Wainrib2013, Sompolinsky1988}. These diverse systems all share a common structure with feed forward DNNs subject to regularisation in that their dynamics balance a disordered term with a quadratic potential. It has remained unexplored, to the best of the authors' knowledge, whether regularised feed forward DNNs undergo a TTT in their loss function (and in fact, we find that they do.)

The goal of this work is to quantitatively assess the similarities between DNNs and structural glass-forming materials by performing comparisons across a wide range of numerical experiments typical of glass formers. Do DNNs undergo glass-like transitions? Do they exhibit the typical glassy behaviours of diverging relaxation times, caging, violation of the Stokes-Einstein relation, dynamic heterogeneity, and ageing? Answering these questions is an important first step in guiding the application of the theory of structural glasses to the field of deep learning. 

\section{\label{sec:setup} Numerical setup}
In this work we use fully connected feedforward networks with no biases, only weights, so that all parameters are of the same type. We use this structure as it is the most general feed forward architecture, while noting that our DNNs differ considerably both in structure and size from common models such as convolutional neural networks and transformers. The output of a DNN is denoted $f(\mathbf{x}, \mathbf{W})$ for weight vector $\mathbf{W}$, of length $N$, and input data $\mathbf{x}$. The activation functions are ReLU, $\sigma(z) = \text{max}(0, z)$. We consider a problem of supervised binary classification, with a training set $\mathcal{T} = \{\mathbf{x}_i, y_i \}_{i=1}^{P_D}$, of size $P_D$, consisting of input examples $\mathbf{x}$ of dimension $d$, and corresponding labels $y\in \{+1, -1\}$. 

Our primary dataset is a preprocessed version of MNIST \cite{MNIST_dataset}, consisting of $P_D=6\times10^4$ images of hand-drawn digits from 0--9, with corresponding labels. We binarise the classification problem by assigning a label of +1 to all even digits, and -1 to all odd digits. Furthermore, we use Principal Component Analysis (PCA) to reduce the dimension of the input data to $d=10$ \cite{Jolliffe2002}, in order to decrease the number of weights in the first layer. Hence all networks have an input layer of width 10 and an output of width 1.  The networks have a fixed depth of 6 hidden layers, all with a constant width. This width is varied to alter the size of the networks, resulting in a total number of weights between 4830 (a width of 30) and 34,522 (a width of 82). We also repeat our analysis with the CIFAR-10 dataset (see Appendix \ref{sec:app_cifar10_tau90_plots}).

The choice of using the full MNIST and CIFAR-10 datasets, which consequently require large DNNs to reach the overparameterised state, was made to maximise the realism of our investigation, thus ensuring that our numerical experiments reflect the true complexity of modern deep learning applications. However, this complexity comes with a significant computational cost. Compounding this problem, glassy measurements often require the networks to be trained for a long time, and averaged over many repeats. Approximately ten thousand networks were involved in this study, many of which were trained for over a hundred CPU hours.

As our loss function for the training of our DNNs, we use the quadratic hinge loss, 
\begin{align}
\mathcal{L}_{\text{data}} =  \frac{1}{P_D} \sum_{i=1}^{P_D} \ell(f(\mathbf{x}_i, \mathbf{W}), y_i) ,\label{eq:quad_hinge_loss}
\end{align}
where $\ell(f, y_i) = \frac{1}{2}\text{max}(0, \Delta(y_i, f))^2$, and $\Delta(y, f) = 1-yf$. This loss was chosen following the example of \cite{Geiger2019}, as it is similar to the energy of a system of particles interacting via a finite-range harmonic potential. When the particles are separated beyond the cut-off distance, all pairwise interactions are 0, and the total energy is 0. When the particles are close together they repel each other and the energy is a sum of quadratic terms. Analogously, when all data is fit exactly, $\ell(f(\mathbf{x}_i, \mathbf{W}), y_i)=0, \: \forall i$, hence $\mathcal{L}_{\text{data}}=0$, whereas when this is not the case, the loss is an average of quadratic terms. However, the loss is quadratic in $f(\mathbf{x}, \mathbf{W})$, a highly non-linear function of the weights (which, in our analogy to structural glasses, may be interpreted as the equivalent of particle positions, i.e. the degrees of freedom.)

We also use L2-regularisation, a common method to avoid overfitting \cite{Ying2019}. The resulting loss function is
\begin{align}
\mathcal{L} = \mathcal{L}_{\text{data}} + \frac{\lambda}{2}|\mathbf{W}|^2,\label{eq:loss}
\end{align}
where $\lambda>0$ is a control parameter. When studying dynamic variables, we train the DNNs with overdamped Langevin dynamics, 
\begin{align}
\partial_t w_i = -\mu \frac{\partial \mathcal{L}}{\partial w_i} + \xi_i,\label{eq:Langevin}
\end{align}
where $w_i$ is a weight, $\mu$ is the learning rate (set to $10^{-3}$ unless otherwise stated), and the noise term obeys $\langle \xi_i \rangle = 0$, and $\langle \xi_i(t) \xi_j(t') \rangle = (2k_BT \mu/P_D)\delta_{ij}\delta(t-t')$, for temperature $T$ and Boltzmann constant $k_B$. For the remainder of the paper we work in natural units where $k_B=1$. In contrast to physical systems a factor of $1/P_D$, the size of the dataset, appears in the noise correlation due to $\mathcal{L}$ being intensive with respect to the dataset size. See Appendix \ref{sec:app_temP_Def} for a more detailed discussion. 

At times we also train networks with SGD, where the dynamics are given by
\begin{align}
\partial_t w_i = -\mu \frac{\partial \tilde{\mathcal{L}}}{\partial w_i}. \label{eq:SGD_def}
\end{align}
Here the loss is approximated by using only a subset of the data (a ``batch'') of size $B$, $\tilde{\mathcal{L}} = 1/B \sum_{i=1}^B \ell(f(\mathbf{x}_i, \mathbf{W}), y_i)$. Although no noise appears explicitly in Eq.~\eqref{eq:SGD_def}, stochasticity is introduced by the random selection of batches. We do not use SGD for dynamic measurements as the noise is highly unphysical -  it is coloured, non-delta correlated, and non-constant in time.

Equation~\eqref{eq:loss} highlights the similarity between DNNs and the disordered elastic model of \cite{Fyodorov2004}, which undergoes a TTT. The Hamiltonian of the elastic system is $\mathcal{H}(x_i)=V(x_i) + \frac{\eta}{2} |\mathbf{x}|^2$, where $V$ is a random potential, and the second term describes the elasticity of the medium. This is also similar to the Hamiltonian of spherical spin glasses \cite{Cugliandolo1995}. The question we wish to address is, in essence, how much glassy behaviour remains when we substitute $\mathcal{H}$ for $\mathcal{L}$.

We will now consider various aspects of glassy dynamics, each of which requires a different protocol to potentially probe it in DNNs. We will discuss both the protocol and the results, starting with the phase diagram in order to identify the transition lines. We will then address relaxation dynamics in the resulting disordered phases, time-temperature superposition, caging and the mean square displacement, the Stokes-Einstein relation, non-Gaussianity, dynamic heterogeneity, and finally end with aging. Concluding comments are given in Sec.\ \ref{sec:Discussion}. The code used to generate and analyse the DNNs is available at \url{https://github.com/mkerrwinter/glass_net_publication}.


\section{\label{sec:numerics} Results}
\subsection{Phase diagram}
\begin{figure}[h]
\includegraphics[width=8.6cm]{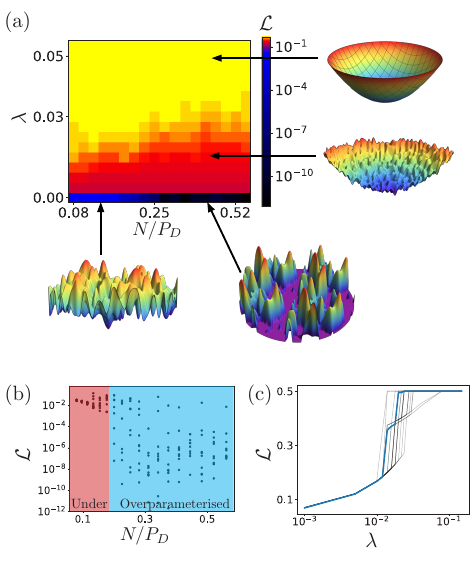}
\caption{Glass-like transitions in neural networks. (a) The final training loss (colour bar) in the $(N/P_D, \lambda)$ plane, where $N$ is the number of weights, $P_D$ the number of training data, and $\lambda$ the regularisation strength. Schematic illustrations of the loss landscape in the resulting four phases are shown around the edge. Note that these sketches are only illustrative and not based on real data. At high $\lambda$ the quadratic L2 regularisation term of Eq.~\eqref{eq:loss} dominates and the loss is a simple well. As $\lambda$ decreases the loss landscape undergoes a TTT and develops many random minima. On the $\lambda=0$ line, at low $N/P_D$ the network is underparameterised, corresponding to another disordered landscape. Finally, at high $N/P_D$ the network is over parameterised and has a loss landscape combining disordered regions with flat plateaus at zero loss. (b) The final training loss either side of the UOT. Each black point is a different initial condition. For this plot $\lambda=0$. (c) The final training loss on either side of the TTT. The translucent lines correspond to different initial conditions, and the blue line is their average. For this plot $N/P_D=0.0805$, corresponding to the left most column in (a).}
\label{fig:phase_diag}
\end{figure}

We first seek to establish a phase diagram to identify two relevant transition lines for our DNNs. The first is a transition in the DNN loss as a function of the ratio of network size, $N$, to dataset size, $P_D$, from underparameterised ($\mathcal{L}>0$) to overparameterised ($\mathcal{L}\approx0$). This transition is analogous to the jamming transition of interacting particles, and was first applied to the perceptron in \cite{Franz2016}, and then to DNNs in \cite{Geiger2019}.
The second transition, the Topology Trivialisation Transition of \cite{Fyodorov2004}, separates a regime with a single energy minimum, from  a disordered energy landscape with many minima. As discussed below, and demonstrated in Fig.\ \ref{fig:phase_diag}, we find the same under-to-overparameterised/jamming transition, and show that an analogous TTT exists in the loss of DNNs as a function of the regularisation strength, $\lambda$. 

In order to obtain the phase diagram, we sample a uniform grid in the $(N, \lambda)$ plane, i.e. construct DNNs of varying size and regularisation, and train these networks with Stochastic Gradient Descent for 7000 epochs with a batch size of 100. We then perform 50 epochs of gradient descent at the original learning rate, and a further 50 epochs of gradient descent after reducing the learning rate by a factor of 10. Convergence was checked manually on a representative subset of DNNs. As such we are sampling (close to) the minima of the landscape. Finally we average over 10 DNNs with different, random initial conditions at each point in the $(N, \lambda)$ plane, resulting in the phase diagram shown in Fig.\ \ref{fig:phase_diag}(a). We use SGD instead of Langevin dynamics as we are studying static properties of the landscape, not the dynamics, and SGD is much faster than Eq.~\eqref{eq:Langevin}. We include illustrative sketches of the resulting phases around the edge of Fig.\ \ref{fig:phase_diag}(a) emphasising their important properties: the absence/presence of flat regions on either side of the UOT, and the emergence of many minima below the TTT.

The UOT is visible on the $\lambda=0$ line, where at low $N/P_D$ the loss is approximately $10^{-2}$ (underparameterised), and at high $N/P_D$ falls to around $10^{-10}$ (overparameterised). There is some ambiguity as to where exactly the $\mathcal{L}=0$ phase appears, but we take the transition point to be $N^\ast/P_D \approx 0.18$. The TTT occurs in the vertical direction, where at high $\lambda$ the loss is 0.5, as is expected for a DNN where all weights are zero (see Eq.~\eqref{eq:quad_hinge_loss}, with $f=0$). At low $\lambda$ the networks are able to find alternative solutions (i.e. with non-zero weights), and hence achieve losses of around 0.1. We take the TTT point to be where the loss falls below 0.5, hence $\lambda^\ast \approx 0.03$. In Appendix \ref{sec:app_TTT}, we demonstrate that the departure of the loss from a value of 0.5 corresponds to the landscape changing from having a single minimum at $w_i=0$ $\forall i$, to one with multiple distinct minima. The transitions are shown separately in Fig.\ \ref{fig:phase_diag}(b) (UOT) and (c) (TTT). Figure \ref{fig:phase_diag}(b) uses the same networks as on the $\lambda=0$ line from Fig.\ \ref{fig:phase_diag}(a), and shows that achieving the $\mathcal{L}\approx0$ state is by no means guaranteed. It is strongly dependent on the initial conditions and training trajectory through weight space, even in the overparameterised phase, in agreement with \cite{Geiger2019}. 

On either side of the UOT we have two disordered phases, underparameterised, and overparameterised, illustrated schematically in Fig.\ \ref{fig:phase_diag}(a). The key difference between these phases is the presence of flat plateaus at zero loss in the overparameterised phase. The UOT does not exist when $\lambda>0$ as the addition of the regularisation term destroys the plateaus. This is similar to the case where the addition of a spherical constraint makes a system of equations unsolvable \cite{Fyodorov2022}. Figure \ref{fig:phase_diag}(c) uses 20 DNNs of width 30, each with different initial conditions, to demonstrate the TTT. Each DNN undergoes a sharp TTT, but the position of the transition varies, resulting in an average with some width. In this transition region a small number of initial conditions find a long-lived, high loss minimum, resulting in a kink in the average curve. The TTT separates a trivial phase, where the landscape is essentially a quadratic bowl (high $\lambda$), from a disordered phase (low $\lambda$).

\subsection{Relaxation with overdamped Langevin dynamics}
We now turn to the time-dependent dynamics of our DNNs to explore potential analogies with the relaxation dynamics of structural glassformers. There are four regions of interest in our phase diagram, on either side of each transition: the regularised phase ($0<\lambda<\lambda^\ast$); the quadratic phase ($\lambda>\lambda^\ast$); the underparameterised phase ($\lambda=0$, $N<N^\ast$); and the overparameterised phase ($\lambda=0$, $N>N^\ast$). In practice, the quadratic phase is trivial as networks rapidly reach steady state at the bottom of the quadratic well ($w_i\approx0$ $\forall i$), hence we do not discuss this phase further. As can be seen from Fig.\ \ref{fig:phase_diag}(a), $\lambda^\ast \approx 0.03$, with some $N$ dependence, and $N^\ast/P_D \approx 0.18$. 

We first measure the dynamics of a network by defining the overlap correlation function,
\begin{align}
Q(t, t_w) = \left \langle \frac{1}{N} \sum_{i=1}^N \left[1-\Theta\left( |w_i(t_w+t)-w_i(t_w)| - \epsilon\right)\right] \right \rangle,\label{eq:VH_def}
\end{align}
where $t_w$ is the waiting time between the system being prepared and the start of the measurement of $Q$, $\Theta$ is the Heaviside function, $\epsilon>0$ is a sensitivity parameter, and $\langle \cdot \rangle$ denotes averaging over both initial conditions and noise realisations. $Q(0, t_w)=1$, and as the weights of a network decorrelate from their initial value $Q$ decreases. For physical systems there will typically be a relevant length scale (e.g.\ the particle diameter) that determines $\epsilon$. No such scale exists for DNN weights, and so the choice of $\epsilon$ is somewhat arbitrary. We define $\epsilon$ with respect to a new control parameter, $\chi$, and the standard deviation of weights at the end of training, $\sigma$, such that $\epsilon = \sigma/\chi$. A more in-depth discussion of our choice of $\epsilon$, and the effect of varying it, is presented in Appendix \ref{sec:app_epsilon}. 

\begin{figure}[h]
\includegraphics[width=9cm]{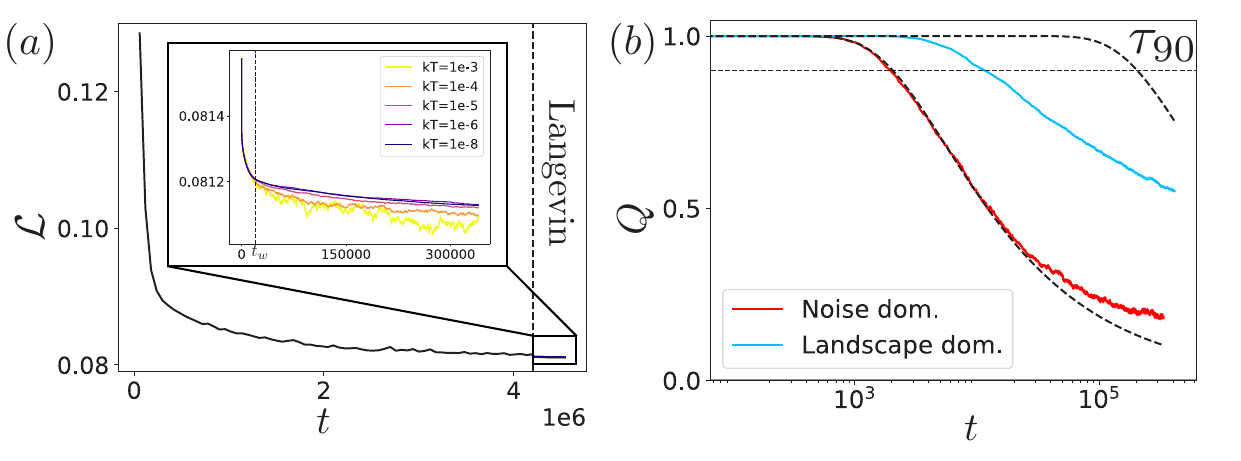}
\caption{(a) An example loss curve demonstrating the measurement procedure. Networks are generated with random weights, and trained for an initial period with SGD. The networks are then trained for a second period with overdamped Langevin dynamics. Measurements start after a waiting time, $t_w$. (b) Examples of the overlap function for networks in the high temperature, noise dominated regime (red, lower curve, $T=10^{-4}$), and in the low temperature, landscape dominated regime (blue, upper curve, $T=10^{-6}$). These measurements were made with $t_w=5\times10^3$, and using $\chi=400$. The analytic predictions of $Q(t)$ for free Brownian particles are shown by the dashed black lines.}
\label{fig:measurement_procedure}
\end{figure}

We perform a set of numerical experiments designed to be qualitatively similar to the quench and anneal protocol used to study glasses (e.g. in \cite{Pihlajamaa2023a,Kob1995} and \cite{Hopkins2010} among many others). Networks are first initialised with random weights drawn from a uniform distribution. Next, the networks are trained with SGD, at batch size 100, to efficiently move them away from their random initial position in the loss landscape, and into a local basin. Ideally this initial training period would be performed with Langevin dynamics, however SGD is significantly faster, and due to the long timescales involved in our measurements computing time was a limiting factor. The networks are then trained for a further period using the overdamped Langevin dynamics of Eq.~\eqref{eq:Langevin}, at a range of temperatures, $T$. After waiting some time, $t_w$, to allow the networks to move beyond their transient response to the new dynamics we start to measure $Q$. Measurements are made on approximately 50 independent initial conditions, although in some cases the network weights become unstable and grow to infinity. These networks are excluded from our analysis. A typical example of the measurement procedure is shown in Fig.\ \ref{fig:measurement_procedure}(a). The decay of $Q$ at two temperatures is shown in Fig.\ \ref{fig:measurement_procedure}(b). As expected, networks decorrelate more quickly at higher temperatures. We take as a typical decorrelation time $\tau_{90}$, the time it takes for $Q$ to reach 90\% of its initial value, $Q(\tau_{90})=0.9$. Glassy systems typically undergo two step relaxation: an initial fast relaxation, followed by a caging plateau, and then slower relaxation in the long time limit. The $e$-fold relaxation time is used to probe the slower relaxation. However, we do not observe two step relaxation in any of our experiments, making the choice of overlap value somewhat arbitrary. As such, we choose $\tau_{90}$ for computational convenience as it is reached quickly. For completeness we repeat our measurements with $\tau_e$ and find no significant differences with $\tau_{90}$. These results are presented in Appendix \ref{sec:app_cifar10_tau90_plots}. 

\begin{figure}[h]
\includegraphics[width=9cm]{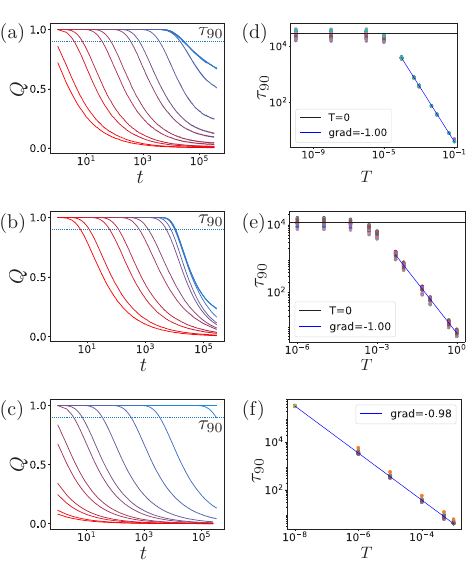}
\caption{(a) The overlap function in the regularised phase, measured with $\chi=300$, at temperatures (from red to blue, left to right) $T=1$, $0.5$, $0.1$, $0.05$, $0.01$, $0.005$, $0.001$, $5\times10^{-4}$, $10^{-4}$, $10^{-5}$, $10^{-6}$, $10^{-8}$, $10^{-10}$, $0$.  (b) The overlap function in the underparameterised phase, measured with $\chi=500$, at temperatures (from red to blue, left to right) $T=1$, $0.5$, $0.1$, $0.05$, $0.01$, $0.005$, $0.001$, $5\times10^{-4}$, $10^{-4}$, $10^{-5}$, $10^{-6}$, $0$. (c) The overlap function in the overparameterised phase, measured with $\chi=6000$, at temperatures (from red to blue, left to right) $T=1$, $0.5$, $0.1$, $0.05$, $0.01$, $0.005$, $0.001$, $5\times10^{-4}$, $10^{-4}$, $10^{-5}$, $10^{-6}$, $10^{-8}$, $10^{-10}$, $0$. (d), (e), (f) The dependence of $\tau_{90}$ on $T$ in the regularised, underparameterised, and overparameterised phases respectively. All phases exhibit a noise-dominated regime where $\tau_{90} \sim 1/T$. The regularised and underparameterised phases also have a landscape-dominated regime where $\tau_{90}$ becomes temperature independent below $T^\ast$. All measurements were made with $t_w=2\times10^4$.}
\label{fig:Langevin_tau90}
\end{figure}

In Fig.\ \ref{fig:Langevin_tau90}(a), (b), and (c), $Q(t)$ is plotted for a range of temperatures in the regularised (width=30, $\lambda=10^{-2}$), underparameterised (width=30, $\lambda=0$), and overparameterised (width=72, $\lambda=0$) phases respectively. The corresponding dependence of $\tau_{90}$ on the temperature, $T$, is shown in (d), (e), and (f). Both the regularised and underparameterised phases display two dynamical regimes. At high $T$, decorrelation is dominated by thermal noise, and we find a power law relationship, $\tau_{90}\sim T^{-1}$, which we call the noise-dominated regime. This is what we would expect from an overdamped, memoryless system with dynamics of the form $\dot{Q} = -\Omega Q$, where the frequency term $\Omega \sim T$ (e.g.\ Eq.~28 in \cite{Reichman2005b} if we neglect the memory kernel). Interestingly, a $1/T$ relationship has also been observed for the decorrelation of the loss of DNNs trained on random data, and using the cross entropy loss function \cite{Jules2023}. At low $T$, the decorrelation process becomes temperature independent, and instead is dominated by the first term of Eq.~\eqref{eq:Langevin}, the deterministic gradients of the loss landscape. The mean value of $\tau_{90}$ at $T=0$ is shown with the horizontal black line. Consequently, the value $T^*$ where plots (d), (e), and (f) exhibit a kink is the point at which the noise term of Eq.~\eqref{eq:Langevin} can effectively be neglected. We hypothesise that this is because the networks become trapped in the basins of the loss landscape. We call this low $T$ behaviour the landscape-dominated regime. In the overparameterised phase only the noise-dominated regime, $\tau_{90}\sim T^{-1}$, is observed. We suggest that this is because the landscape consists of flat plateaus at $\mathcal{L}\approx0$ where there are no gradients and hence the noise term of Eq.~\eqref{eq:Langevin} always dominates. Furthermore, the observed divergence in Fig.\ \ref{fig:Langevin_tau90}(f) of $\tau_{90}$ as $T \to 0$ is to be expected, as both terms in Eq.~\eqref{eq:Langevin} disappear.

Note the parameter $\epsilon$, that defines the overlap region in Eq.~\eqref{eq:VH_def}, is not the same between phases in Fig.\ \ref{fig:Langevin_tau90}, hence we cannot make quantitative comparisons between the measurements for the regularised, under-, and overparameterised networks. This stems from the lack of a typical scale for the weights. Choosing a fixed $\chi$ would also not result in comparable results as the standard deviation of the weights, $\sigma$, varies between phases. However, quantitative comparisons can be made between different temperatures in the same phase, as $\chi$ is fixed and $\sigma$ varies by around 1\%. Furthermore, the key conclusions that two regimes exist (noise and landscape dominated), and that they behave as a $T^{-1}$ power law and a temperature independent regime, are robust to the changes in $\epsilon$ between phases. We cannot rule out that there exists some $\epsilon$ small enough that a landscape dominated regime emerges in the overparameterised phase. However, using the smallest $\epsilon$ possible without introducing issues with floating point precision (at approximately $\epsilon=10^{-7}$), we do not observe any deviation from the noise dominated regime.  

For the noise-dominated regime we can predict the decorrelation curve by treating the weights as independent Brownian particles. The corresponding probability density function of weight displacements $\Delta w$ obeys the diffusion equation and has solution
\begin{align}
P(\Delta w, t) = \frac{\Theta(t)}{\sqrt{4\pi D t}} e^{-\frac{\Delta w^2}{4Dt}},
\end{align}
with diffusion constant $D=\frac{\mu T}{P}$. The overlap is equivalent to integrating $P(\Delta w)$ over the interval $[-\epsilon, \epsilon]$, resulting in
\begin{align}
Q(t) = \int_{-\epsilon}^\epsilon P(\Delta w) d \Delta w = \frac{2\Theta(t)}{\sqrt{\pi}} \text{erf}\left( \frac{\epsilon}{\sqrt{4 D t}} \right),\label{eq:analytic_decorr}
\end{align}
where $\text{erf}(x)$ is the error function. This analytic curve is shown in the dashed line of Fig.\ \ref{fig:measurement_procedure}(b) where, for a network in the noise-dominated regime, the prediction closely follows the DNN measurement at short and intermediate times. In the landscape-dominated regime, the DNNs decorrelate faster than would be predicted by Eq.~ \eqref{eq:analytic_decorr} for Brownian weights at the equivalent temperature. Weights in a harmonic basin subjected to thermal noise decorrelate faster the steeper the basin (see Appendix \ref{sec:app_overlap} for details), hence, although we cannot perform analytic calculations with the true loss landscape, it is to be expected that DNNs in the landscape-dominated regime decorrelate faster than the prediction of Eq.~\eqref{eq:analytic_decorr}.

\subsection{The long-time limit of $Q(t)$}
\begin{figure}[h]
\includegraphics[width=8.8cm]{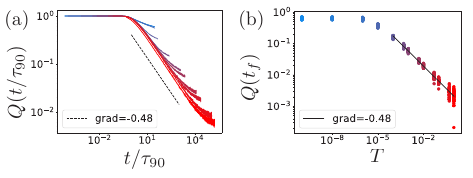}
\caption{(a) $Q(t)$ curves collapse onto a master curve for short and intermediate times, when time is rescaled by $\tau_{90}$. Higher temperatures (redder colours) leave the master curve at longer times, and lower values of $Q(t)$. The measurements were made with the same parameters as Fig.\ \ref{fig:Langevin_tau90}(a) and the displayed temperatures are $T=1$, $0.5$, $0.1$, $0.05$, $0.01$, $0.005$, $0.001$, $5\times10^{-4}$, $10^{-4}$, $10^{-5}$, $10^{-6}$, $10^{-8}$, $10^{-10}$, $0$. (b) The final value of the overlap, $Q(t_f=3\times10^5)$ vs $T$.}
\label{fig:TTSP}
\end{figure}

By considering the long-time behaviour of $Q$, we are able to make quantitative comparisons with the theory of structural glasses. For this comparison we would like to measure $Q_\infty = \lim_{t \to \infty} Q(t)$. An approximate measurement of $Q_\infty$ can be performed in the regularised phase by observing that the $Q(t)$ curves at different temperatures collapse onto a master curve when time is rescaled in a temperature-dependent way, $\hat{t} = t/\tau_{90}$, as shown in Fig.\ \ref{fig:TTSP}(a). This is similar to the time-temperature superposition principle (TTSP) of glasses \cite{Olsen2001, Li2000, Billon2023}, however here the collapse occurs at short and intermediate times, instead of long times as found in standard TTSP \cite{Kob1995}. The resulting master curve obeys a power law $Q(t) \sim t^{-0.48}$. This is in close agreement with the prediction for the intermediate scattering function, $\phi \sim t^{-0.5}$, from the F1 schematic version of Mode-Coupling Theory (MCT), a highly idealised model of structural glasses \cite{Schnyder2011,Mandal2017a,Spanner2013}. This power law is also observed in the under and overparameterised phases (see Appendix \ref{sec:app_q_infty}).

Different temperatures leave the master curve and begin to plateau at different times, with higher temperatures doing so at longer times and lower values of $Q(t)$. Note that networks at $T<10^{-5}$ did not yet have time to leave the master curve and so behave in a temperature-independent way. We take the final value of each curve in Fig.\ \ref{fig:TTSP}(a) that leaves the master curve as an approximate measurement of $Q_\infty$, while acknowledging that in all cases this is an overestimate. Plotting $Q_\infty$ against temperature, as in Fig.\ \ref{fig:TTSP}(b), we find a power law, $Q_\infty \sim T^{-0.48}$. At first glance this appears to be another point of quantitative agreement with MCT (specifically the F2 version), which predicts a power-law scaling of the non-ergodicity parameter (analogous to $Q_\infty$) with exponent $-0.5$ \cite{Gotze1999, Leutheusser1984}. However, caution is required. The weights of a regularised network are constrained by the quadratic term in the loss. As shown in Appendix \ref{sec:app_q_infty}, for a landscape of the form $\mathcal{L} = \frac{\lambda}{2} |\mathbf{W}|^2$, and the appropriate parameter values from our DNNs, the resulting values of $Q_\infty$ are similar to those measured in Fig.\ \ref{fig:TTSP}(b). As such, the $Q_\infty \sim T^{-0.48}$ power law is a simple result of L2-regularisation, not a signature of glassy behaviour.

A similar analysis cannot be performed for the over- or underparameterised phases. $Q(t)$ in the overparameterised phase does not leave the master curve during our experimental window. This is to be expected, as we believe it is the effect of gradients in the landscape that breaks the TTSP in this system. In the underparameterised phase $Q(t)$ does leave the master curve, but not in a well controlled way, as shown in Appendix \ref{sec:app_q_infty}.

\subsection{Caging}
Caging is a fundamental aspect of the glass transition where relaxation is temporarily arrested \cite{Charbonneau2012,Charbonneau2014,Weeks2002,Jack2005,Biroli2021}. This process leads to a typical signature in the mean square displacement (MSD) of glassy particles where at short times the ballistic MSD goes as $t^2$, then there is an intermediate-time plateau from the cage effect, followed by a long-time diffusive regime which goes as $t^1$ (or $t^1$, plateau, $t^1$ in the case of Brownian simulations \cite{Tokuyama2008}) \cite{Schroder2020}. The signature plateau of caging is not unique to the MSD and can also be seen in many other correlation functions. In structural glasses, the intuitive explanation of caging is where the motion of a particle becomes constrained by steric repulsion with its neighbours \cite{Janssen2018}. Interestingly, temporarily arrested relaxation is also observed in spin glasses, despite the lack of steric interactions \cite{Crisanti2015,Ferrari2012,Baity-Jesi2019}. This has been explained by arguing that the dynamics of a spin glass exhibit a dramatic slow down near the glass transition temperature as the system navigates saddle points with very few unstable directions \cite{Cavagna2003,Castellani2005}. In this picture, the glass transition corresponds to a transition from a landscape dominated by saddles ($T>T_g$) to one dominated by minima ($T<T_g$) \cite{Cavagna2001}. 

\begin{figure}[h]
\includegraphics[width=8.8cm]{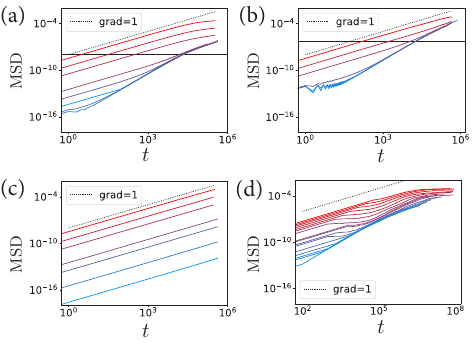}
\caption{The MSD of (a) regularised, (b) underparameterised, and (c) overparameterised networks. The dotted black line shows a gradient of 1, i.e. normal diffusion. The solid black line is the MSD corresponding to the value of $\epsilon$ used in Fig.\ \ref{fig:Langevin_tau90}. The network and loss parameters are the same as those in Fig.\ \ref{fig:Langevin_tau90}(a), (b), and (c) respectively, and $t_w=2\times10^4$. The temperatures used in (a) are $T=1$, $0.5$, $0.1$, $0.05$, $0.01$, $0.005$, $0.001$, $5\times10^{-4}$, $10^{-4}$, $10^{-5}$, $10^{-6}$, $10^{-8}$, $10^{-10}$, $0$. In (b) they are $T=1$, $0.5$, $0.1$, $0.05$, $0.01$, $0.005$, $0.001$, $5\times10^{-4}$, $10^{-4}$, $10^{-5}$, $10^{-6}$, $0$. In (c) they are $T=1$, $0.5$, $0.1$, $0.05$, $0.01$, $0.005$, $0.001$, $5\times10^{-4}$, $10^{-4}$, $10^{-5}$, $10^{-6}$, $10^{-8}$, $10^{-10}$. In all three cases the temperatures run from high (red, top curves) to low (blue, bottom curves). Panel (d) shows the MSD in the regularised phase (same parameters as (a), but $t_w=6\times10^4$) for networks trained with SGD. The localisation of the networks at long times is visible. The batch sizes used are $b=1$, $2$, $4$, $8$, $16$, $32$, $64$, $100$, $128$, $300$, $600$, $1000$, $3000$, $6000$, $10000$, $60000$. Small batch sizes, corresponding to high noise, are the high curves (red), and large batch sizes are the low (blue) curves.}
\label{fig:MSD}
\end{figure}

Once again, we ask the question: Does the typical caging behaviour of glasses occur in DNNs? We define the weight MSD in an analogous way to particle positions as
\begin{align}
\text{MSD}(t, t_w) = \left \langle \frac{1}{N}\sum_{i=1}^N [w_i(t+t_w)-w_i(t_w)]^2 \right \rangle,\label{eq:MSD} 
\end{align}
where $t_w$ is the waiting time since the system was initialised, as defined for Eq.~\eqref{eq:VH_def}. As shown in Fig.\ \ref{fig:MSD}, we see no evidence of caging. In Fig.\ \ref{fig:MSD} (a), (b), and (c) the MSD is plotted for regularised, underparameterised, and overparameterised networks respectively, subject to overdamped Langevin dynamics. In all three cases, at high temperatures and intermediate times the MSD demonstrates simple diffusive behaviour with a slope of $1$. In the regularised and underparameterised phases the MSD curves begin to collapse onto a single curve below $T^\ast$, i.e., the inflection temperature from the decorrelation plots in Fig.\ \ref{fig:Langevin_tau90}. For comparison with the decorrelation plots an MSD equivalent to $\epsilon^2$ is shown by the horizontal line. Below $T^\ast$ the dynamics are dominated by the gradients of the loss landscape, hence the MSD becomes temperature insensitive, and no longer follows a simple power law in $t$. 

Landscape effects are not present in the overparameterised phase as the networks exist in the flat, $\mathcal{L}\approx0$ plateaus. As such the stochastic part of the dynamics always dominates, and the MSD is always diffusive. Furthermore, overparameterised networks trained at $T=0$ undergo almost no motion in weight space, achieving an MSD of around $10^{-16}$ after $10^5$ timesteps. We do not include this data in Fig.\ \ref{fig:MSD}(c) as it falls below floating point precision. 

At low temperatures and short times there are oscillations in the MSD in the regularised and underparameterised phases. This is a numerical artefact due to gradients in the loss varying over length scales of a similar size to the learning rate (i.e. step size) used for training. The oscillations disappear when a smaller learning rate is used (not shown), however the long time sections of the curves then become unobtainable. It is also noteworthy that at low temperatures and short times there is a small plateau in Fig.\ \ref{fig:MSD}(a) and (b), reminiscent of the cage escape plateau. However, the plateau is also present at $T=0$, hence we conclude the plateau is an effect of the landscape dynamics, not thermal noise driving the networks over a barrier. 

At long times and high temperatures the MSD becomes sub-diffusive in the regularised phase. We can study the long-time limit further by using SGD instead of overdamped Langevin dynamics. The use of batches in SGD drastically speeds up training, however the price we pay is that batch noise is quite dissimilar to thermal noise \cite{Simsekli2019,Xie2020}, hence none of the curves are straight lines. With this caveat in mind, the SGD MSD of regularised DNNs is shown in Fig.\ \ref{fig:MSD}(d). Localisation occurs at $\text{MSD}\approx10^{-4}$, which is what we would expect from the quadratic term in Eq.~\eqref{eq:loss} (see Appendix \ref{sec:quad_MSD}). As such, it is unsurprising that the localisation effect is absent in both under and overparameterised DNNs. 

We hypothesise that the root cause of the lack of caging in DNNs is due to the geometrical complexity of the basins of attraction in the loss landscape. The energy landscape of a structural glass is made up of topologically simple, distinct basins that are roughly spherical in shape as they are formed by the repulsive interactions of the particles acting on length scales of approximately one particle diameter. These basins have a typical associated length scale, and the barriers separating them a typical height. Consequently the signature plateau of caging is visible in a predictable way when the dynamics are resolved on length scales at least as small as the cage size, and measurements are made over timescales at least as long as the cage-escape time. The situation with the DNN loss landscape is much more complicated. There is no reason for the basins of attraction to have a simple structure, typical length scale, or typical escape time. They are likely to be geometrically complex (e.g.\ being much larger in some dimensions than others) as well as containing topologically non-trivial level sets. The recent work of \cite{Draxler2018} demonstrating the existence of nearly flat paths between distant loss minima, as well as that of \cite{Annesi2023} mapping the high degree of connectivity between solutions of the spherical negative perceptron, also suggest the basins have a highly extended shape. 

What about caging in spin glasses? Here the energy landscape is divided into a high temperature region of saddle points, and a low temperature region of minima, with caging occurring at saddles near the boundary as the number of unstable modes gets very small. Figure \ref{fig:MSD} suggest this picture does not apply to DNNs and raises the interesting question of whether the topology of the DNN loss and glass energy landscapes are different, or whether geometric features are sufficient to explain the lack of caging. 

\subsection{The Stokes-Einstein relation and Dynamic Heterogeneity}
The Stokes-Einstein relation is the most famous example of a fluctuation-dissipation theorem, and relates the diffusion constant of a colloidal tracer particle, $D$, to the viscosity, $\eta$, of the surrounding fluid, according to
\begin{align}
D \propto \frac{k_BT}{\eta} \label{eq:Stokes-Einstein},
\end{align}
where the denominator comes from the Stokes drag equation, and the exact constant of proportionality is system dependent \cite{Einstein1905}. In colloidal glasses (which obey overdamped Langevin dynamics like Eq.~\eqref{eq:Langevin}) the effective viscosity of the fluid-colloid mixture, $\eta_{\text{eff}}$, is used instead of the bare viscosity of the background fluid \cite{Banchio1999}. It is a well known signature of the glass transition that the Stokes-Einstein relation begins to be violated as the system enters the supercooled regime (the temperature regime where the system begins to fall out of equilibrium, but has not yet cooled sufficiently to be considered a true glass) \cite{Tarjus1995,Shi2013,Mei2019}. The cause of this violation is thought to be associated with non-Gaussian fluctuations of the particles' displacement distribution, as measured by the non-Gaussian parameter (or excess kurtosis), $\alpha_2$ \cite{Flenner2005a,Kob1997a,Donati1998}, and the emergence of dynamic heterogeneity (DH). DH is another typical glassy phenomenon, whereby the system begins to separate into spatially localised populations of mobile and non-mobile particles \cite{Mei2019,Flenner2014,Sengupta2013}. The mobile particles dominate the diffusion, whereas the non-mobile particles dominate the viscosity, resulting in the decoupling of these two processes, and the violation of Eq.~\eqref{eq:Stokes-Einstein}.

\begin{figure}[h]
\includegraphics[width=8.8cm]{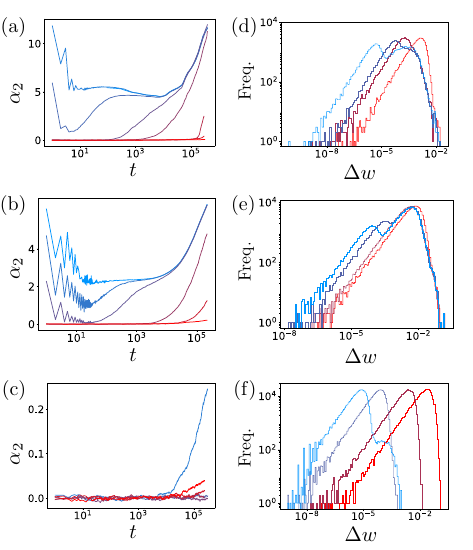}
\caption{The non-Gaussian parameter, $\alpha_2$, deviates from 0 and hence demonstrates non-Gaussian behaviour in the weight displacements of the regularised (a), underparameterised (b), and overparameterised (c) phases. The measurement parameters are the same as in Fig.\ \ref{fig:Langevin_tau90}. The temperatures used in (a) are (going from red, lowest curves, to blue, highest curves) $T=0.1$, $0.01$, $0.001$, $10^{-5}$, $10^{-6}$, $10^{-8}$, $10^{-10}$, $0$, in (b) they are $T=0.1$, $0.01$, $0.001$, $10^{-5}$, $10^{-6}$, $0$, and in (c) they are $T=0.1$, $0.01$, $0.001$, $10^{-5}$, $10^{-6}$, $10^{-8}$. Histograms of the weight displacements are plotted in the right-hand column, for the regularised (d), underparameterised (e), and overparameterised (f) phases. All phases exhibit bimodal distributions at low temperatures. The temperatures (going from red, right-most curves, to blue, left-most curves) in (d) are $T=10^{-3}$, $10^{-5}$, $10^{-6}$, $10^{-8}$, in (e) they are $T=10^{-2}$, $10^{-3}$, $10^{-5}$, $10^{-6}$, and in (f) they are $T=10^{-1}$, $10^{-3}$, $10^{-6}$, $10^{-8}$. For the distributions in (d), (e), and (f), $\Delta w(t, t_w)$ was evaluated at $t=2.2\times10^5$. For all plots $t_w=2\times10^4$.}
\label{fig:SE}
\end{figure}

To what extent does this phenomenology also occur in DNNs? The overdamped Langevin dynamics of Eq.~\eqref{eq:Langevin} result in a Fokker-Planck equation with diffusion constant $D=\mu kT/P_D$ \cite{Risken_FP_book}, as any Brownian process does. This expression can be rewritten replacing the mobility, $\mu$, with an ``effective'' drag, $\gamma = 1/\mu$, resulting in the (trivial) recovery of the fluctuation-dissipation theorem $D=kT/\gamma P_D$. However, there is no real justification for introducing the concept of a viscosity like in Eq.~\eqref{eq:Stokes-Einstein}, to a system with no colloidal particle, and no background fluid. Despite this, we can still look for deviations from Gaussian behaviour by measuring the non-Gaussian parameter,
\begin{align}
\alpha_2(t, t_w) = \frac{\langle \Delta w^4(t, t_w) \rangle}{3\langle \Delta w^2(t, t_w) \rangle^2}  - 1,
\end{align}
where $\Delta w(t, t_w) = |w(t) - w(t_w)|$ are the weight displacements \cite{Weeks2000}. $\alpha_2$ is 0 for Gaussian fluctuations in $\Delta w$. As can be seen in Fig.\ \ref{fig:SE} (left-hand column), $\alpha_2$ does in fact deviate from 0 for our DNNs, with the deviation starting at long times for high temperatures, and coming to dominate the whole time window as the temperature decreases. This is to be expected as it is the landscape term in Eq.~\eqref{eq:Langevin} that causes the non-Gaussian behaviour. The $\alpha_2$ curves are quite different from those measured in structural glasses, where $\alpha_2$ falls again at long times as cage hopping restores Brownian motion \cite{Flenner2005b}. Our data does not reach long enough times to see if this occurs in DNNs, however given the lack of caging we suggest it is unlikely.

We next test for the presence of DH by plotting the distribution of weight displacements, $\Delta w$, at a sequence of decreasing $T$ values. For structural glasses, a bimodal distribution of mobile and non-mobile particles emerges as $T$ approaches the glass temperature \cite{Flenner2005a}. As shown in Fig.\ \ref{fig:SE} (right-hand column), a bimodal distribution also emerges for DNNs in all three phases. The presence of DH in DNNs is surprising, as DH in structural glasses is typically understood as a result of spatially correlated movements of groups of particles interacting via steric repulsion between neighbours \cite{Scalliet2022,Flenner2014,Flenner2010}. In contrast, the weights of a DNN interact by complicated, indirect couplings, and their neighbour relations (or equivalently the topology of the space they inhabit) differ considerably from particles packed in 3D Euclidean space.  

\subsection{Ageing}\label{sec:aging}
\begin{figure}[h]
\includegraphics[width=8cm]{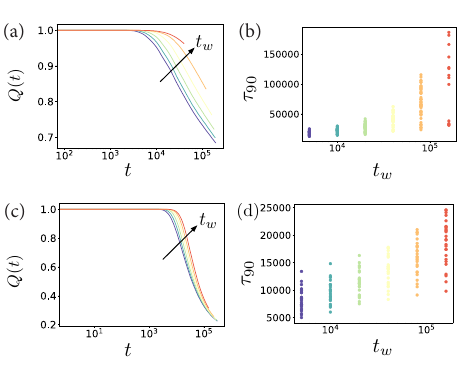}
\caption{The overlap function of networks as they decorrelate in (a) the regularised phase, and (c) the underparameterised phase. The arrows denote increasing $t_w$. In panels (b) and (d) $\tau_{90}$ is plotted against the waiting time $t_w$ at $T=10^{-6}$ (regularised) and $T=10^{-4}$ (underparameterised). The featured waiting times in both phases are $t_w\in \{2\times10^4, 5\times10^4, 8\times10^4, 10^5, 1.5\times10^5 \}$. All other parameters are the same as in Fig.\ \ref{fig:Langevin_tau90}.}
\label{fig:aging}
\end{figure}

Ageing, the phenomenon where the behaviour of a system depends explicitly on its age or the waiting time $t_w$ (e.g.\ the time following a quench) is another hallmark of glassy systems. As $T$ decreases towards the glass transition temperature, glass-forming systems fall out of equilibrium, and the resulting broken ergodicity produces ageing effects. The effect of ageing on $Q$ is shown for the regularised and underparameterised phases in Fig.\ \ref{fig:aging} (a) and (c) respectively, as well as the dependence of $\tau_{90}$ on $t_w$ in those phases in (b) and (d). Both the regularised and underparameterised regimes exhibit ageing in the landscape-dominated regime $(T<T^\ast)$, though the effect is stronger in the regularised phase. In both phases a longer $t_w$ results in slower dynamics, as it does in glasses. At $T>T^\ast$ the networks behave like an equilibrium system and there is no ageing (not shown). The overparameterised phase does not have a transition from noise-dominated to landscape-dominated dynamics as the networks exist in flat plateaus. As such, overparameterised networks never exhibit ageing, as demonstrated in Fig.\ \ref{fig:aging_overparam}. With the current data it is impossible to make a quantitative statement about the ageing behaviour (e.g.\ the power law ageing documented in glasses \cite{Kob1997, Hodge1995, Angelini2013}), and varying $t_w$ over several more orders of magnitude is computationally unfeasible. However, it is interesting to note that the authors of \cite{Segura2022a} measure sub-aging behaviour in DNNs trained with SGD, where much larger training times are achievable.

\section{\label{sec:Discussion} Discussion}
At first glance, glasses and DNNs share many important similarities, and yet DNNs fail to exhibit an unambiguous glass phase. Despite the existence of two glass-like transitions, the TTT and UOT, DNNs do not have diverging relaxation times - the defining characteristic of a glass. This is despite the existence of a critical temperature below which the dynamics of the DNNs fall out of equilibrium, are dominated by the landscape, and exhibit non-Gaussian fluctuations, DH, and ageing. We conclude that the disorder inherent in the loss landscape of these DNNs is not sufficient to induce a dramatic slowdown in the dynamics. In fact, our measurements suggest that the differences in the detailed structure of the glass and DNN landscapes, in particular the lack of a cage length in the latter, is enough to eliminate crucial glassy phenomena. This latter point is particularly relevant to the wider field of deep learning as this study was conducted with real image data, and using realistically large networks (consisting of tens of thousands of weights), and as such we expect the investigated loss landscapes to be representative of other ``real-life'' loss landscapes. 

The presence of DH in DNNs raises the intriguing question of whether there is some ``spatial'' organisation of the mobile weights, as there is in structural glasses. Such structure could have important practical applications in the sub-field of DNN pruning. Pruning, i.e. reducing a network's size with minimal reduction in its performance, is an area of growing importance as the immense number of parameters in modern networks makes them impractical for use on mobile phones or smart devices, as well as consuming worrying amounts of energy \cite{LeCun1990,Han2015,Blalock2020}. The presence of two distinct populations of weights, coupled with the insights of disordered physics, could lead to innovative new approaches to pruning. 

A challenge inherent in all numerical deep learning investigations, including our own, is the large number of hyperparameters, architectures, minimisation algorithms, and datasets in use across the field. Furthermore, due to the significant computational cost of training large networks, we can only ever hope to sample a small region of this experimental space. For example, we chose not to address finite size effects (e.g. by varying the network and dataset size while keeping $N/P_D$ constant) as the present study was already at the computational limits of what can be achieved on a university computing cluster. Where possible we have used physical arguments to choose particular parameters (such as the similarity between the quadratic hinge loss and short-range repulsive interactions), and have provided evidence as to the generality of our findings by repeating them on the CIFAR10 dataset. We hope our work stimulates further research on this topic, extending it to a broader range of deep learning systems.

The relationship between deep learning and structural glasses is complicated, with both surprising points of agreement (e.g.\ DH) and disagreement (e.g.\ the lack of diverging timescales). Looking beyond structural glasses, the wider field of the physics of disordered systems potentially has a lot to offer as a theoretical blueprint for deep learning. The Random Lorentz Gas (RLG) and Anderson Localisation both suggest themselves as useful, and as of yet unexplored, physical analogies to DNNs as their complex landscapes result in a localisation transition as opposed to the caging seen in structural glasses. Furthermore, the link between the localisation transition of the RLG and percolation theory is well known, as is the existence of a fractal void space near the percolation transition  \cite{Elam1984,Hofling2006,Spanner2016,Hofling2008,Bauer2010}. Perhaps similar phenomena can be discovered in the connectedness of plateaus in the overparameterised phase. Exploring such topics further would be a highly novel direction of research, and further enrich the exciting work currently being done at the interface of deep learning and statistical physics. 

\section{Acknowledgements}
We warmly thank Ilian Pihlajamaa and Corentin C. L. Laudicina for their detailed reading of this manuscript, and Thomas Voigtmann, Kees Storm, Wouter Ellenbroek, and Sophie de Hont for their insightful discussions and advice. We acknowledge the Dutch Research Council (NWO) for its financial support through a START-UP grant.

\appendix

\renewcommand{\thefigure}{A\arabic{figure}}

\setcounter{figure}{0}

\section{Temperature in networks with overdamped Langevin dynamics}\label{sec:app_temP_Def}
We wish to construct the overdamped Langevin dynamics of a network to be as similar as possible to that of a physical system. The degrees of freedom, the weights, are analogous to the particle positions. The energy of a system is an extensive quantity, whereas the loss, being an average over the data, is intensive. Consequently, the quantity $P_D \mathcal{L}$ plays the role of the energy. 

We can see this argument more clearly by considering the overdamped Langevin equation for the weights of a DNN,
\begin{align}
\partial_t w_i = -\mu \partial_{w_i}\mathcal{L} + \xi_i,\label{eq:app_Langevin}
\end{align}
where $\langle \xi_i(t)\xi_j(t') = 2D \delta_{ij} \delta(t-t')$. At this stage we do not specify how the diffusion constant, $D$, relates to temperature. The probability distribution of weights that obey such an overdamped Langevin equation, $\rho$, evolves according to the Fokker-Planck equation \cite{Risken_FP_book},
\begin{align}
\partial_t \rho = -\nabla (\mu \rho \nabla \mathcal{L} - D\nabla \rho)\label{eq:app_Fokker-Planck}.
\end{align} 
The steady state of this equation is given by
\begin{align}
\frac{d\rho}{d\mathcal{L}} = -\frac{\mu}{D}\rho,
\end{align}
with solution $\rho=\rho_0 \exp\left(-\mu \mathcal{L}/D\right)$. In order to mimic physics, we wish to choose the noise on Eq.~\eqref{eq:app_Langevin} such that the steady state solution of Eq.~\eqref{eq:app_Fokker-Planck} is the Boltzmann distribution with ``energy'' $U=P_D\mathcal{L}$. This results in
\begin{align}
e^{-\frac{\mu \mathcal{L}}{D}} = e^{-\frac{U}{k_B T}} = e^{-\frac{P_D\mathcal{L}}{k_B T}},
\end{align}
and hence we conclude that $D = \frac{\mu kT}{P_D}$, a version of the familiar fluctuation dissipation relation but taking into account the intensive nature of $\mathcal{L}$, and $\langle \xi_i(t)\xi_j(t') \rangle = (2 \mu k_B T/P_D)\delta_{ij}\delta(t-t')$, as stated in the main body of the text.

\section{Networks above and below the TTT}\label{sec:app_TTT}
\begin{figure}[h]
\includegraphics[width=8cm]{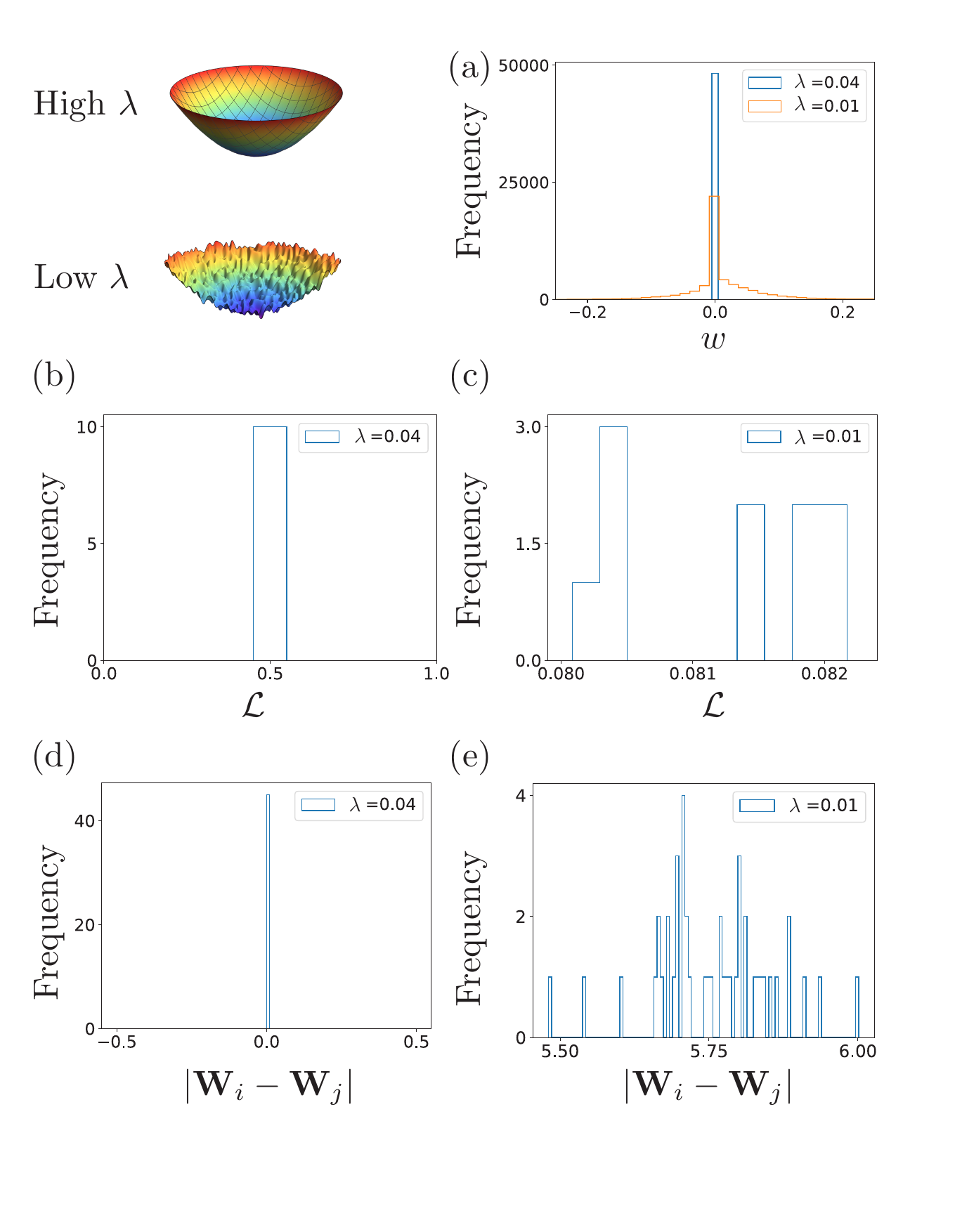}
\caption{Networks above and below the TTT. (a) The distribution of weights. (b), (c) The distribution of final training loss values above and below the TTT respectively. (d), (e) The distribution of Euclidean distances between pairs of networks in weight space, above and below the TTT respectively.}
\label{fig:TTT_exploration}
\end{figure}

In Fig.~\ref{fig:TTT_exploration} we measure several characteristics of DNNs above and below the TTT. We train 10 networks at $\lambda=0.04$ (high regularisation, above the TTT), and 10 at $\lambda=0.01$ (low regularisation, below the TTT), initialised with weights randomly drawn from a uniform distribution of width $2/\sqrt{N_L}$, where $N_L$ is the layer width (this is the default PyTorch initialisation). All networks were trained for 7000 epochs using SGD with a learning rate $\mu=10^{-3}$ and batch size 100, followed by 100 epochs of gradient descent to reach their local minima. In Fig.~\ref{fig:TTT_exploration}(a) we plot the distribution of weights, demonstrating that above the TTT $w_i\approx0$ $\forall i$, whereas below the TTT, the weights adopt a broad range of values. We plot the distribution of final training loss values above the TTT in Fig.~\ref{fig:TTT_exploration}(b), and below it in Fig.~\ref{fig:TTT_exploration}(c). These plots demonstrate that above the TTT $\mathcal{L}\approx 0.5$, as expected for the quadratic hinge loss (Eq.~\eqref{eq:quad_hinge_loss}) evaluated at $f=0$, whereas below the TTT the loss adopts a spread of different values. Finally, in Fig.~\ref{fig:TTT_exploration}(d) and (e), we calculate the Euclidean distance in weight space between all pairs of networks, $|\mathbf{W}_i-\mathbf{W}_j|$, where $\mathbf{W}_i$ is the vector of weights of network $i$. Above the TTT (Fig.~\ref{fig:TTT_exploration}(d)) all networks exist in the $\mathbf{W}=\mathbf{0}$ state, hence the distance between them is zero. Below the TTT, the networks fall into distinct minima, and the distances between them adopt a spread of values. In the main text we use $\mathcal{L}\approx0.5$ to indicate that the DNNs are in the quadratic state (i.e. above the TTT), as the loss is a convenient and computationally efficient observable. It is highly unlikely, though not impossible, that an ensemble of networks not in the quadratic state, could also all have $\mathcal{L}\approx0.5$.

\section{Choosing $\epsilon$}\label{sec:app_epsilon}
The overlap function defined in Eq.~\eqref{eq:VH_def} relies on an appropriate definition of the parameter $\epsilon$. There is no natural length scale associated with the weights of a DNN, hence we use the standard deviation of weights at the end of training, $\sigma$, and define $\epsilon$ as
\begin{align}
\epsilon = \frac{\sigma}{\chi},
\end{align}
where the value of the parameter $\chi > 0$ can be chosen freely within some reasonable bounds. We have verified experimentally that $\sigma$ varies by at most roughly 1\% between networks in the same phase, although it does vary more between phases. Choosing a smaller $\chi$ slows down the decorrelation of $Q(t)$, hence we are not so interested in a lower bound for $\chi$ stricter than $\chi>0$. Choosing a larger $\chi$ speeds up decorrelation, which has significant practical advantages in that we do not need to train our DNNs for such a long time. What is the maximum value of $\chi$ we can reasonably choose? The values in a Pytorch DNN are stored as single precision floating point numbers with a resolution of approximately $1\times10^{-7}$, i.e. the next number after 1 is approximately $1+10^{-7}$. Consequently we place an upper bound on $\chi$ such that $\epsilon>10^{-7}$.  

\begin{figure}[h]
\includegraphics[width=8cm]{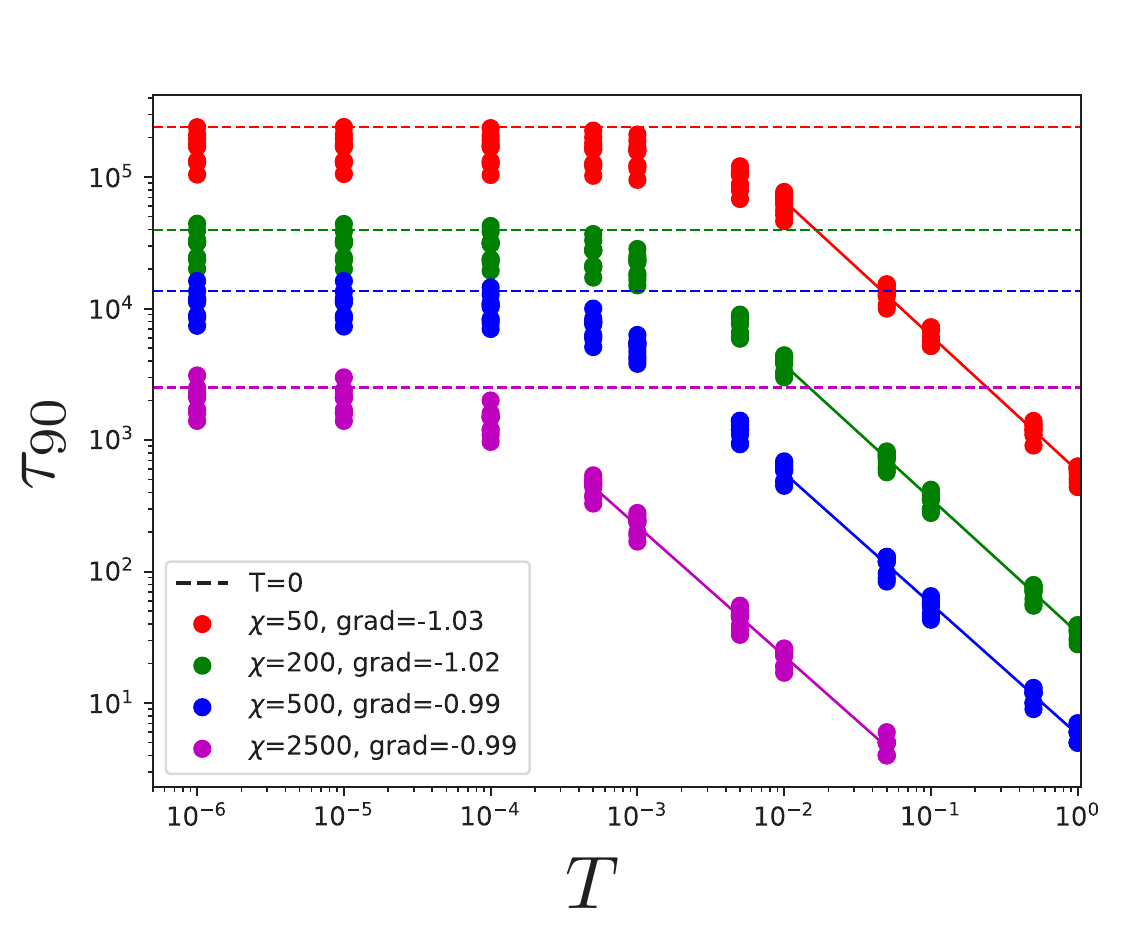}
\caption{The dependence of the decorrelation time on $\chi$. As $\chi$ is increased the networks decorrelate more quickly, and the transition point between noise and landscape-dominated dynamics occurs at lower temperatures. However, the overall shape of the curves is unaffected.}
\label{fig:varying_epsilon}
\end{figure}

How does the choice of $\epsilon$ affect our measurements of $Q(t)$, and in particular the identification of two distinct regimes in the decorrelation times of overdamped Langevin networks? A very large $\epsilon$ increases $\tau_{90}$ beyond the maximum training time of our networks, whereas a very small $\epsilon$ decreases $\tau_{90}$ below the resolution at which we save data (which varies from every epoch at short times to every 100 epochs at long times). Between these two extremes varying $\epsilon$ simply shifts the $\tau_{90}$ vs $kT$ curve as shown in Fig.\ \ref{fig:varying_epsilon}. In this figure we vary $\chi$ by a factor of 50 without changing the shape of the curve. The critical value of $kT$ at which networks transition between the two dynamic regimes changes by about an order of magnitude, so whenever we make quantitative comparisons we do so at a fixed $\chi$.

In general the overlap function in spin glasses is not defined with respect to a particular length scale, and yet avoids introducing a parameter like $\epsilon$. This is because the spins are Ising-like binary variables, hence simply taking an inner product like $\mathbf{s}_1 \cdot \mathbf{s}_2$ results in fast decorrelation. The exception is the p-spin model, which uses continuous spins. However, in this model the strict spherical constraint, $|\mathbf{s}|=1$, also ensures fast decorrelation. In our case of DNN weights, defining the overlap as the inner product of the weights results in very slow decorrelation, and as such we turn to the definition in the main text. 

\section{Transitions with CIFAR10}
\begin{figure}[h]
\includegraphics[width=8.8cm]{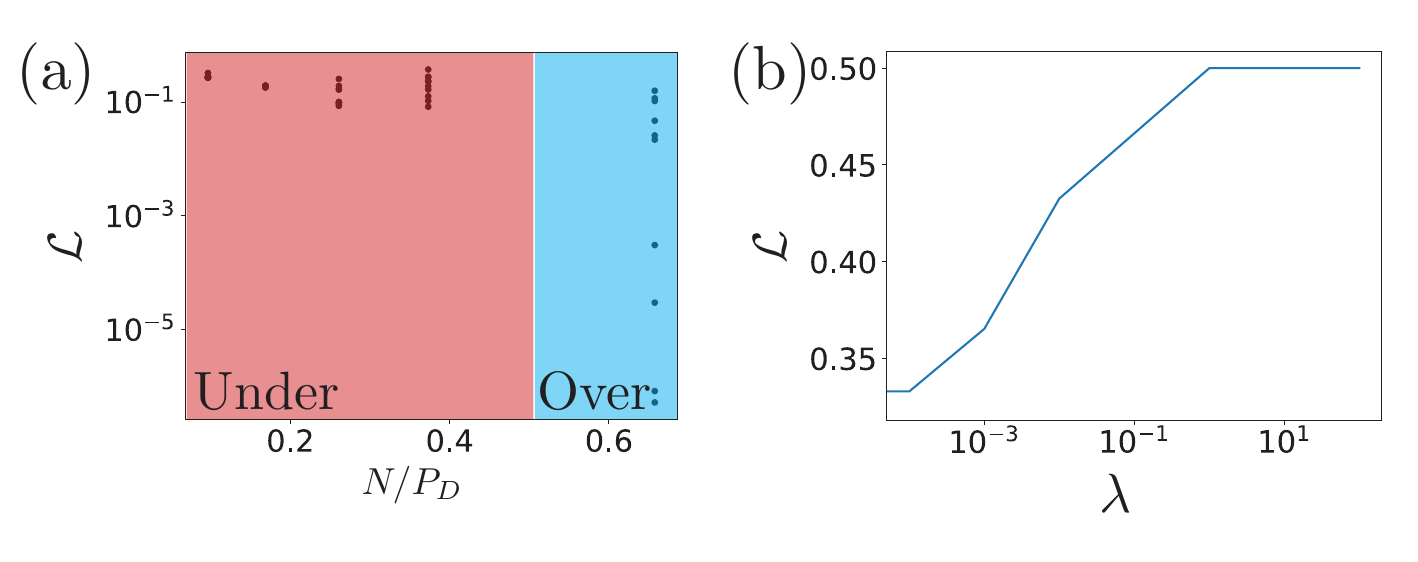}
\caption{(a) The UOT for CIFAR10 data. (b) The TTT for CIFAR10 data.}
\label{fig:cifar10_trans}
\end{figure}

The UOT is shown for CIFAR10 data in Fig.\ \ref{fig:cifar10_trans}(a), and the TTT is shown in Fig.\ \ref{fig:cifar10_trans}(b). 
 
\section{Further decorrelation plots}\label{sec:app_cifar10_tau90_plots}
\begin{figure}[h]
\includegraphics[width=8cm]{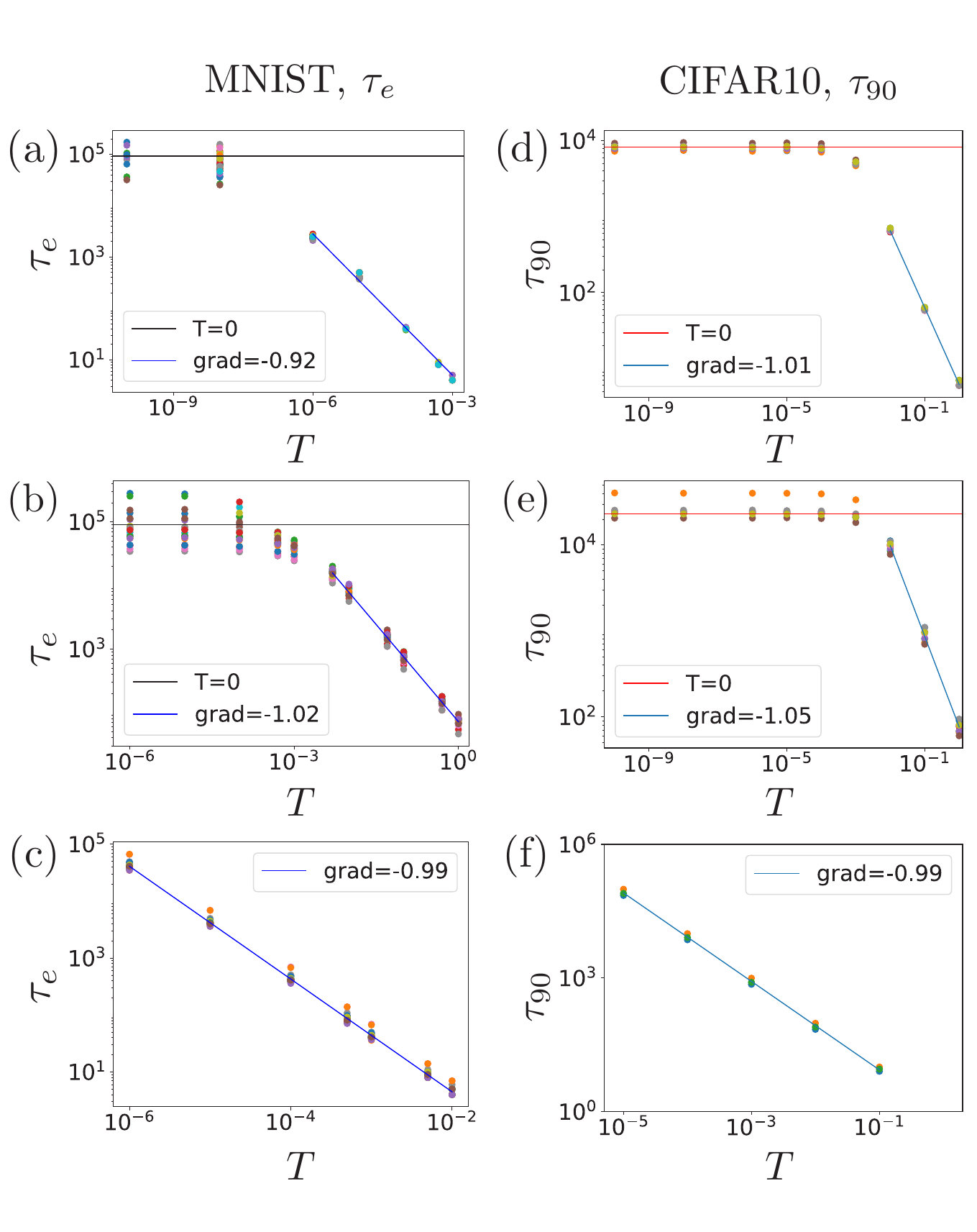}
\caption{Left-hand column: The e-fold decorrelation time vs temperature in (a) regularised, (b) underparameterised, and (c) overparameterised networks trained on MNIST. Right-hand column: The dependence of $\tau_{90}$ on temperature in (d) regularised, (e) underparameterised, and (f) overparameterised networks trained on CIFAR10.}
\label{fig:taue_and_cifar10}
\end{figure}

We repeat the measurements of the decorrelation time as a function of temperature shown in Fig.\ \ref{fig:Langevin_tau90}. We first plot the e-fold relaxation time, $\tau_e$, as a function of $T$ for regularised, underparameterised, and overparameterised networks, shown in Fig.\ \ref{fig:taue_and_cifar10}. As with $\tau_{90}$, we see two dynamical phases in regularised and underparameterised networks. At high temperatures $\tau_e\sim 1/T$, and at low temperatures $\tau_e$ becomes $T$ independent. 

These findings are replicated by networks trained on the CIFAR-10 dataset \cite{CIFAR10_dataset}. Following the same steps as with MNIST, we convert CIFAR-10 to a binary classification problem by assigning labels of +1 to the categories `airplane', `bird', `deer', `frog', and `ship', and assigning -1 to `automobile', `cat', `dog', `horse', and `truck'. We also use PCA to reduce the dimension of the images to the first 10 PCA components. The $90\%$ relaxation times are shown in Fig.\ \ref{fig:taue_and_cifar10}.

\section{Calculation of $Q(t)$ for weights in a quadratic basin}\label{sec:app_overlap}
Here we consider the case of DNNs trained with overdamped Langevin dynamics within a loss landscape of a simple harmonic well, $\mathcal{L}=\frac{1}{2}k\sum_{i=1}^N w_i^2$, with some ``spring constant'' $k$. The resulting overdamped Langevin equation is
\begin{align}
\partial_t w_i = -\mu k w_i + \xi_i,
\end{align}
where the noise term obeys $\langle \xi(t) \rangle = 0$, $\langle \xi_i(t) \xi_j(t') \rangle = 2 \mu T \delta_{ij} \delta(t-t')$. Note that here we do not define the noise with a $1/P_D$ term as in the main text, as this loss function is independent of the dataset size. Such an overdamped Langevin equation results in the following Fokker-Planck equation for the weight probability distribution, 
\begin{align}
\partial_t P(w, t) = \partial_w \left(\mu k w P \right) + \mu T \partial_w^2 P,
\end{align}
with solution
\begin{align}
P(w, t) = A \exp \left[ \frac{-k\left( w-w(0) e^{-k\mu t} \right)^2}{2T\left( 1-e^{-2k\mu t} \right)}\right],
\end{align}
where 
\begin{align}
A = \left[ \frac{k}{2\pi T\left( 1-e^{-2\mu k t} \right)} \right]^{\frac{1}{2}}.
\end{align}
The overlap function is given by integrating over a region of size $2\epsilon$ around the initial condition $w(0)$,
\begin{align}
Q(t) &= \int_{w(0)-\epsilon}^{w(0)+\epsilon} P(w, t) dw,
\end{align}
which evaluates to
\begin{align}
Q(t)= \frac{1}{2}\Bigg \{ &\text{erf}\left[ \frac{w(0) \left( 1-e^{-k\mu t} \right) + \epsilon}{\sqrt{c}} \right] \nonumber \\ &- \text{erf}\left[ \frac{w(0) \left( 1-e^{-k\mu t} \right) - \epsilon}{\sqrt{c}} \right] \Bigg \},\label{eq:app_Qt}
\end{align}
where 
\begin{align}
\frac{1}{\sqrt{c}} = \left[ \frac{k}{2T \left( 1-e^{-2\mu kt} \right)} \right]^{\frac{1}{2}}.
\end{align}

\begin{figure}[h]
\includegraphics[width=8cm]{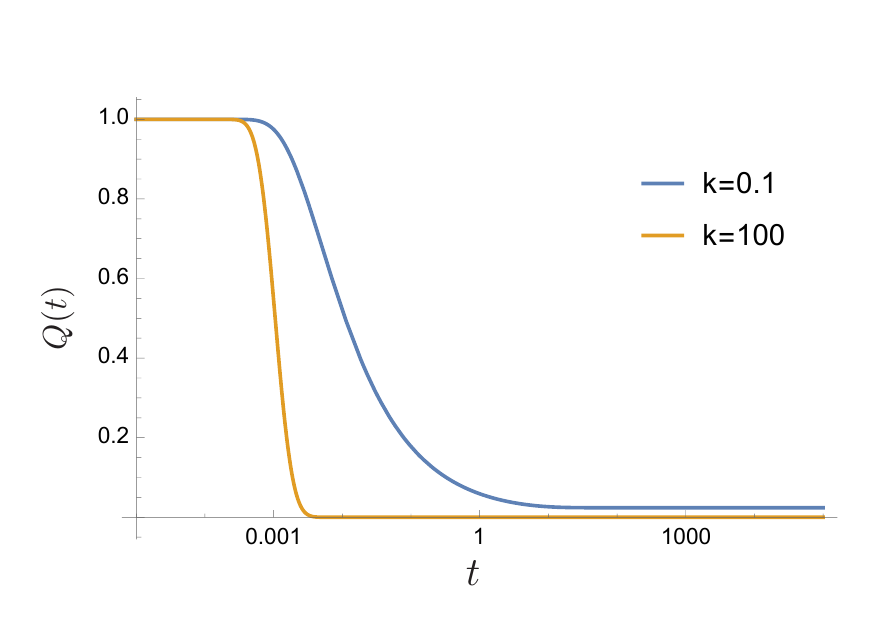}
\caption{The overlap function of Brownian weights in a quadratic well with high ($k=100$) and low ($k=0.01$) spring constant.}
\label{fig:quad_overlap}
\end{figure}

The effect of varying the spring constant of the quadratic well is dependent on $w(0)$. The case where $w(0)\gg \epsilon$, which is the relevant regime for modelling landscape-dominated dynamics which have fallen out of equilibrium,  is shown in Fig.\ \ref{fig:quad_overlap}, where a higher $k$ results in faster decorrelation. In Fig.\ \ref{fig:quad_overlap}, $w(0) = 1$, $\epsilon=0.1$, $2T=1$, and $\mu=1$.

\section{Further details of $\lim_{t \to \infty} Q(t)$}\label{sec:app_q_infty}
\begin{figure}[h]
\includegraphics[width=8cm]{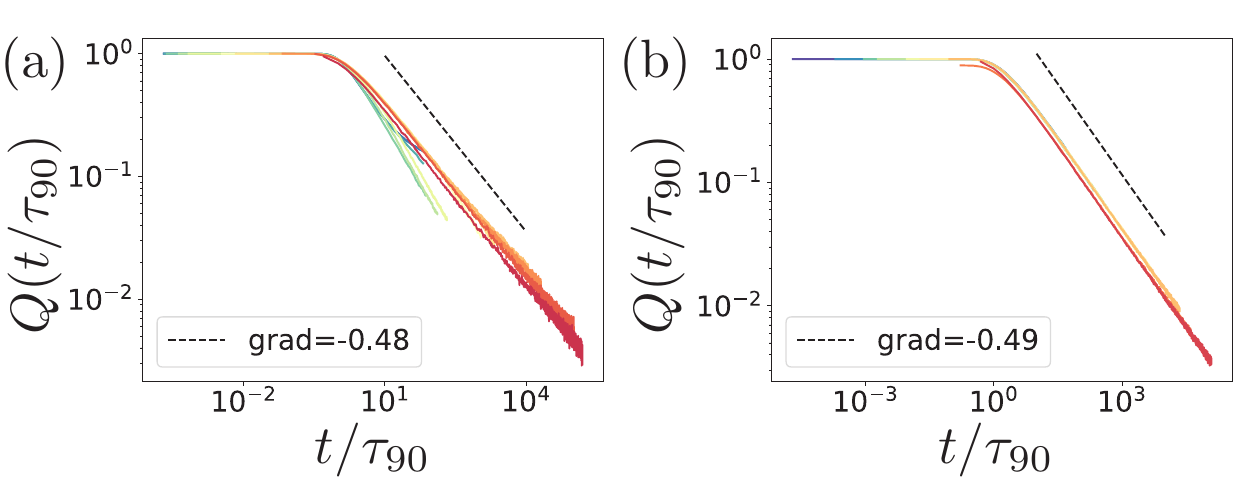}
\caption{Time-temperature superposition of $Q(t)$ in the underparameterised (a) and overparameterised (b) phases.}
\label{fig:app_TTSP}
\end{figure}

Although both the under and overparameterised phases also exhibit time-temperature superposition, as shown in Fig.\ \ref{fig:app_TTSP}, their behaviour is markedly different to the regularised phase. In the underparameterised phase, some temperatures eventually leave the master curve, and some do not. Those that do, do so in an irregular fashion, with individual curves leaving both above and below the master curve, in a temperature non-monotonic manner. As such, we cannot use the late time values of $Q(t)$ as an approximate measure of $Q_\infty$.

As for the overparameterised phase, the individual temperatures never leave the master curve. We believe this is because the breaking of the TTSP is due to landscape effects in the dynamics, and overparameterised DNNs exist in a gradient-less landscape.

An important question raised by the measurement of a power-law in Fig.\ \ref{fig:TTSP}(b) is, could the large, quadratic basin of the regularisation term cause the observed behaviour in $Q_\infty$? Starting with the results from the previous section, modifying the temperature like $T \to T/P_D$ to replicate the convention of the main text, and taking the long time limit of Eq.~\eqref{eq:app_Qt}, we get
\begin{align}
Q_\infty = \frac{1}{2}\Bigg \{& \text{erf}\left[ \sqrt{\frac{kP_D}{2T}} (w(0)+\epsilon) \right] \nonumber \\ & -\text{erf}\left[ \sqrt{\frac{kP_D}{2T}} (w(0)-\epsilon) \right] \Bigg\}.\label{eq:Q_infty}
\end{align}

\begin{figure}[h]
\includegraphics[width=8cm]{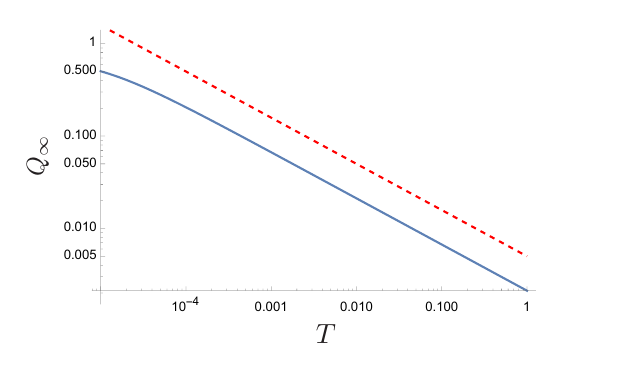}
\caption{The analytic result of $Q_\infty$ vs $T$ for weights in a quadratic bowl. The dashed red line is $T^{-0.5}$.}
\label{fig:app_Qf_analytic}
\end{figure}

Setting $k$, the spring constant, to the regularisation parameter $\lambda=0.01$, and using $w(0)$ and $\epsilon$ measured from the relevant DNNs, we can plot $Q_\infty$ vs $T$, as shown in Fig.\ \ref{fig:app_Qf_analytic}. We do indeed recover the $T^{-0.5}$ power-law, and similar numerical values to those measured in Fig.\ \ref{fig:TTSP}(b).

\section{MSD of weights in a quadratic well}\label{sec:quad_MSD}
Consider the case of weights evolving according to Eq.~\eqref{eq:Langevin}, in a quadratic well, $\mathcal{L}=\frac{1}{2}k\sum_{i=1}^N w_i^2$. The Fokker-Planck equation for the weights is
\begin{align}
\partial_t P = \partial_w(\mu k w P) + \frac{\mu T}{P_{d}} \partial_w^2 P,
\end{align}
where $P$ is the probability density of weights. The steady state distribution is
\begin{align}
P_{ss} = \sqrt{\frac{k P_D}{2\pi T}}\exp \left(-\frac{ k P_D w^2}{2T} \right).
\end{align}
We can then calculate the long time limit MSD as
\begin{align}
\text{MSD} = \int_{-\infty}^\infty P_{ss}(w) w^2 dw,
\end{align}
resulting in
\begin{align}
\text{MSD} = \frac{T}{k P_D}.
\end{align}

Subbing in $T=0.1$, $k=\lambda =10^{-2}$, and $P_D=6\times10^4$, which are the relevant values for the $T=0.1$ curve in Fig.\ \ref{fig:MSD}(a), results in an MSD of $1.6\times10^{-4}$, close to the value at which the MSD plateaus in Fig.\ \ref{fig:MSD}(d). So we conclude the weights are localised by the quadratic basin. This is in agreement with our calculation of $Q_\infty$.

\section{Non-Gaussianity and Dynamic Heterogeneity in CIFAR-10} \label{sec:app_cifar10_DH}
Using the DNNs trained on CIFAR-10 from Sec. \ref{sec:app_cifar10_tau90_plots}, we measure the non-Gaussian parameter, $\alpha_2$, and the distribution of weight motilities at a range of temperatures. These networks exhibit the same behaviour as the MNIST examples in the main text. All three phases have $\alpha_2 \neq 0$, starting at long times for high temperatures, then occurring at all times for low temperatures. This behaviour is less pronounced for the overparameterised phase. Note that the pronounced oscillations up to $t=100$ are due to the fact that the network states are saved at a higher frequency at shorter times. The weight motilities follow bimodal distributions, i.e. demonstrate DH, at low temperatures for the regularised and underparameterised, but not overparameterised, phases. 

\begin{figure}[h]
\includegraphics[width=8cm]{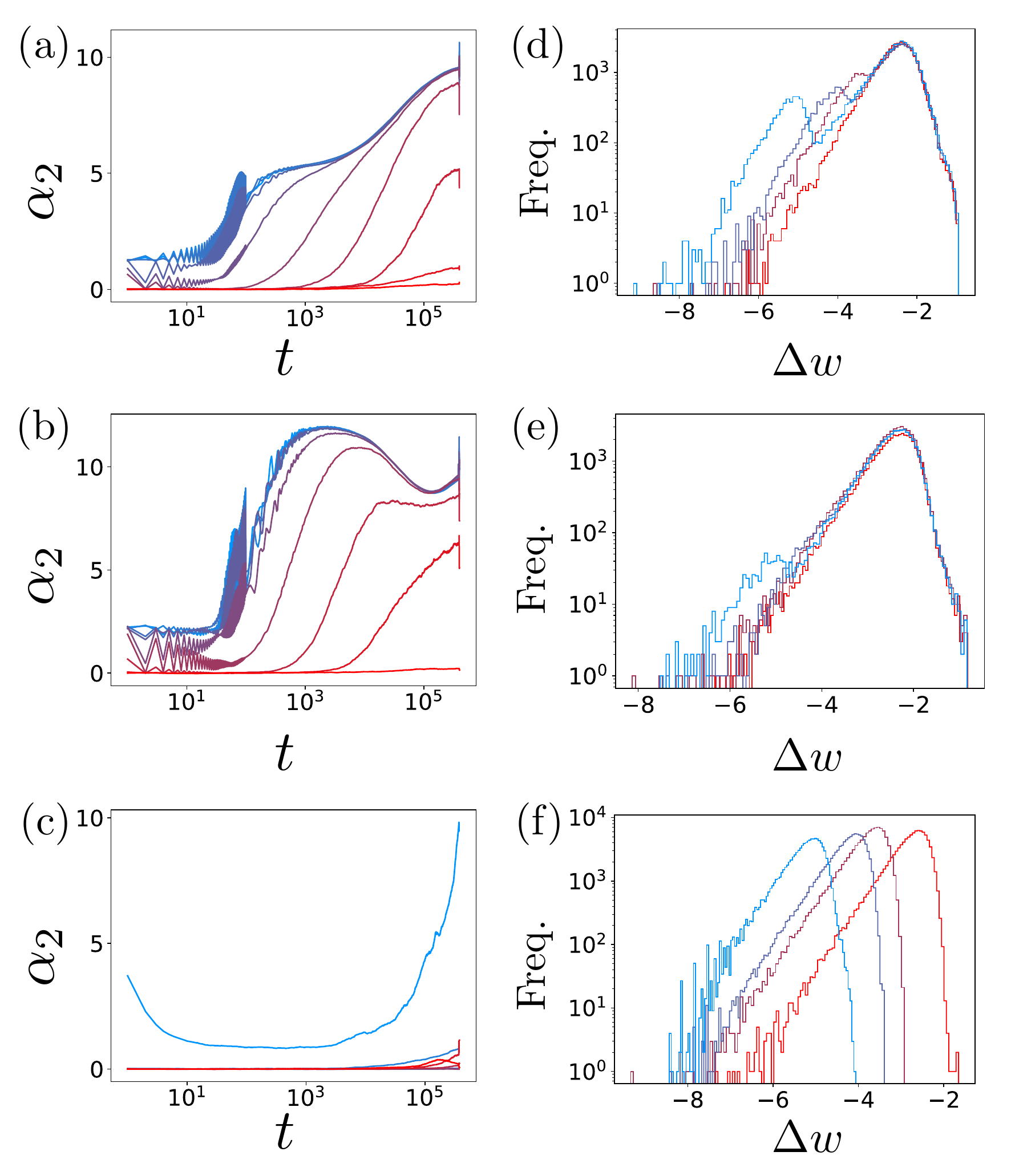}
\caption{The non-Gaussian parameter, $\alpha_2$, and weight motility distributions for DNNs trained on CIFAR-10. Non-Gaussian behaviour, corresponding to $\alpha_2 \neq 0$ occurs at long times for high temperatures, and all times for low temperatures in the regularised (a), underparameterised (b) and overparameterised (c) phases. The temperatures in (a) and (b), going from blue (high curves) to red (low curves) are $T=10^{-10}$, $10^{-8}$, $10^{-6}$, $10^{-5}$, $10^{-4}$, $10^{-3}$, $10^{-2}$, $10^{-1}$, $1$. The temperatures in (c) are $T=10^{-10}$, $10^{-8}$, $10^{-6}$, $10^{-5}$, $10^{-4}$, $10^{-3}$, $10^{-2}$, $1$. The regularised and underparameterised networks also have bimodal weight motility distributions (panels (d) and (e) respectively), unlike the overparameterised phase (f). The temperatures used in (d), (e), and (f), going from blue (left curves) to red (right curves) are $T=10^{-8}$, $10^{-6}$, $10^{-5}$, $10^{-3}$.}
\label{fig:cifar10_DH}
\end{figure}

\section{Aging effects} \label{sec:aging_appendix}
\begin{figure}[h]
\includegraphics[width=8cm]{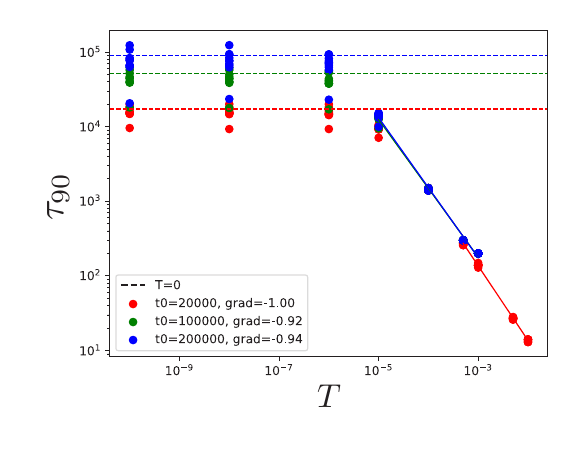}
\caption{The effect of waiting time on the $\tau_{90}$ vs $T$ curve in the regularised phase.}
\label{fig:aging_tau90}
\end{figure}

\begin{figure}[h]
\includegraphics[width=8cm]{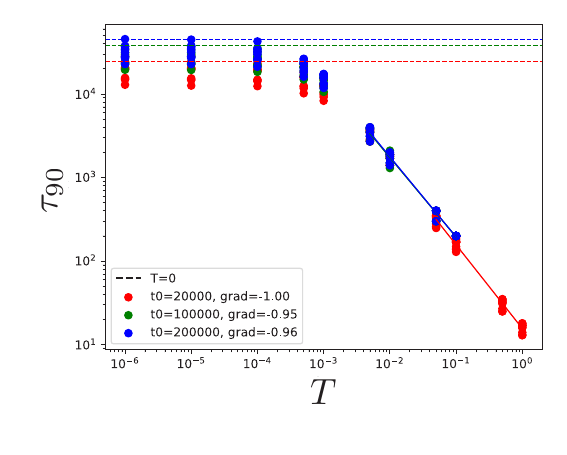}
\caption{The effect of waiting time on the $\tau_{90}$ vs $T$ curve in the underparameterised phase.}
\label{fig:aging_tau90}
\end{figure}

The relaxation behaviour does not change significantly at different waiting times. In Fig.\ \ref{fig:aging_tau90} (a) and (b), $t_w$ is varied over an order of magnitude for regularised and underparameterised networks respectively. In all cases, $\tau_{90}$ displays the same behaviour of a noise-dominated regime with a gradient $\approx -1$, and a temperature independent landscape-dominated regime. In both phases, increasing $t_w$ slows down the decorrelation process, without a noticeable effect on $T^\ast$. The ageing effect is stronger in the regularised phase than the underparameterised phase. The gradients are slightly less than 1 when $t_w \neq 20000$. This is due to a data sampling effect. In general the network states are saved every 100 timesteps. However, for $t\in[20000, 20100]$ the networks are saved every timestep, and for $t\in[20100, 21000]$ every 10 timesteps. This is to allow high resolution measurements of short decorrelation times when $t_w=20000$, as it is for the majority of our measurements. Consequently, the time resolution of measurements at $t_w\neq20000$ is limited to every 100 timesteps, which introduces some inaccuracy in the gradient measurements. 

\begin{figure}[h]
\includegraphics[width=8cm]{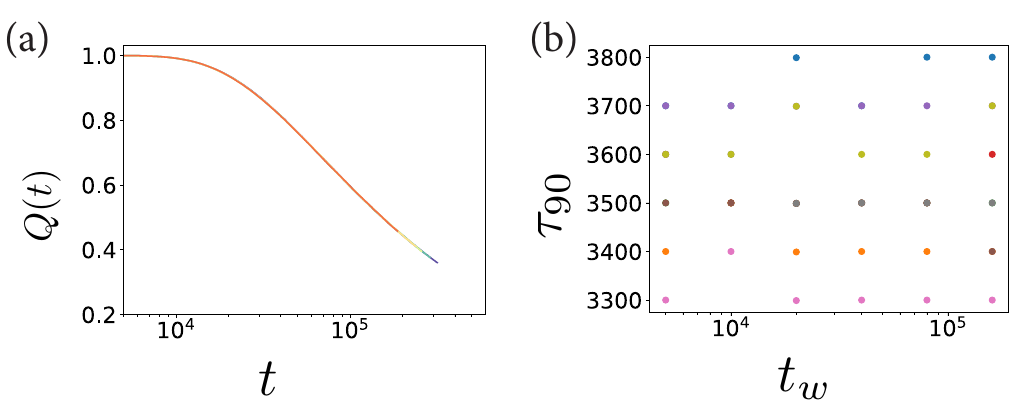}
\caption{(a) The overlap function at various $t_w$, and (b) the (lack of) dependence of $\tau_{90}$ on $t_w$, in the overparameterised phase.}
\label{fig:aging_overparam}
\end{figure}

The overparameterised phase does not undergo a transition from noise-dominated to landscape-dominated dynamics, hence does not experience ageing. This is demonstrated in Fig.\ \ref{fig:aging_overparam}.

\bibliography{how_glassy_Langevin.bib}

\begin{thebibliography}{111}%
\makeatletter
\providecommand \@ifxundefined [1]{%
 \@ifx{#1\undefined}
}%
\providecommand \@ifnum [1]{%
 \ifnum #1\expandafter \@firstoftwo
 \else \expandafter \@secondoftwo
 \fi
}%
\providecommand \@ifx [1]{%
 \ifx #1\expandafter \@firstoftwo
 \else \expandafter \@secondoftwo
 \fi
}%
\providecommand \natexlab [1]{#1}%
\providecommand \enquote  [1]{``#1''}%
\providecommand \bibnamefont  [1]{#1}%
\providecommand \bibfnamefont [1]{#1}%
\providecommand \citenamefont [1]{#1}%
\providecommand \href@noop [0]{\@secondoftwo}%
\providecommand \href [0]{\begingroup \@sanitize@url \@href}%
\providecommand \@href[1]{\@@startlink{#1}\@@href}%
\providecommand \@@href[1]{\endgroup#1\@@endlink}%
\providecommand \@sanitize@url [0]{\catcode `\\12\catcode `\$12\catcode
  `\&12\catcode `\#12\catcode `\^12\catcode `\_12\catcode `\%12\relax}%
\providecommand \@@startlink[1]{}%
\providecommand \@@endlink[0]{}%
\providecommand \url  [0]{\begingroup\@sanitize@url \@url }%
\providecommand \@url [1]{\endgroup\@href {#1}{\urlprefix }}%
\providecommand \urlprefix  [0]{URL }%
\providecommand \Eprint [0]{\href }%
\providecommand \doibase [0]{http://dx.doi.org/}%
\providecommand \selectlanguage [0]{\@gobble}%
\providecommand \bibinfo  [0]{\@secondoftwo}%
\providecommand \bibfield  [0]{\@secondoftwo}%
\providecommand \translation [1]{[#1]}%
\providecommand \BibitemOpen [0]{}%
\providecommand \bibitemStop [0]{}%
\providecommand \bibitemNoStop [0]{.\EOS\space}%
\providecommand \EOS [0]{\spacefactor3000\relax}%
\providecommand \BibitemShut  [1]{\csname bibitem#1\endcsname}%
\let\auto@bib@innerbib\@empty
\bibitem [{\citenamefont {Bahri}\ \emph {et~al.}(2020)\citenamefont {Bahri},
  \citenamefont {Kadmon}, \citenamefont {Pennington}, \citenamefont
  {Schoenholz}, \citenamefont {Sohl-Dickstein},\ and\ \citenamefont
  {Ganguli}}]{Bahri2020}%
  \BibitemOpen
  \bibfield  {author} {\bibinfo {author} {\bibfnamefont {Y.}~\bibnamefont
  {Bahri}}, \bibinfo {author} {\bibfnamefont {J.}~\bibnamefont {Kadmon}},
  \bibinfo {author} {\bibfnamefont {J.}~\bibnamefont {Pennington}}, \bibinfo
  {author} {\bibfnamefont {S.~S.}\ \bibnamefont {Schoenholz}}, \bibinfo
  {author} {\bibfnamefont {J.}~\bibnamefont {Sohl-Dickstein}}, \ and\ \bibinfo
  {author} {\bibfnamefont {S.}~\bibnamefont {Ganguli}},\ }\href {\doibase
  10.1146/annurev-conmatphys-031119-050745} {\bibfield  {journal} {\bibinfo
  {journal} {Annual Review of Condensed Matter Physics}\ }\textbf {\bibinfo
  {volume} {11}},\ \bibinfo {pages} {501} (\bibinfo {year} {2020})}\BibitemShut
  {NoStop}%
\bibitem [{\citenamefont {Li}\ and\ \citenamefont
  {Sompolinsky}(2021)}]{Li2020}%
  \BibitemOpen
  \bibfield  {author} {\bibinfo {author} {\bibfnamefont {Q.}~\bibnamefont
  {Li}}\ and\ \bibinfo {author} {\bibfnamefont {H.}~\bibnamefont
  {Sompolinsky}},\ }\href {\doibase 10.1103/PhysRevX.11.031059} {\bibfield
  {journal} {\bibinfo  {journal} {Physical Review X}\ }\textbf {\bibinfo
  {volume} {11}},\ \bibinfo {pages} {031059} (\bibinfo {year} {2021})},\
  \Eprint {http://arxiv.org/abs/2012.04030} {arXiv:2012.04030} \BibitemShut
  {NoStop}%
\bibitem [{\citenamefont {Goldt}\ \emph {et~al.}(2020)\citenamefont {Goldt},
  \citenamefont {M{\'{e}}zard}, \citenamefont {Krzakala},\ and\ \citenamefont
  {Zdeborov{\'{a}}}}]{Goldt2020}%
  \BibitemOpen
  \bibfield  {author} {\bibinfo {author} {\bibfnamefont {S.}~\bibnamefont
  {Goldt}}, \bibinfo {author} {\bibfnamefont {M.}~\bibnamefont {M{\'{e}}zard}},
  \bibinfo {author} {\bibfnamefont {F.}~\bibnamefont {Krzakala}}, \ and\
  \bibinfo {author} {\bibfnamefont {L.}~\bibnamefont {Zdeborov{\'{a}}}},\
  }\href {\doibase 10.1103/PhysRevX.10.041044} {\bibfield  {journal} {\bibinfo
  {journal} {Physical Review X}\ }\textbf {\bibinfo {volume} {10}},\ \bibinfo
  {pages} {041044} (\bibinfo {year} {2020})},\ \Eprint
  {http://arxiv.org/abs/1909.11500} {arXiv:1909.11500} \BibitemShut {NoStop}%
\bibitem [{\citenamefont {Geiger}\ \emph {et~al.}(2020)\citenamefont {Geiger},
  \citenamefont {Jacot}, \citenamefont {Spigler}, \citenamefont {Gabriel},
  \citenamefont {Sagun}, \citenamefont {D'Ascoli}, \citenamefont {Biroli},
  \citenamefont {Hongler},\ and\ \citenamefont {Wyart}}]{Geiger2020}%
  \BibitemOpen
  \bibfield  {author} {\bibinfo {author} {\bibfnamefont {M.}~\bibnamefont
  {Geiger}}, \bibinfo {author} {\bibfnamefont {A.}~\bibnamefont {Jacot}},
  \bibinfo {author} {\bibfnamefont {S.}~\bibnamefont {Spigler}}, \bibinfo
  {author} {\bibfnamefont {F.}~\bibnamefont {Gabriel}}, \bibinfo {author}
  {\bibfnamefont {L.}~\bibnamefont {Sagun}}, \bibinfo {author} {\bibfnamefont
  {S.}~\bibnamefont {D'Ascoli}}, \bibinfo {author} {\bibfnamefont
  {G.}~\bibnamefont {Biroli}}, \bibinfo {author} {\bibfnamefont
  {C.}~\bibnamefont {Hongler}}, \ and\ \bibinfo {author} {\bibfnamefont
  {M.}~\bibnamefont {Wyart}},\ }\href {\doibase
  10.1088/1742-5468/2011/07/L07002} {\bibfield  {journal} {\bibinfo  {journal}
  {Journal of Statistical Mechanics: Theory and Experiment}\ }\textbf {\bibinfo
  {volume} {2020}},\ \bibinfo {pages} {023401} (\bibinfo {year}
  {2020})}\BibitemShut {NoStop}%
\bibitem [{\citenamefont {Zhang}\ \emph {et~al.}(2018)\citenamefont {Zhang},
  \citenamefont {Saxe}, \citenamefont {Advani},\ and\ \citenamefont
  {Lee}}]{Zhang2018}%
  \BibitemOpen
  \bibfield  {author} {\bibinfo {author} {\bibfnamefont {Y.}~\bibnamefont
  {Zhang}}, \bibinfo {author} {\bibfnamefont {A.~M.}\ \bibnamefont {Saxe}},
  \bibinfo {author} {\bibfnamefont {M.~S.}\ \bibnamefont {Advani}}, \ and\
  \bibinfo {author} {\bibfnamefont {A.~A.}\ \bibnamefont {Lee}},\ }\href
  {\doibase 10.1080/00268976.2018.1483535} {\bibfield  {journal} {\bibinfo
  {journal} {Molecular Physics}\ }\textbf {\bibinfo {volume} {116}},\ \bibinfo
  {pages} {3214} (\bibinfo {year} {2018})},\ \Eprint
  {http://arxiv.org/abs/1803.01927} {arXiv:1803.01927} \BibitemShut {NoStop}%
\bibitem [{\citenamefont {Hopfield}(1982)}]{Hopfield1982}%
  \BibitemOpen
  \bibfield  {author} {\bibinfo {author} {\bibfnamefont {J.~J.}\ \bibnamefont
  {Hopfield}},\ }\href {\doibase 10.1073/pnas.79.8.2554} {\bibfield  {journal}
  {\bibinfo  {journal} {Proceedings of the National Academy of Sciences of the
  United States of America}\ }\textbf {\bibinfo {volume} {79}},\ \bibinfo
  {pages} {2554} (\bibinfo {year} {1982})}\BibitemShut {NoStop}%
\bibitem [{\citenamefont {Little}(1974)}]{Little1974}%
  \BibitemOpen
  \bibfield  {author} {\bibinfo {author} {\bibfnamefont {W.~A.}\ \bibnamefont
  {Little}},\ }\href {\doibase 10.1016/0025-5564(74)90031-5} {\bibfield
  {journal} {\bibinfo  {journal} {Mathematical Biosciences}\ }\textbf {\bibinfo
  {volume} {19}},\ \bibinfo {pages} {101} (\bibinfo {year} {1974})}\BibitemShut
  {NoStop}%
\bibitem [{\citenamefont {Amit}\ \emph
  {et~al.}(1985{\natexlab{a}})\citenamefont {Amit}, \citenamefont {Gutfreund},\
  and\ \citenamefont {Sompolinsky}}]{Amit1985}%
  \BibitemOpen
  \bibfield  {author} {\bibinfo {author} {\bibfnamefont {D.~J.}\ \bibnamefont
  {Amit}}, \bibinfo {author} {\bibfnamefont {H.}~\bibnamefont {Gutfreund}}, \
  and\ \bibinfo {author} {\bibfnamefont {H.}~\bibnamefont {Sompolinsky}},\
  }\href {\doibase 10.1103/PhysRevLett.55.1530} {\bibfield  {journal} {\bibinfo
   {journal} {Physical Review Letters}\ }\textbf {\bibinfo {volume} {55}},\
  \bibinfo {pages} {1530} (\bibinfo {year} {1985}{\natexlab{a}})}\BibitemShut
  {NoStop}%
\bibitem [{\citenamefont {Amit}\ \emph
  {et~al.}(1985{\natexlab{b}})\citenamefont {Amit}, \citenamefont {Gutfreund},\
  and\ \citenamefont {Sompolinsky}}]{Amit1985a}%
  \BibitemOpen
  \bibfield  {author} {\bibinfo {author} {\bibfnamefont {D.~J.}\ \bibnamefont
  {Amit}}, \bibinfo {author} {\bibfnamefont {H.}~\bibnamefont {Gutfreund}}, \
  and\ \bibinfo {author} {\bibfnamefont {H.}~\bibnamefont {Sompolinsky}},\
  }\href {\doibase 10.1103/PhysRevA.34.3435} {\bibfield  {journal} {\bibinfo
  {journal} {Physical Review A}\ }\textbf {\bibinfo {volume} {32}},\ \bibinfo
  {pages} {1007} (\bibinfo {year} {1985}{\natexlab{b}})}\BibitemShut {NoStop}%
\bibitem [{\citenamefont {M{\'{e}}zard}\ \emph {et~al.}(1986)\citenamefont
  {M{\'{e}}zard}, \citenamefont {Nadal},\ and\ \citenamefont
  {Toulouse}}]{Mezard1986}%
  \BibitemOpen
  \bibfield  {author} {\bibinfo {author} {\bibfnamefont {M.}~\bibnamefont
  {M{\'{e}}zard}}, \bibinfo {author} {\bibfnamefont {J.~P.}\ \bibnamefont
  {Nadal}}, \ and\ \bibinfo {author} {\bibfnamefont {G.}~\bibnamefont
  {Toulouse}},\ }\href@noop {} {\bibfield  {journal} {\bibinfo  {journal}
  {Journal de Physique}\ }\textbf {\bibinfo {volume} {47}},\ \bibinfo {pages}
  {1457} (\bibinfo {year} {1986})}\BibitemShut {NoStop}%
\bibitem [{\citenamefont {Krauth}\ and\ \citenamefont
  {Mezard}(1987)}]{Krauth1987}%
  \BibitemOpen
  \bibfield  {author} {\bibinfo {author} {\bibfnamefont {W.}~\bibnamefont
  {Krauth}}\ and\ \bibinfo {author} {\bibfnamefont {M.}~\bibnamefont
  {Mezard}},\ }\href {\doibase 10.1088/0305-4470/20/11/013} {\bibfield
  {journal} {\bibinfo  {journal} {Journal of Physics A: Mathematical and
  General}\ }\textbf {\bibinfo {volume} {20}},\ \bibinfo {pages} {L745}
  (\bibinfo {year} {1987})}\BibitemShut {NoStop}%
\bibitem [{\citenamefont {Gardner}(1988)}]{Gardner1988}%
  \BibitemOpen
  \bibfield  {author} {\bibinfo {author} {\bibfnamefont {E.}~\bibnamefont
  {Gardner}},\ }\href {\doibase 10.1142/S0129183116500674} {\bibfield
  {journal} {\bibinfo  {journal} {Journal of Physics A: Mathematical and
  General}\ }\textbf {\bibinfo {volume} {21}},\ \bibinfo {pages} {257}
  (\bibinfo {year} {1988})}\BibitemShut {NoStop}%
\bibitem [{\citenamefont {Sompolinsky}\ \emph {et~al.}(1988)\citenamefont
  {Sompolinsky}, \citenamefont {Crisanti},\ and\ \citenamefont
  {Sommers}}]{Sompolinsky1988}%
  \BibitemOpen
  \bibfield  {author} {\bibinfo {author} {\bibfnamefont {H.}~\bibnamefont
  {Sompolinsky}}, \bibinfo {author} {\bibfnamefont {A.}~\bibnamefont
  {Crisanti}}, \ and\ \bibinfo {author} {\bibfnamefont {H.~J.}\ \bibnamefont
  {Sommers}},\ }\href {\doibase 10.1103/PhysRevLett.61.259} {\bibfield
  {journal} {\bibinfo  {journal} {Physical Review Letters}\ }\textbf {\bibinfo
  {volume} {61}},\ \bibinfo {pages} {259} (\bibinfo {year} {1988})}\BibitemShut
  {NoStop}%
\bibitem [{\citenamefont {Barra}\ \emph {et~al.}(2012)\citenamefont {Barra},
  \citenamefont {Genovese}, \citenamefont {Guerra},\ and\ \citenamefont
  {Tantari}}]{Barra2012}%
  \BibitemOpen
  \bibfield  {author} {\bibinfo {author} {\bibfnamefont {A.}~\bibnamefont
  {Barra}}, \bibinfo {author} {\bibfnamefont {G.}~\bibnamefont {Genovese}},
  \bibinfo {author} {\bibfnamefont {F.}~\bibnamefont {Guerra}}, \ and\ \bibinfo
  {author} {\bibfnamefont {D.}~\bibnamefont {Tantari}},\ }\href {\doibase
  10.1088/1742-5468/2012/07/P07009} {\bibfield  {journal} {\bibinfo  {journal}
  {Journal of Statistical Mechanics: Theory and Experiment}\ }\textbf {\bibinfo
  {volume} {2012}},\ \bibinfo {pages} {P07009} (\bibinfo {year} {2012})},\
  \Eprint {http://arxiv.org/abs/1205.3900} {arXiv:1205.3900} \BibitemShut
  {NoStop}%
\bibitem [{\citenamefont {Ghio}\ \emph {et~al.}(2023)\citenamefont {Ghio},
  \citenamefont {Dandi}, \citenamefont {Krzakala},\ and\ \citenamefont
  {Zdeborov{\'{a}}}}]{Ghio2023}%
  \BibitemOpen
  \bibfield  {author} {\bibinfo {author} {\bibfnamefont {D.}~\bibnamefont
  {Ghio}}, \bibinfo {author} {\bibfnamefont {Y.}~\bibnamefont {Dandi}},
  \bibinfo {author} {\bibfnamefont {F.}~\bibnamefont {Krzakala}}, \ and\
  \bibinfo {author} {\bibfnamefont {L.}~\bibnamefont {Zdeborov{\'{a}}}},\
  }\href {http://arxiv.org/abs/2308.14085} {\bibfield  {journal} {\bibinfo
  {journal} {arXiv}\ ,\ \bibinfo {pages} {1}} (\bibinfo {year} {2023})},\
  \Eprint {http://arxiv.org/abs/2308.14085} {arXiv:2308.14085} \BibitemShut
  {NoStop}%
\bibitem [{\citenamefont {Biroli}\ \emph {et~al.}(2024)\citenamefont {Biroli},
  \citenamefont {Bonnaire}, \citenamefont {de~Bortoli},\ and\ \citenamefont
  {M{\'{e}}zard}}]{Biroli2024}%
  \BibitemOpen
  \bibfield  {author} {\bibinfo {author} {\bibfnamefont {G.}~\bibnamefont
  {Biroli}}, \bibinfo {author} {\bibfnamefont {T.}~\bibnamefont {Bonnaire}},
  \bibinfo {author} {\bibfnamefont {V.}~\bibnamefont {de~Bortoli}}, \ and\
  \bibinfo {author} {\bibfnamefont {M.}~\bibnamefont {M{\'{e}}zard}},\ }\href
  {http://arxiv.org/abs/2402.18491} {\bibfield  {journal} {\bibinfo  {journal}
  {arXiv}\ } (\bibinfo {year} {2024})},\ \Eprint
  {http://arxiv.org/abs/2402.18491} {arXiv:2402.18491} \BibitemShut {NoStop}%
\bibitem [{\citenamefont {Mignacco}\ and\ \citenamefont
  {Urbani}(2022)}]{Mignacco2022}%
  \BibitemOpen
  \bibfield  {author} {\bibinfo {author} {\bibfnamefont {F.}~\bibnamefont
  {Mignacco}}\ and\ \bibinfo {author} {\bibfnamefont {P.}~\bibnamefont
  {Urbani}},\ }\href {\doibase 10.1088/1742-5468/ac841d} {\bibfield  {journal}
  {\bibinfo  {journal} {Journal of Statistical Mechanics: Theory and
  Experiment}\ }\textbf {\bibinfo {volume} {2022}} (\bibinfo {year} {2022}),\
  10.1088/1742-5468/ac841d},\ \Eprint {http://arxiv.org/abs/2112.10852}
  {arXiv:2112.10852} \BibitemShut {NoStop}%
\bibitem [{\citenamefont {Becker}\ \emph {et~al.}(2020)\citenamefont {Becker},
  \citenamefont {Zhang},\ and\ \citenamefont {Lee}}]{Becker2020}%
  \BibitemOpen
  \bibfield  {author} {\bibinfo {author} {\bibfnamefont {S.}~\bibnamefont
  {Becker}}, \bibinfo {author} {\bibfnamefont {Y.}~\bibnamefont {Zhang}}, \
  and\ \bibinfo {author} {\bibfnamefont {A.~A.}\ \bibnamefont {Lee}},\ }\href
  {\doibase 10.1103/PhysRevLett.124.108301} {\bibfield  {journal} {\bibinfo
  {journal} {Physical Review Letters}\ }\textbf {\bibinfo {volume} {124}},\
  \bibinfo {pages} {108301} (\bibinfo {year} {2020})},\ \Eprint
  {http://arxiv.org/abs/1808.00408} {arXiv:1808.00408} \BibitemShut {NoStop}%
\bibitem [{\citenamefont {Choromanska}\ \emph {et~al.}(2015)\citenamefont
  {Choromanska}, \citenamefont {Henaff}, \citenamefont {Mathieu}, \citenamefont
  {{Ben Arous}},\ and\ \citenamefont {LeCun}}]{Choromanska2015}%
  \BibitemOpen
  \bibfield  {author} {\bibinfo {author} {\bibfnamefont {A.}~\bibnamefont
  {Choromanska}}, \bibinfo {author} {\bibfnamefont {M.}~\bibnamefont {Henaff}},
  \bibinfo {author} {\bibfnamefont {M.}~\bibnamefont {Mathieu}}, \bibinfo
  {author} {\bibfnamefont {G.}~\bibnamefont {{Ben Arous}}}, \ and\ \bibinfo
  {author} {\bibfnamefont {Y.}~\bibnamefont {LeCun}},\ }\href {\doibase
  10.1016/0040-8166(86)90068-6} {\bibfield  {journal} {\bibinfo  {journal}
  {Proceedings of the 18th International Conference on Artificial Intelligence
  and Statistics (AISTATS) 2015}\ } (\bibinfo {year} {2015}),\
  10.1016/0040-8166(86)90068-6}\BibitemShut {NoStop}%
\bibitem [{\citenamefont {Kawaguchi}(2016)}]{Kawaguchi2016}%
  \BibitemOpen
  \bibfield  {author} {\bibinfo {author} {\bibfnamefont {K.}~\bibnamefont
  {Kawaguchi}},\ }\href@noop {} {\bibfield  {journal} {\bibinfo  {journal}
  {Advances in Neural Information Processing Systems}\ }\textbf {\bibinfo
  {volume} {30}} (\bibinfo {year} {2016})},\ \Eprint
  {http://arxiv.org/abs/1605.07110} {arXiv:1605.07110} \BibitemShut {NoStop}%
\bibitem [{\citenamefont {Soudry}\ and\ \citenamefont
  {Carmon}(2016)}]{Soudry2016}%
  \BibitemOpen
  \bibfield  {author} {\bibinfo {author} {\bibfnamefont {D.}~\bibnamefont
  {Soudry}}\ and\ \bibinfo {author} {\bibfnamefont {Y.}~\bibnamefont
  {Carmon}},\ }\href {http://arxiv.org/abs/1605.08361} {\bibfield  {journal}
  {\bibinfo  {journal} {arXiv}\ } (\bibinfo {year} {2016})},\ \Eprint
  {http://arxiv.org/abs/1605.08361} {arXiv:1605.08361} \BibitemShut {NoStop}%
\bibitem [{\citenamefont {Kawaguchi}\ and\ \citenamefont
  {Kaelbling}(2020)}]{Kawaguchi2020}%
  \BibitemOpen
  \bibfield  {author} {\bibinfo {author} {\bibfnamefont {K.}~\bibnamefont
  {Kawaguchi}}\ and\ \bibinfo {author} {\bibfnamefont {L.~P.}\ \bibnamefont
  {Kaelbling}},\ }\href@noop {} {\bibfield  {journal} {\bibinfo  {journal}
  {Proceedings of Machine Learning Research}\ }\textbf {\bibinfo {volume}
  {108}},\ \bibinfo {pages} {853} (\bibinfo {year} {2020})},\ \Eprint
  {http://arxiv.org/abs/1901.00279} {arXiv:1901.00279} \BibitemShut {NoStop}%
\bibitem [{\citenamefont {Chen}\ \emph {et~al.}(2022)\citenamefont {Chen},
  \citenamefont {Qu},\ and\ \citenamefont {Gong}}]{Chen2022}%
  \BibitemOpen
  \bibfield  {author} {\bibinfo {author} {\bibfnamefont {G.}~\bibnamefont
  {Chen}}, \bibinfo {author} {\bibfnamefont {C.~K.}\ \bibnamefont {Qu}}, \ and\
  \bibinfo {author} {\bibfnamefont {P.}~\bibnamefont {Gong}},\ }\href {\doibase
  10.1016/j.neunet.2022.01.019} {\bibfield  {journal} {\bibinfo  {journal}
  {Neural Networks}\ }\textbf {\bibinfo {volume} {149}},\ \bibinfo {pages} {18}
  (\bibinfo {year} {2022})},\ \Eprint {http://arxiv.org/abs/2009.10588}
  {arXiv:2009.10588} \BibitemShut {NoStop}%
\bibitem [{\citenamefont {Baity-Jesi}\ \emph {et~al.}(2019)\citenamefont
  {Baity-Jesi}, \citenamefont {Sagun}, \citenamefont {Geiger}, \citenamefont
  {Spigler}, \citenamefont {{Ben Arous}}, \citenamefont {Cammarota},
  \citenamefont {LeCun}, \citenamefont {Wyart},\ and\ \citenamefont
  {Biroli}}]{Baity-Jesi2019}%
  \BibitemOpen
  \bibfield  {author} {\bibinfo {author} {\bibfnamefont {M.}~\bibnamefont
  {Baity-Jesi}}, \bibinfo {author} {\bibfnamefont {L.}~\bibnamefont {Sagun}},
  \bibinfo {author} {\bibfnamefont {M.}~\bibnamefont {Geiger}}, \bibinfo
  {author} {\bibfnamefont {S.}~\bibnamefont {Spigler}}, \bibinfo {author}
  {\bibfnamefont {G.}~\bibnamefont {{Ben Arous}}}, \bibinfo {author}
  {\bibfnamefont {C.}~\bibnamefont {Cammarota}}, \bibinfo {author}
  {\bibfnamefont {Y.}~\bibnamefont {LeCun}}, \bibinfo {author} {\bibfnamefont
  {M.}~\bibnamefont {Wyart}}, \ and\ \bibinfo {author} {\bibfnamefont
  {G.}~\bibnamefont {Biroli}},\ }\href {\doibase 10.1088/1742-5468/ab3281}
  {\bibfield  {journal} {\bibinfo  {journal} {Journal of Statistical Mechanics:
  Theory and Experiment}\ }\textbf {\bibinfo {volume} {2019}},\ \bibinfo
  {pages} {124013} (\bibinfo {year} {2019})},\ \Eprint
  {http://arxiv.org/abs/1803.06969} {arXiv:1803.06969} \BibitemShut {NoStop}%
\bibitem [{\citenamefont {Baity-Jesi}\ and\ \citenamefont
  {Mart{\'{i}}n-Mayor}(2019)}]{Baity-Jesi2019b}%
  \BibitemOpen
  \bibfield  {author} {\bibinfo {author} {\bibfnamefont {M.}~\bibnamefont
  {Baity-Jesi}}\ and\ \bibinfo {author} {\bibfnamefont {V.}~\bibnamefont
  {Mart{\'{i}}n-Mayor}},\ }\href {\doibase 10.1088/1742-5468/ab333c} {\bibfield
   {journal} {\bibinfo  {journal} {Journal of Statistical Mechanics: Theory and
  Experiment}\ ,\ \bibinfo {pages} {084016}} (\bibinfo {year} {2019})},\
  \Eprint {http://arxiv.org/abs/1901.05581} {arXiv:1901.05581} \BibitemShut
  {NoStop}%
\bibitem [{\citenamefont {Kirkpatrick}\ \emph {et~al.}(1989)\citenamefont
  {Kirkpatrick}, \citenamefont {Thirumalai},\ and\ \citenamefont
  {Wolynes}}]{Kirkpatrick1989}%
  \BibitemOpen
  \bibfield  {author} {\bibinfo {author} {\bibfnamefont {T.~R.}\ \bibnamefont
  {Kirkpatrick}}, \bibinfo {author} {\bibfnamefont {D.}~\bibnamefont
  {Thirumalai}}, \ and\ \bibinfo {author} {\bibfnamefont {P.~G.}\ \bibnamefont
  {Wolynes}},\ }\href {\doibase 10.1103/PhysRevA.40.1045} {\bibfield  {journal}
  {\bibinfo  {journal} {Physical Review A}\ }\textbf {\bibinfo {volume} {40}},\
  \bibinfo {pages} {1045} (\bibinfo {year} {1989})}\BibitemShut {NoStop}%
\bibitem [{\citenamefont {Charbonneau}\ \emph
  {et~al.}(2014{\natexlab{a}})\citenamefont {Charbonneau}, \citenamefont
  {Kurchan}, \citenamefont {Parisi}, \citenamefont {Urbani},\ and\
  \citenamefont {Zamponi}}]{Charbonneau2014a}%
  \BibitemOpen
  \bibfield  {author} {\bibinfo {author} {\bibfnamefont {P.}~\bibnamefont
  {Charbonneau}}, \bibinfo {author} {\bibfnamefont {J.}~\bibnamefont
  {Kurchan}}, \bibinfo {author} {\bibfnamefont {G.}~\bibnamefont {Parisi}},
  \bibinfo {author} {\bibfnamefont {P.}~\bibnamefont {Urbani}}, \ and\ \bibinfo
  {author} {\bibfnamefont {F.}~\bibnamefont {Zamponi}},\ }\href {\doibase
  10.1038/ncomms4725} {\bibfield  {journal} {\bibinfo  {journal} {Nature
  Communications}\ }\textbf {\bibinfo {volume} {5}},\ \bibinfo {pages} {3725}
  (\bibinfo {year} {2014}{\natexlab{a}})},\ \Eprint
  {http://arxiv.org/abs/1404.6809} {arXiv:1404.6809} \BibitemShut {NoStop}%
\bibitem [{\citenamefont {Folena}\ \emph {et~al.}(2020)\citenamefont {Folena},
  \citenamefont {Franz},\ and\ \citenamefont {Ricci-Tersenghi}}]{Folena2020}%
  \BibitemOpen
  \bibfield  {author} {\bibinfo {author} {\bibfnamefont {G.}~\bibnamefont
  {Folena}}, \bibinfo {author} {\bibfnamefont {S.}~\bibnamefont {Franz}}, \
  and\ \bibinfo {author} {\bibfnamefont {F.}~\bibnamefont {Ricci-Tersenghi}},\
  }\href {\doibase 10.1103/PhysRevX.10.031045} {\bibfield  {journal} {\bibinfo
  {journal} {Physical Review X}\ }\textbf {\bibinfo {volume} {10}},\ \bibinfo
  {pages} {031045} (\bibinfo {year} {2020})},\ \Eprint
  {http://arxiv.org/abs/1903.01421} {arXiv:1903.01421} \BibitemShut {NoStop}%
\bibitem [{\citenamefont {Laudicina}\ \emph {et~al.}(2024)\citenamefont
  {Laudicina}, \citenamefont {Charbonneau}, \citenamefont {Hu}, \citenamefont
  {Janssen}, \citenamefont {Morse}, \citenamefont {Pihlajamaa},\ and\
  \citenamefont {Szamel}}]{Laudicina2024}%
  \BibitemOpen
  \bibfield  {author} {\bibinfo {author} {\bibfnamefont {C.~C.~L.}\
  \bibnamefont {Laudicina}}, \bibinfo {author} {\bibfnamefont {P.}~\bibnamefont
  {Charbonneau}}, \bibinfo {author} {\bibfnamefont {Y.}~\bibnamefont {Hu}},
  \bibinfo {author} {\bibfnamefont {L.~M.~C.}\ \bibnamefont {Janssen}},
  \bibinfo {author} {\bibfnamefont {P.~K.}\ \bibnamefont {Morse}}, \bibinfo
  {author} {\bibfnamefont {I.}~\bibnamefont {Pihlajamaa}}, \ and\ \bibinfo
  {author} {\bibfnamefont {G.}~\bibnamefont {Szamel}},\ }\href
  {http://arxiv.org/abs/2408.06933} {\ ,\ \bibinfo {pages} {1} (\bibinfo {year}
  {2024})},\ \Eprint {http://arxiv.org/abs/2408.06933} {arXiv:2408.06933}
  \BibitemShut {NoStop}%
\bibitem [{\citenamefont {Janssen}(2018)}]{Janssen2018}%
  \BibitemOpen
  \bibfield  {author} {\bibinfo {author} {\bibfnamefont {L.~M.~C.}\
  \bibnamefont {Janssen}},\ }\href {\doibase 10.3389/fphy.2018.00097}
  {\bibfield  {journal} {\bibinfo  {journal} {Frontiers in Physics}\ }\textbf
  {\bibinfo {volume} {6}},\ \bibinfo {pages} {97} (\bibinfo {year} {2018})},\
  \Eprint {http://arxiv.org/abs/1806.01369} {arXiv:1806.01369} \BibitemShut
  {NoStop}%
\bibitem [{\citenamefont {Das}(2004)}]{Das2004}%
  \BibitemOpen
  \bibfield  {author} {\bibinfo {author} {\bibfnamefont {S.~P.}\ \bibnamefont
  {Das}},\ }\href {\doibase 10.1103/RevModPhys.76.785} {\bibfield  {journal}
  {\bibinfo  {journal} {Reviews of Modern Physics}\ }\textbf {\bibinfo {volume}
  {76}},\ \bibinfo {pages} {785} (\bibinfo {year} {2004})}\BibitemShut
  {NoStop}%
\bibitem [{\citenamefont {Micoulaut}(2016)}]{Micoulaut2016}%
  \BibitemOpen
  \bibfield  {author} {\bibinfo {author} {\bibfnamefont {M.}~\bibnamefont
  {Micoulaut}},\ }\href {\doibase 10.1088/0034-4885/79/6/066504} {\bibfield
  {journal} {\bibinfo  {journal} {Reports on Progress in Physics}\ }\textbf
  {\bibinfo {volume} {79}},\ \bibinfo {pages} {066504} (\bibinfo {year}
  {2016})},\ \Eprint {http://arxiv.org/abs/1605.07403} {arXiv:1605.07403}
  \BibitemShut {NoStop}%
\bibitem [{\citenamefont {Berthier}\ and\ \citenamefont
  {Biroli}(2011)}]{Berthier2011}%
  \BibitemOpen
  \bibfield  {author} {\bibinfo {author} {\bibfnamefont {L.}~\bibnamefont
  {Berthier}}\ and\ \bibinfo {author} {\bibfnamefont {G.}~\bibnamefont
  {Biroli}},\ }\href {\doibase 10.1103/RevModPhys.83.587} {\bibfield  {journal}
  {\bibinfo  {journal} {Reviews of Modern Physics}\ }\textbf {\bibinfo {volume}
  {83}},\ \bibinfo {pages} {587} (\bibinfo {year} {2011})},\ \Eprint
  {http://arxiv.org/abs/1011.2578} {arXiv:1011.2578} \BibitemShut {NoStop}%
\bibitem [{\citenamefont {Parisi}\ and\ \citenamefont
  {Zamponi}(2010)}]{Parisi2010}%
  \BibitemOpen
  \bibfield  {author} {\bibinfo {author} {\bibfnamefont {G.}~\bibnamefont
  {Parisi}}\ and\ \bibinfo {author} {\bibfnamefont {F.}~\bibnamefont
  {Zamponi}},\ }\href {\doibase 10.1103/RevModPhys.82.789} {\bibfield
  {journal} {\bibinfo  {journal} {Reviews of Modern Physics}\ }\textbf
  {\bibinfo {volume} {82}},\ \bibinfo {pages} {789} (\bibinfo {year} {2010})},\
  \Eprint {http://arxiv.org/abs/0802.2180} {arXiv:0802.2180} \BibitemShut
  {NoStop}%
\bibitem [{\citenamefont {Geiger}\ \emph {et~al.}(2019)\citenamefont {Geiger},
  \citenamefont {Spigler}, \citenamefont {D'Ascoli}, \citenamefont {Sagun},
  \citenamefont {Baity-Jesi}, \citenamefont {Biroli},\ and\ \citenamefont
  {Wyart}}]{Geiger2019}%
  \BibitemOpen
  \bibfield  {author} {\bibinfo {author} {\bibfnamefont {M.}~\bibnamefont
  {Geiger}}, \bibinfo {author} {\bibfnamefont {S.}~\bibnamefont {Spigler}},
  \bibinfo {author} {\bibfnamefont {S.}~\bibnamefont {D'Ascoli}}, \bibinfo
  {author} {\bibfnamefont {L.}~\bibnamefont {Sagun}}, \bibinfo {author}
  {\bibfnamefont {M.}~\bibnamefont {Baity-Jesi}}, \bibinfo {author}
  {\bibfnamefont {G.}~\bibnamefont {Biroli}}, \ and\ \bibinfo {author}
  {\bibfnamefont {M.}~\bibnamefont {Wyart}},\ }\href {\doibase
  10.1103/PhysRevE.100.012115} {\bibfield  {journal} {\bibinfo  {journal}
  {Physical Review E}\ }\textbf {\bibinfo {volume} {100}},\ \bibinfo {pages}
  {012115} (\bibinfo {year} {2019})},\ \Eprint
  {http://arxiv.org/abs/1809.09349} {arXiv:1809.09349} \BibitemShut {NoStop}%
\bibitem [{\citenamefont {Spigler}\ \emph {et~al.}(2019)\citenamefont
  {Spigler}, \citenamefont {Geiger}, \citenamefont {D'Ascoli}, \citenamefont
  {Sagun}, \citenamefont {Biroli},\ and\ \citenamefont {Wyart}}]{Spigler2019}%
  \BibitemOpen
  \bibfield  {author} {\bibinfo {author} {\bibfnamefont {S.}~\bibnamefont
  {Spigler}}, \bibinfo {author} {\bibfnamefont {M.}~\bibnamefont {Geiger}},
  \bibinfo {author} {\bibfnamefont {S.}~\bibnamefont {D'Ascoli}}, \bibinfo
  {author} {\bibfnamefont {L.}~\bibnamefont {Sagun}}, \bibinfo {author}
  {\bibfnamefont {G.}~\bibnamefont {Biroli}}, \ and\ \bibinfo {author}
  {\bibfnamefont {M.}~\bibnamefont {Wyart}},\ }\href {\doibase
  10.1088/1751-8121/ab4c8b} {\bibfield  {journal} {\bibinfo  {journal} {Journal
  of Physics A: Mathematical and Theoretical}\ }\textbf {\bibinfo {volume}
  {52}},\ \bibinfo {pages} {474001} (\bibinfo {year} {2019})}\BibitemShut
  {NoStop}%
\bibitem [{\citenamefont {Franz}\ \emph {et~al.}(2019)\citenamefont {Franz},
  \citenamefont {Hwang},\ and\ \citenamefont {Urbani}}]{Franz2019}%
  \BibitemOpen
  \bibfield  {author} {\bibinfo {author} {\bibfnamefont {S.}~\bibnamefont
  {Franz}}, \bibinfo {author} {\bibfnamefont {S.}~\bibnamefont {Hwang}}, \ and\
  \bibinfo {author} {\bibfnamefont {P.}~\bibnamefont {Urbani}},\ }\href
  {\doibase 10.1103/PhysRevLett.123.160602} {\bibfield  {journal} {\bibinfo
  {journal} {Physical Review Letters}\ }\textbf {\bibinfo {volume} {123}},\
  \bibinfo {pages} {160602} (\bibinfo {year} {2019})},\ \Eprint
  {http://arxiv.org/abs/1809.09945} {arXiv:1809.09945} \BibitemShut {NoStop}%
\bibitem [{\citenamefont {Franz}\ and\ \citenamefont
  {Parisi}(2016)}]{Franz2016}%
  \BibitemOpen
  \bibfield  {author} {\bibinfo {author} {\bibfnamefont {S.}~\bibnamefont
  {Franz}}\ and\ \bibinfo {author} {\bibfnamefont {G.}~\bibnamefont {Parisi}},\
  }\href {\doibase 10.1088/1751-8113/49/14/145001} {\bibfield  {journal}
  {\bibinfo  {journal} {Journal of Physics A: Mathematical and Theoretical}\
  }\textbf {\bibinfo {volume} {49}},\ \bibinfo {pages} {145001} (\bibinfo
  {year} {2016})},\ \Eprint {http://arxiv.org/abs/1501.03397}
  {arXiv:1501.03397} \BibitemShut {NoStop}%
\bibitem [{\citenamefont {Urbani}(2023{\natexlab{a}})}]{Urbani2023}%
  \BibitemOpen
  \bibfield  {author} {\bibinfo {author} {\bibfnamefont {P.}~\bibnamefont
  {Urbani}},\ }\href {\doibase 10.1088/1751-8121/acb742} {\bibfield  {journal}
  {\bibinfo  {journal} {Journal of Physics A: Mathematical and Theoretical}\
  }\textbf {\bibinfo {volume} {56}},\ \bibinfo {pages} {115003} (\bibinfo
  {year} {2023}{\natexlab{a}})},\ \Eprint {http://arxiv.org/abs/2208.11730}
  {arXiv:2208.11730} \BibitemShut {NoStop}%
\bibitem [{\citenamefont {Urbani}(2023{\natexlab{b}})}]{Urbani2023a}%
  \BibitemOpen
  \bibfield  {author} {\bibinfo {author} {\bibfnamefont {P.}~\bibnamefont
  {Urbani}},\ }\href {http://arxiv.org/abs/2301.11236} {\bibfield  {journal}
  {\bibinfo  {journal} {arXiv}\ } (\bibinfo {year} {2023}{\natexlab{b}})},\
  \Eprint {http://arxiv.org/abs/2301.11236} {arXiv:2301.11236} \BibitemShut
  {NoStop}%
\bibitem [{\citenamefont {Fyodorov}(2004)}]{Fyodorov2004}%
  \BibitemOpen
  \bibfield  {author} {\bibinfo {author} {\bibfnamefont {Y.~V.}\ \bibnamefont
  {Fyodorov}},\ }\href {\doibase 10.1103/PhysRevLett.92.240601} {\bibfield
  {journal} {\bibinfo  {journal} {Physical Review Letters}\ }\textbf {\bibinfo
  {volume} {92}},\ \bibinfo {pages} {240601} (\bibinfo {year}
  {2004})}\BibitemShut {NoStop}%
\bibitem [{\citenamefont {Ros}\ \emph {et~al.}(2019)\citenamefont {Ros},
  \citenamefont {{Ben Arous}}, \citenamefont {Biroli},\ and\ \citenamefont
  {Cammarota}}]{Ros2019}%
  \BibitemOpen
  \bibfield  {author} {\bibinfo {author} {\bibfnamefont {V.}~\bibnamefont
  {Ros}}, \bibinfo {author} {\bibfnamefont {G.}~\bibnamefont {{Ben Arous}}},
  \bibinfo {author} {\bibfnamefont {G.}~\bibnamefont {Biroli}}, \ and\ \bibinfo
  {author} {\bibfnamefont {C.}~\bibnamefont {Cammarota}},\ }\href {\doibase
  10.1103/PhysRevX.9.011003} {\bibfield  {journal} {\bibinfo  {journal}
  {Physical Review X}\ }\textbf {\bibinfo {volume} {9}},\ \bibinfo {pages}
  {011003} (\bibinfo {year} {2019})},\ \Eprint
  {http://arxiv.org/abs/1804.02686} {arXiv:1804.02686} \BibitemShut {NoStop}%
\bibitem [{\citenamefont {Subag}\ and\ \citenamefont
  {Zeitouni}(2021)}]{Subag2021}%
  \BibitemOpen
  \bibfield  {author} {\bibinfo {author} {\bibfnamefont {E.}~\bibnamefont
  {Subag}}\ and\ \bibinfo {author} {\bibfnamefont {O.}~\bibnamefont
  {Zeitouni}},\ }\href {\doibase 10.1063/5.0070582} {\bibfield  {journal}
  {\bibinfo  {journal} {Journal of Mathematical Physics}\ }\textbf {\bibinfo
  {volume} {62}},\ \bibinfo {pages} {123301} (\bibinfo {year}
  {2021})}\BibitemShut {NoStop}%
\bibitem [{\citenamefont {Naumis}(2005)}]{Naumis2005}%
  \BibitemOpen
  \bibfield  {author} {\bibinfo {author} {\bibfnamefont {G.~G.}\ \bibnamefont
  {Naumis}},\ }\href {\doibase 10.1103/PhysRevE.71.026114} {\bibfield
  {journal} {\bibinfo  {journal} {Physical Review E}\ }\textbf {\bibinfo
  {volume} {71}},\ \bibinfo {pages} {026114} (\bibinfo {year} {2005})},\
  \Eprint {http://arxiv.org/abs/0410484} {arXiv:0410484 [cond-mat]}
  \BibitemShut {NoStop}%
\bibitem [{\citenamefont {Wainrib}\ and\ \citenamefont
  {Touboul}(2013)}]{Wainrib2013}%
  \BibitemOpen
  \bibfield  {author} {\bibinfo {author} {\bibfnamefont {G.}~\bibnamefont
  {Wainrib}}\ and\ \bibinfo {author} {\bibfnamefont {J.}~\bibnamefont
  {Touboul}},\ }\href {\doibase 10.1103/PhysRevLett.110.118101} {\bibfield
  {journal} {\bibinfo  {journal} {Physical Review Letters}\ }\textbf {\bibinfo
  {volume} {110}},\ \bibinfo {pages} {118101} (\bibinfo {year} {2013})},\
  \Eprint {http://arxiv.org/abs/1210.5082} {arXiv:1210.5082} \BibitemShut
  {NoStop}%
\bibitem [{\citenamefont {{Sarao Mannelli}}\ \emph {et~al.}(2020)\citenamefont
  {{Sarao Mannelli}}, \citenamefont {Biroli}, \citenamefont {Cammarota},
  \citenamefont {Krzakala}, \citenamefont {Urbani},\ and\ \citenamefont
  {Zdeborov{\'{a}}}}]{SaraoMannelli2020}%
  \BibitemOpen
  \bibfield  {author} {\bibinfo {author} {\bibfnamefont {S.}~\bibnamefont
  {{Sarao Mannelli}}}, \bibinfo {author} {\bibfnamefont {G.}~\bibnamefont
  {Biroli}}, \bibinfo {author} {\bibfnamefont {C.}~\bibnamefont {Cammarota}},
  \bibinfo {author} {\bibfnamefont {F.}~\bibnamefont {Krzakala}}, \bibinfo
  {author} {\bibfnamefont {P.}~\bibnamefont {Urbani}}, \ and\ \bibinfo {author}
  {\bibfnamefont {L.}~\bibnamefont {Zdeborov{\'{a}}}},\ }\href {\doibase
  10.1103/PhysRevX.10.011057} {\bibfield  {journal} {\bibinfo  {journal}
  {Physical Review X}\ }\textbf {\bibinfo {volume} {10}},\ \bibinfo {pages}
  {011057} (\bibinfo {year} {2020})},\ \Eprint
  {http://arxiv.org/abs/1812.09066} {arXiv:1812.09066} \BibitemShut {NoStop}%
\bibitem [{\citenamefont {Mannelli}\ \emph {et~al.}(2019)\citenamefont
  {Mannelli}, \citenamefont {Biroli}, \citenamefont {Cammarota}, \citenamefont
  {Krzakala},\ and\ \citenamefont {Zdeborov{\'{a}}}}]{Mannelli2019}%
  \BibitemOpen
  \bibfield  {author} {\bibinfo {author} {\bibfnamefont {S.~S.}\ \bibnamefont
  {Mannelli}}, \bibinfo {author} {\bibfnamefont {G.}~\bibnamefont {Biroli}},
  \bibinfo {author} {\bibfnamefont {C.}~\bibnamefont {Cammarota}}, \bibinfo
  {author} {\bibfnamefont {F.}~\bibnamefont {Krzakala}}, \ and\ \bibinfo
  {author} {\bibfnamefont {L.}~\bibnamefont {Zdeborov{\'{a}}}},\ }\href@noop {}
  {\bibfield  {journal} {\bibinfo  {journal} {Advances in Neural Information
  Processing Systems}\ }\textbf {\bibinfo {volume} {33}} (\bibinfo {year}
  {2019})},\ \Eprint {http://arxiv.org/abs/1907.08226} {arXiv:1907.08226}
  \BibitemShut {NoStop}%
\bibitem [{\citenamefont {Cugliandolo}\ \emph {et~al.}(1996)\citenamefont
  {Cugliandolo}, \citenamefont {Kurchan},\ and\ \citenamefont {{Le
  Doussal}}}]{Cugliandolo1996}%
  \BibitemOpen
  \bibfield  {author} {\bibinfo {author} {\bibfnamefont {L.~F.}\ \bibnamefont
  {Cugliandolo}}, \bibinfo {author} {\bibfnamefont {J.}~\bibnamefont
  {Kurchan}}, \ and\ \bibinfo {author} {\bibfnamefont {P.}~\bibnamefont {{Le
  Doussal}}},\ }\href {\doibase 10.1103/PhysRevLett.76.2390} {\bibfield
  {journal} {\bibinfo  {journal} {Physical Review Letters}\ }\textbf {\bibinfo
  {volume} {76}},\ \bibinfo {pages} {2390} (\bibinfo {year}
  {1996})}\BibitemShut {NoStop}%
\bibitem [{\citenamefont {Cugliandolo}\ and\ \citenamefont {{Le
  Doussal}}(1996)}]{Cugliandolo1996a}%
  \BibitemOpen
  \bibfield  {author} {\bibinfo {author} {\bibfnamefont {L.~F.}\ \bibnamefont
  {Cugliandolo}}\ and\ \bibinfo {author} {\bibfnamefont {P.}~\bibnamefont {{Le
  Doussal}}},\ }\href {\doibase 10.1103/PhysRevE.53.1525} {\bibfield  {journal}
  {\bibinfo  {journal} {Physical Review E}\ }\textbf {\bibinfo {volume} {53}},\
  \bibinfo {pages} {1525} (\bibinfo {year} {1996})}\BibitemShut {NoStop}%
\bibitem [{\citenamefont {Altieri}\ \emph {et~al.}(2021)\citenamefont
  {Altieri}, \citenamefont {Roy}, \citenamefont {Cammarota},\ and\
  \citenamefont {Biroli}}]{Altieri2021}%
  \BibitemOpen
  \bibfield  {author} {\bibinfo {author} {\bibfnamefont {A.}~\bibnamefont
  {Altieri}}, \bibinfo {author} {\bibfnamefont {F.}~\bibnamefont {Roy}},
  \bibinfo {author} {\bibfnamefont {C.}~\bibnamefont {Cammarota}}, \ and\
  \bibinfo {author} {\bibfnamefont {G.}~\bibnamefont {Biroli}},\ }\href
  {\doibase 10.1103/PhysRevLett.126.258301} {\bibfield  {journal} {\bibinfo
  {journal} {Physical Review Letters}\ }\textbf {\bibinfo {volume} {126}},\
  \bibinfo {pages} {258301} (\bibinfo {year} {2021})},\ \Eprint
  {http://arxiv.org/abs/2009.10565} {arXiv:2009.10565} \BibitemShut {NoStop}%
\bibitem [{\citenamefont {Arous}\ \emph {et~al.}(2021)\citenamefont {Arous},
  \citenamefont {Fyodorov},\ and\ \citenamefont {Khoruzhenko}}]{Arous2021}%
  \BibitemOpen
  \bibfield  {author} {\bibinfo {author} {\bibfnamefont {G.~B.}\ \bibnamefont
  {Arous}}, \bibinfo {author} {\bibfnamefont {Y.~V.}\ \bibnamefont {Fyodorov}},
  \ and\ \bibinfo {author} {\bibfnamefont {B.~A.}\ \bibnamefont
  {Khoruzhenko}},\ }\href {\doibase 10.1073/pnas.2023719118} {\bibfield
  {journal} {\bibinfo  {journal} {Proceedings of the National Academy of
  Sciences of the United States of America}\ }\textbf {\bibinfo {volume} {118}}
  (\bibinfo {year} {2021}),\ 10.1073/pnas.2023719118},\ \Eprint
  {http://arxiv.org/abs/2008.00690} {arXiv:2008.00690} \BibitemShut {NoStop}%
\bibitem [{MNI()}]{MNIST_dataset}%
  \BibitemOpen
  \href {http://yann.lecun.com/exdb/mnist/} {\enquote {\bibinfo {title} {{MNIST
  dataset, http://yann.lecun.com/exdb/mnist/}},}\ }\BibitemShut {NoStop}%
\bibitem [{\citenamefont {Jolliffe}(2002)}]{Jolliffe2002}%
  \BibitemOpen
  \bibfield  {author} {\bibinfo {author} {\bibfnamefont {I.~T.}\ \bibnamefont
  {Jolliffe}},\ }\href@noop {} {\emph {\bibinfo {title} {{Principal Component
  Analysis}}}},\ \bibinfo {edition} {2nd}\ ed.\ (\bibinfo  {publisher}
  {Springer New York},\ \bibinfo {year} {2002})\BibitemShut {NoStop}%
\bibitem [{\citenamefont {Ying}(2019)}]{Ying2019}%
  \BibitemOpen
  \bibfield  {author} {\bibinfo {author} {\bibfnamefont {X.}~\bibnamefont
  {Ying}},\ }\href@noop {} {\bibfield  {journal} {\bibinfo  {journal} {Journal
  of Physics: Conference Series}\ }\textbf {\bibinfo {volume} {1168}},\
  \bibinfo {pages} {022022} (\bibinfo {year} {2019})}\BibitemShut {NoStop}%
\bibitem [{\citenamefont {Cugliandolo}\ and\ \citenamefont
  {Dean}(1995)}]{Cugliandolo1995}%
  \BibitemOpen
  \bibfield  {author} {\bibinfo {author} {\bibfnamefont {L.~F.}\ \bibnamefont
  {Cugliandolo}}\ and\ \bibinfo {author} {\bibfnamefont {D.~S.}\ \bibnamefont
  {Dean}},\ }\href {\doibase 10.1088/0305-4470/28/15/003} {\bibfield  {journal}
  {\bibinfo  {journal} {Journal of Physics A: Mathematical and General}\
  }\textbf {\bibinfo {volume} {28}},\ \bibinfo {pages} {4213} (\bibinfo {year}
  {1995})}\BibitemShut {NoStop}%
\bibitem [{\citenamefont {Fyodorov}\ and\ \citenamefont
  {Tublin}(2022)}]{Fyodorov2022}%
  \BibitemOpen
  \bibfield  {author} {\bibinfo {author} {\bibfnamefont {Y.~V.}\ \bibnamefont
  {Fyodorov}}\ and\ \bibinfo {author} {\bibfnamefont {R.}~\bibnamefont
  {Tublin}},\ }\href {\doibase 10.1088/1751-8121/ac6d8e} {\bibfield  {journal}
  {\bibinfo  {journal} {Journal of Physics A: Mathematical and Theoretical}\
  }\textbf {\bibinfo {volume} {55}},\ \bibinfo {pages} {244008} (\bibinfo
  {year} {2022})}\BibitemShut {NoStop}%
\bibitem [{\citenamefont {Pihlajamaa}\ \emph {et~al.}(2023)\citenamefont
  {Pihlajamaa}, \citenamefont {Debets}, \citenamefont {Laudicina},\ and\
  \citenamefont {Janssen}}]{Pihlajamaa2023a}%
  \BibitemOpen
  \bibfield  {author} {\bibinfo {author} {\bibfnamefont {I.}~\bibnamefont
  {Pihlajamaa}}, \bibinfo {author} {\bibfnamefont {V.~E.}\ \bibnamefont
  {Debets}}, \bibinfo {author} {\bibfnamefont {C.~C.~L.}\ \bibnamefont
  {Laudicina}}, \ and\ \bibinfo {author} {\bibfnamefont {L.~M.~C.}\
  \bibnamefont {Janssen}},\ }\href {http://arxiv.org/abs/2307.03443} {\bibfield
   {journal} {\bibinfo  {journal} {SciPost Physics}\ }\textbf {\bibinfo
  {volume} {Submission}},\ \bibinfo {pages} {1} (\bibinfo {year} {2023})},\
  \Eprint {http://arxiv.org/abs/2307.03443} {arXiv:2307.03443} \BibitemShut
  {NoStop}%
\bibitem [{\citenamefont {Kob}\ and\ \citenamefont {Andersen}(1995)}]{Kob1995}%
  \BibitemOpen
  \bibfield  {author} {\bibinfo {author} {\bibfnamefont {W.}~\bibnamefont
  {Kob}}\ and\ \bibinfo {author} {\bibfnamefont {H.~C.}\ \bibnamefont
  {Andersen}},\ }\href {\doibase 10.1103/PhysRevE.52.4134} {\bibfield
  {journal} {\bibinfo  {journal} {Physical Review E}\ }\textbf {\bibinfo
  {volume} {51}},\ \bibinfo {pages} {4626} (\bibinfo {year} {1995})},\ \Eprint
  {http://arxiv.org/abs/9505118} {arXiv:9505118 [cond-mat]} \BibitemShut
  {NoStop}%
\bibitem [{\citenamefont {Hopkins}\ \emph {et~al.}(2010)\citenamefont
  {Hopkins}, \citenamefont {Fortini}, \citenamefont {Archer},\ and\
  \citenamefont {Schmidt}}]{Hopkins2010}%
  \BibitemOpen
  \bibfield  {author} {\bibinfo {author} {\bibfnamefont {P.}~\bibnamefont
  {Hopkins}}, \bibinfo {author} {\bibfnamefont {A.}~\bibnamefont {Fortini}},
  \bibinfo {author} {\bibfnamefont {A.~J.}\ \bibnamefont {Archer}}, \ and\
  \bibinfo {author} {\bibfnamefont {M.}~\bibnamefont {Schmidt}},\ }\href
  {\doibase 10.1063/1.3511719} {\bibfield  {journal} {\bibinfo  {journal}
  {Journal of Chemical Physics}\ }\textbf {\bibinfo {volume} {133}} (\bibinfo
  {year} {2010}),\ 10.1063/1.3511719},\ \Eprint
  {http://arxiv.org/abs/1010.2124} {arXiv:1010.2124} \BibitemShut {NoStop}%
\bibitem [{\citenamefont {Reichman}\ and\ \citenamefont
  {Charbonneau}(2005)}]{Reichman2005b}%
  \BibitemOpen
  \bibfield  {author} {\bibinfo {author} {\bibfnamefont {D.~R.}\ \bibnamefont
  {Reichman}}\ and\ \bibinfo {author} {\bibfnamefont {P.}~\bibnamefont
  {Charbonneau}},\ }\href {\doibase 10.1088/1742-5468/2005/05/P05013}
  {\bibfield  {journal} {\bibinfo  {journal} {Journal of Statistical Mechanics:
  Theory and Experiment}\ ,\ \bibinfo {pages} {P05013}} (\bibinfo {year}
  {2005})}\BibitemShut {NoStop}%
\bibitem [{\citenamefont {Jules}\ \emph {et~al.}()\citenamefont {Jules},
  \citenamefont {Brener}, \citenamefont {Kachman}, \citenamefont {Levi},\ and\
  \citenamefont {Bar-sinai}}]{Jules2023}%
  \BibitemOpen
  \bibfield  {author} {\bibinfo {author} {\bibfnamefont {T.}~\bibnamefont
  {Jules}}, \bibinfo {author} {\bibfnamefont {G.}~\bibnamefont {Brener}},
  \bibinfo {author} {\bibfnamefont {T.}~\bibnamefont {Kachman}}, \bibinfo
  {author} {\bibfnamefont {N.}~\bibnamefont {Levi}}, \ and\ \bibinfo {author}
  {\bibfnamefont {Y.}~\bibnamefont {Bar-sinai}},\ }\href@noop {} {\ }\Eprint
  {http://arxiv.org/abs/2304.01335v2} {arXiv:2304.01335v2} \BibitemShut
  {NoStop}%
\bibitem [{\citenamefont {Olsen}\ \emph {et~al.}(2001)\citenamefont {Olsen},
  \citenamefont {Christensen},\ and\ \citenamefont {Dyre}}]{Olsen2001}%
  \BibitemOpen
  \bibfield  {author} {\bibinfo {author} {\bibfnamefont {N.~B.}\ \bibnamefont
  {Olsen}}, \bibinfo {author} {\bibfnamefont {T.}~\bibnamefont {Christensen}},
  \ and\ \bibinfo {author} {\bibfnamefont {J.~C.}\ \bibnamefont {Dyre}},\
  }\href {\doibase 10.1103/PhysRevLett.86.1271} {\bibfield  {journal} {\bibinfo
   {journal} {Physical Review Letters}\ }\textbf {\bibinfo {volume} {86}},\
  \bibinfo {pages} {1271} (\bibinfo {year} {2001})},\ \Eprint
  {http://arxiv.org/abs/0006165} {arXiv:0006165 [cond-mat]} \BibitemShut
  {NoStop}%
\bibitem [{\citenamefont {Li}(2000)}]{Li2000}%
  \BibitemOpen
  \bibfield  {author} {\bibinfo {author} {\bibfnamefont {R.}~\bibnamefont
  {Li}},\ }\href {\doibase 10.1016/s0921-5093(99)00602-4} {\bibfield  {journal}
  {\bibinfo  {journal} {Materials Science and Engineering: A}\ }\textbf
  {\bibinfo {volume} {278}},\ \bibinfo {pages} {36} (\bibinfo {year}
  {2000})}\BibitemShut {NoStop}%
\bibitem [{\citenamefont {Billon}\ \emph {et~al.}(2023)\citenamefont {Billon},
  \citenamefont {Federico}, \citenamefont {Rival}, \citenamefont {Bouvard},\
  and\ \citenamefont {Burr}}]{Billon2023}%
  \BibitemOpen
  \bibfield  {author} {\bibinfo {author} {\bibfnamefont {N.}~\bibnamefont
  {Billon}}, \bibinfo {author} {\bibfnamefont {C.~E.}\ \bibnamefont
  {Federico}}, \bibinfo {author} {\bibfnamefont {G.}~\bibnamefont {Rival}},
  \bibinfo {author} {\bibfnamefont {J.~L.}\ \bibnamefont {Bouvard}}, \ and\
  \bibinfo {author} {\bibfnamefont {A.}~\bibnamefont {Burr}},\ }\href {\doibase
  10.3390/ijms24043944} {\bibfield  {journal} {\bibinfo  {journal}
  {International Journal of Molecular Sciences}\ }\textbf {\bibinfo {volume}
  {24}},\ \bibinfo {pages} {3944} (\bibinfo {year} {2023})}\BibitemShut
  {NoStop}%
\bibitem [{\citenamefont {Schnyder}\ \emph {et~al.}(2011)\citenamefont
  {Schnyder}, \citenamefont {H{\"{o}}fling}, \citenamefont {Franosch},\ and\
  \citenamefont {Voigtmann}}]{Schnyder2011}%
  \BibitemOpen
  \bibfield  {author} {\bibinfo {author} {\bibfnamefont {S.~K.}\ \bibnamefont
  {Schnyder}}, \bibinfo {author} {\bibfnamefont {F.}~\bibnamefont
  {H{\"{o}}fling}}, \bibinfo {author} {\bibfnamefont {T.}~\bibnamefont
  {Franosch}}, \ and\ \bibinfo {author} {\bibfnamefont {T.}~\bibnamefont
  {Voigtmann}},\ }\href {\doibase 10.1088/0953-8984/23/23/234121} {\bibfield
  {journal} {\bibinfo  {journal} {Journal of Physics: Condensed Matter}\
  }\textbf {\bibinfo {volume} {23}},\ \bibinfo {pages} {234121} (\bibinfo
  {year} {2011})}\BibitemShut {NoStop}%
\bibitem [{\citenamefont {Mandal}\ \emph {et~al.}(2017)\citenamefont {Mandal},
  \citenamefont {Spanner-Denzer}, \citenamefont {Leitmann},\ and\ \citenamefont
  {Franosch}}]{Mandal2017a}%
  \BibitemOpen
  \bibfield  {author} {\bibinfo {author} {\bibfnamefont {S.}~\bibnamefont
  {Mandal}}, \bibinfo {author} {\bibfnamefont {M.}~\bibnamefont
  {Spanner-Denzer}}, \bibinfo {author} {\bibfnamefont {S.}~\bibnamefont
  {Leitmann}}, \ and\ \bibinfo {author} {\bibfnamefont {T.}~\bibnamefont
  {Franosch}},\ }\href {\doibase 10.1140/epjst/e2017-70077-5} {\bibfield
  {journal} {\bibinfo  {journal} {European Physical Journal: Special Topics}\
  }\textbf {\bibinfo {volume} {226}},\ \bibinfo {pages} {3129} (\bibinfo {year}
  {2017})}\BibitemShut {NoStop}%
\bibitem [{\citenamefont {Spanner}\ \emph {et~al.}(2013)\citenamefont
  {Spanner}, \citenamefont {Schnyder}, \citenamefont {H{\"{o}}fling},
  \citenamefont {Voigtmann},\ and\ \citenamefont {Franosch}}]{Spanner2013}%
  \BibitemOpen
  \bibfield  {author} {\bibinfo {author} {\bibfnamefont {M.}~\bibnamefont
  {Spanner}}, \bibinfo {author} {\bibfnamefont {S.~K.}\ \bibnamefont
  {Schnyder}}, \bibinfo {author} {\bibfnamefont {F.}~\bibnamefont
  {H{\"{o}}fling}}, \bibinfo {author} {\bibfnamefont {T.}~\bibnamefont
  {Voigtmann}}, \ and\ \bibinfo {author} {\bibfnamefont {T.}~\bibnamefont
  {Franosch}},\ }\href {\doibase 10.1039/c2sm27060a} {\bibfield  {journal}
  {\bibinfo  {journal} {Soft Matter}\ }\textbf {\bibinfo {volume} {9}},\
  \bibinfo {pages} {1604} (\bibinfo {year} {2013})}\BibitemShut {NoStop}%
\bibitem [{\citenamefont {G{\"{o}}tze}(1999)}]{Gotze1999}%
  \BibitemOpen
  \bibfield  {author} {\bibinfo {author} {\bibfnamefont {W.}~\bibnamefont
  {G{\"{o}}tze}},\ }\href {\doibase 10.1088/0953-8984/11/10A/002} {\bibfield
  {journal} {\bibinfo  {journal} {Journal of Physics Condensed Matter}\
  }\textbf {\bibinfo {volume} {11}},\ \bibinfo {pages} {A1} (\bibinfo {year}
  {1999})}\BibitemShut {NoStop}%
\bibitem [{\citenamefont {Leutheusser}(1984)}]{Leutheusser1984}%
  \BibitemOpen
  \bibfield  {author} {\bibinfo {author} {\bibfnamefont {E.}~\bibnamefont
  {Leutheusser}},\ }\href {\doibase 10.1103/PhysRevA.29.2765} {\bibfield
  {journal} {\bibinfo  {journal} {Physical Review A}\ }\textbf {\bibinfo
  {volume} {29}},\ \bibinfo {pages} {2765} (\bibinfo {year}
  {1984})}\BibitemShut {NoStop}%
\bibitem [{\citenamefont {Charbonneau}\ \emph {et~al.}(2012)\citenamefont
  {Charbonneau}, \citenamefont {Ikeda}, \citenamefont {Parisi},\ and\
  \citenamefont {Zamponi}}]{Charbonneau2012}%
  \BibitemOpen
  \bibfield  {author} {\bibinfo {author} {\bibfnamefont {P.}~\bibnamefont
  {Charbonneau}}, \bibinfo {author} {\bibfnamefont {A.}~\bibnamefont {Ikeda}},
  \bibinfo {author} {\bibfnamefont {G.}~\bibnamefont {Parisi}}, \ and\ \bibinfo
  {author} {\bibfnamefont {F.}~\bibnamefont {Zamponi}},\ }\href {\doibase
  10.1073/pnas.1211825109} {\bibfield  {journal} {\bibinfo  {journal}
  {Proceedings of the National Academy of Sciences of the United States of
  America}\ }\textbf {\bibinfo {volume} {109}},\ \bibinfo {pages} {13939}
  (\bibinfo {year} {2012})}\BibitemShut {NoStop}%
\bibitem [{\citenamefont {Charbonneau}\ \emph
  {et~al.}(2014{\natexlab{b}})\citenamefont {Charbonneau}, \citenamefont {Jin},
  \citenamefont {Parisi},\ and\ \citenamefont {Zamponi}}]{Charbonneau2014}%
  \BibitemOpen
  \bibfield  {author} {\bibinfo {author} {\bibfnamefont {P.}~\bibnamefont
  {Charbonneau}}, \bibinfo {author} {\bibfnamefont {Y.}~\bibnamefont {Jin}},
  \bibinfo {author} {\bibfnamefont {G.}~\bibnamefont {Parisi}}, \ and\ \bibinfo
  {author} {\bibfnamefont {F.}~\bibnamefont {Zamponi}},\ }\href {\doibase
  10.1073/pnas.1417182111} {\bibfield  {journal} {\bibinfo  {journal}
  {Proceedings of the National Academy of Sciences of the United States of
  America}\ }\textbf {\bibinfo {volume} {111}},\ \bibinfo {pages} {15025}
  (\bibinfo {year} {2014}{\natexlab{b}})},\ \Eprint
  {http://arxiv.org/abs/1407.5677} {arXiv:1407.5677} \BibitemShut {NoStop}%
\bibitem [{\citenamefont {Weeks}\ and\ \citenamefont
  {Weitz}(2002)}]{Weeks2002}%
  \BibitemOpen
  \bibfield  {author} {\bibinfo {author} {\bibfnamefont {E.~R.}\ \bibnamefont
  {Weeks}}\ and\ \bibinfo {author} {\bibfnamefont {D.~A.}\ \bibnamefont
  {Weitz}},\ }\href {\doibase 10.1103/PhysRevLett.89.095704} {\bibfield
  {journal} {\bibinfo  {journal} {Physical Review Letters}\ }\textbf {\bibinfo
  {volume} {89}},\ \bibinfo {pages} {095704} (\bibinfo {year} {2002})},\
  \Eprint {http://arxiv.org/abs/0107279} {arXiv:0107279 [cond-mat]}
  \BibitemShut {NoStop}%
\bibitem [{\citenamefont {Jack}\ and\ \citenamefont
  {Garrahan}(2005)}]{Jack2005}%
  \BibitemOpen
  \bibfield  {author} {\bibinfo {author} {\bibfnamefont {R.~L.}\ \bibnamefont
  {Jack}}\ and\ \bibinfo {author} {\bibfnamefont {J.~P.}\ \bibnamefont
  {Garrahan}},\ }\href {\doibase 10.1063/1.2075067} {\bibfield  {journal}
  {\bibinfo  {journal} {Journal of Chemical Physics}\ }\textbf {\bibinfo
  {volume} {123}},\ \bibinfo {pages} {164508} (\bibinfo {year} {2005})},\
  \Eprint {http://arxiv.org/abs/0507370} {arXiv:0507370 [cond-mat]}
  \BibitemShut {NoStop}%
\bibitem [{\citenamefont {Biroli}\ \emph {et~al.}(2021)\citenamefont {Biroli},
  \citenamefont {Charbonneau}, \citenamefont {Hu}, \citenamefont {Ikeda},
  \citenamefont {Szamel},\ and\ \citenamefont {Zamponi}}]{Biroli2021}%
  \BibitemOpen
  \bibfield  {author} {\bibinfo {author} {\bibfnamefont {G.}~\bibnamefont
  {Biroli}}, \bibinfo {author} {\bibfnamefont {P.}~\bibnamefont {Charbonneau}},
  \bibinfo {author} {\bibfnamefont {Y.}~\bibnamefont {Hu}}, \bibinfo {author}
  {\bibfnamefont {H.}~\bibnamefont {Ikeda}}, \bibinfo {author} {\bibfnamefont
  {G.}~\bibnamefont {Szamel}}, \ and\ \bibinfo {author} {\bibfnamefont
  {F.}~\bibnamefont {Zamponi}},\ }\href {\doibase 10.1021/acs.jpcb.1c02067}
  {\bibfield  {journal} {\bibinfo  {journal} {Journal of Physical Chemistry B}\
  }\textbf {\bibinfo {volume} {125}},\ \bibinfo {pages} {6244} (\bibinfo {year}
  {2021})}\BibitemShut {NoStop}%
\bibitem [{\citenamefont {Tokuyama}\ and\ \citenamefont
  {Kimura}(2008)}]{Tokuyama2008}%
  \BibitemOpen
  \bibfield  {author} {\bibinfo {author} {\bibfnamefont {M.}~\bibnamefont
  {Tokuyama}}\ and\ \bibinfo {author} {\bibfnamefont {Y.}~\bibnamefont
  {Kimura}},\ }\href {\doibase 10.1016/j.physa.2008.04.021} {\bibfield
  {journal} {\bibinfo  {journal} {Physica A: Statistical Mechanics and its
  Applications}\ }\textbf {\bibinfo {volume} {387}},\ \bibinfo {pages} {4749}
  (\bibinfo {year} {2008})}\BibitemShut {NoStop}%
\bibitem [{\citenamefont {Schr{\o}der}\ and\ \citenamefont
  {Dyre}(2020)}]{Schroder2020}%
  \BibitemOpen
  \bibfield  {author} {\bibinfo {author} {\bibfnamefont {T.~B.}\ \bibnamefont
  {Schr{\o}der}}\ and\ \bibinfo {author} {\bibfnamefont {J.~C.}\ \bibnamefont
  {Dyre}},\ }\href {\doibase 10.1063/5.0004093} {\bibfield  {journal} {\bibinfo
   {journal} {Journal of Chemical Physics}\ }\textbf {\bibinfo {volume}
  {152}},\ \bibinfo {pages} {141101} (\bibinfo {year} {2020})},\ \Eprint
  {http://arxiv.org/abs/1905.11514} {arXiv:1905.11514} \BibitemShut {NoStop}%
\bibitem [{\citenamefont {Crisanti}\ and\ \citenamefont
  {Leuzzi}(2015)}]{Crisanti2015}%
  \BibitemOpen
  \bibfield  {author} {\bibinfo {author} {\bibfnamefont {A.}~\bibnamefont
  {Crisanti}}\ and\ \bibinfo {author} {\bibfnamefont {L.}~\bibnamefont
  {Leuzzi}},\ }\href {\doibase 10.1016/j.jnoncrysol.2014.07.048} {\bibfield
  {journal} {\bibinfo  {journal} {Journal of Non-Crystalline Solids}\ }\textbf
  {\bibinfo {volume} {407}},\ \bibinfo {pages} {110} (\bibinfo {year}
  {2015})}\BibitemShut {NoStop}%
\bibitem [{\citenamefont {Ferrari}\ \emph {et~al.}(2012)\citenamefont
  {Ferrari}, \citenamefont {Leuzzi}, \citenamefont {Parisi},\ and\
  \citenamefont {Rizzo}}]{Ferrari2012}%
  \BibitemOpen
  \bibfield  {author} {\bibinfo {author} {\bibfnamefont {U.}~\bibnamefont
  {Ferrari}}, \bibinfo {author} {\bibfnamefont {L.}~\bibnamefont {Leuzzi}},
  \bibinfo {author} {\bibfnamefont {G.}~\bibnamefont {Parisi}}, \ and\ \bibinfo
  {author} {\bibfnamefont {T.}~\bibnamefont {Rizzo}},\ }\href {\doibase
  10.1103/PhysRevB.86.014204} {\bibfield  {journal} {\bibinfo  {journal}
  {Physical Review B - Condensed Matter and Materials Physics}\ }\textbf
  {\bibinfo {volume} {86}},\ \bibinfo {pages} {014204} (\bibinfo {year}
  {2012})}\BibitemShut {NoStop}%
\bibitem [{\citenamefont {Cavagna}\ \emph {et~al.}(2003)\citenamefont
  {Cavagna}, \citenamefont {Giardina},\ and\ \citenamefont
  {Grigera}}]{Cavagna2003}%
  \BibitemOpen
  \bibfield  {author} {\bibinfo {author} {\bibfnamefont {A.}~\bibnamefont
  {Cavagna}}, \bibinfo {author} {\bibfnamefont {I.}~\bibnamefont {Giardina}}, \
  and\ \bibinfo {author} {\bibfnamefont {T.~S.}\ \bibnamefont {Grigera}},\
  }\href {\doibase 10.1088/0305-4470/36/43/004} {\bibfield  {journal} {\bibinfo
   {journal} {Journal of Physics A: Mathematical and General}\ }\textbf
  {\bibinfo {volume} {36}},\ \bibinfo {pages} {10721} (\bibinfo {year}
  {2003})},\ \Eprint {http://arxiv.org/abs/0212438} {arXiv:0212438 [cond-mat]}
  \BibitemShut {NoStop}%
\bibitem [{\citenamefont {Castellani}\ and\ \citenamefont
  {Cavagna}(2005)}]{Castellani2005}%
  \BibitemOpen
  \bibfield  {author} {\bibinfo {author} {\bibfnamefont {T.}~\bibnamefont
  {Castellani}}\ and\ \bibinfo {author} {\bibfnamefont {A.}~\bibnamefont
  {Cavagna}},\ }\href {\doibase 10.1088/1742-5468/2005/05/P05012} {\bibfield
  {journal} {\bibinfo  {journal} {Journal of Statistical Mechanics: Theory and
  Experiment}\ ,\ \bibinfo {pages} {P05012}} (\bibinfo {year} {2005})},\
  \Eprint {http://arxiv.org/abs/0505032} {arXiv:0505032 [cond-mat]}
  \BibitemShut {NoStop}%
\bibitem [{\citenamefont {Cavagna}\ \emph {et~al.}(2001)\citenamefont
  {Cavagna}, \citenamefont {Giardina},\ and\ \citenamefont
  {Parisi}}]{Cavagna2001}%
  \BibitemOpen
  \bibfield  {author} {\bibinfo {author} {\bibfnamefont {A.}~\bibnamefont
  {Cavagna}}, \bibinfo {author} {\bibfnamefont {I.}~\bibnamefont {Giardina}}, \
  and\ \bibinfo {author} {\bibfnamefont {G.}~\bibnamefont {Parisi}},\ }\href
  {\doibase 10.1088/0305-4470/34/26/302} {\bibfield  {journal} {\bibinfo
  {journal} {Journal of Physics A: Mathematical and General}\ }\textbf
  {\bibinfo {volume} {34}},\ \bibinfo {pages} {5317} (\bibinfo {year}
  {2001})},\ \Eprint {http://arxiv.org/abs/0104537} {arXiv:0104537 [cond-mat]}
  \BibitemShut {NoStop}%
\bibitem [{\citenamefont {Şimşekli}\ \emph {et~al.}(2019)\citenamefont
  {Şimşekli}, \citenamefont {Sagun},\ and\ \citenamefont
  {Gurbuzbalaban}}]{Simsekli2019}%
  \BibitemOpen
  \bibfield  {author} {\bibinfo {author} {\bibfnamefont {U.}~\bibnamefont
  {Şimşekli}}, \bibinfo {author} {\bibfnamefont {L.}~\bibnamefont {Sagun}}, \
  and\ \bibinfo {author} {\bibfnamefont {M.}~\bibnamefont {Gurbuzbalaban}},\
  }\href@noop {} {\bibfield  {journal} {\bibinfo  {journal} {36th International
  Conference on Machine Learning,}\ }\textbf {\bibinfo {volume} {97}},\
  \bibinfo {pages} {5827} (\bibinfo {year} {2019})},\ \Eprint
  {http://arxiv.org/abs/1901.06053} {arXiv:1901.06053} \BibitemShut {NoStop}%
\bibitem [{\citenamefont {Xie}\ \emph {et~al.}(2021)\citenamefont {Xie},
  \citenamefont {Sato},\ and\ \citenamefont {Sugiyama}}]{Xie2020}%
  \BibitemOpen
  \bibfield  {author} {\bibinfo {author} {\bibfnamefont {Z.}~\bibnamefont
  {Xie}}, \bibinfo {author} {\bibfnamefont {I.}~\bibnamefont {Sato}}, \ and\
  \bibinfo {author} {\bibfnamefont {M.}~\bibnamefont {Sugiyama}},\ }in\ \href
  {http://arxiv.org/abs/2002.03495} {\emph {\bibinfo {booktitle} {International
  Conference on Learning Representations}}}\ (\bibinfo {year} {2021})\ pp.\
  \bibinfo {pages} {1--28},\ \Eprint {http://arxiv.org/abs/2002.03495}
  {arXiv:2002.03495} \BibitemShut {NoStop}%
\bibitem [{\citenamefont {Draxler}\ \emph {et~al.}(2018)\citenamefont
  {Draxler}, \citenamefont {Veschgini}, \citenamefont {Salmhofer},\ and\
  \citenamefont {Hamprecht}}]{Draxler2018}%
  \BibitemOpen
  \bibfield  {author} {\bibinfo {author} {\bibfnamefont {F.}~\bibnamefont
  {Draxler}}, \bibinfo {author} {\bibfnamefont {K.}~\bibnamefont {Veschgini}},
  \bibinfo {author} {\bibfnamefont {M.}~\bibnamefont {Salmhofer}}, \ and\
  \bibinfo {author} {\bibfnamefont {F.~A.}\ \bibnamefont {Hamprecht}},\ }in\
  \href@noop {} {\emph {\bibinfo {booktitle} {35th International Conference on
  Machine Learning,}}}\ (\bibinfo {year} {2018})\ p.\ \bibinfo {pages} {1309},\
  \Eprint {http://arxiv.org/abs/1803.00885} {arXiv:1803.00885} \BibitemShut
  {NoStop}%
\bibitem [{\citenamefont {Annesi}\ \emph {et~al.}(2023)\citenamefont {Annesi},
  \citenamefont {Lauditi}, \citenamefont {Lucibello}, \citenamefont
  {Malatesta}, \citenamefont {Perugini}, \citenamefont {Pittorino},\ and\
  \citenamefont {Saglietti}}]{Annesi2023}%
  \BibitemOpen
  \bibfield  {author} {\bibinfo {author} {\bibfnamefont {B.~L.}\ \bibnamefont
  {Annesi}}, \bibinfo {author} {\bibfnamefont {C.}~\bibnamefont {Lauditi}},
  \bibinfo {author} {\bibfnamefont {C.}~\bibnamefont {Lucibello}}, \bibinfo
  {author} {\bibfnamefont {E.~M.}\ \bibnamefont {Malatesta}}, \bibinfo {author}
  {\bibfnamefont {G.}~\bibnamefont {Perugini}}, \bibinfo {author}
  {\bibfnamefont {F.}~\bibnamefont {Pittorino}}, \ and\ \bibinfo {author}
  {\bibfnamefont {L.}~\bibnamefont {Saglietti}},\ }\href {\doibase
  10.1103/PhysRevLett.131.227301} {\bibfield  {journal} {\bibinfo  {journal}
  {Physical Review Letters}\ }\textbf {\bibinfo {volume} {131}},\ \bibinfo
  {pages} {227301} (\bibinfo {year} {2023})},\ \Eprint
  {http://arxiv.org/abs/2305.10623} {arXiv:2305.10623} \BibitemShut {NoStop}%
\bibitem [{\citenamefont {Einstein}(1905)}]{Einstein1905}%
  \BibitemOpen
  \bibfield  {author} {\bibinfo {author} {\bibfnamefont {A.}~\bibnamefont
  {Einstein}},\ }\href {\doibase 10.1002/andp.19053220806} {\bibfield
  {journal} {\bibinfo  {journal} {Annalen der Physik}\ }\textbf {\bibinfo
  {volume} {322}},\ \bibinfo {pages} {549} (\bibinfo {year}
  {1905})}\BibitemShut {NoStop}%
\bibitem [{\citenamefont {Banchio}\ \emph {et~al.}(1999)\citenamefont
  {Banchio}, \citenamefont {N{\"{a}}gele},\ and\ \citenamefont
  {Bergenholtz}}]{Banchio1999}%
  \BibitemOpen
  \bibfield  {author} {\bibinfo {author} {\bibfnamefont {A.~J.}\ \bibnamefont
  {Banchio}}, \bibinfo {author} {\bibfnamefont {G.}~\bibnamefont
  {N{\"{a}}gele}}, \ and\ \bibinfo {author} {\bibfnamefont {J.}~\bibnamefont
  {Bergenholtz}},\ }\href {\doibase 10.1063/1.480212} {\bibfield  {journal}
  {\bibinfo  {journal} {The Journal of Chemical Physics}\ }\textbf {\bibinfo
  {volume} {111}},\ \bibinfo {pages} {8721} (\bibinfo {year}
  {1999})}\BibitemShut {NoStop}%
\bibitem [{\citenamefont {Tarjus}\ and\ \citenamefont
  {Kivelson}(1995)}]{Tarjus1995}%
  \BibitemOpen
  \bibfield  {author} {\bibinfo {author} {\bibfnamefont {G.}~\bibnamefont
  {Tarjus}}\ and\ \bibinfo {author} {\bibfnamefont {D.}~\bibnamefont
  {Kivelson}},\ }\href {\doibase 10.4028/www.scientific.net/JMNM.20-21.541}
  {\bibfield  {journal} {\bibinfo  {journal} {The Journal of Chemical Physics}\
  }\textbf {\bibinfo {volume} {103}},\ \bibinfo {pages} {3071} (\bibinfo {year}
  {1995})}\BibitemShut {NoStop}%
\bibitem [{\citenamefont {Shi}\ \emph {et~al.}(2013)\citenamefont {Shi},
  \citenamefont {Debenedetti},\ and\ \citenamefont {Stillinger}}]{Shi2013}%
  \BibitemOpen
  \bibfield  {author} {\bibinfo {author} {\bibfnamefont {Z.}~\bibnamefont
  {Shi}}, \bibinfo {author} {\bibfnamefont {P.~G.}\ \bibnamefont
  {Debenedetti}}, \ and\ \bibinfo {author} {\bibfnamefont {F.~H.}\ \bibnamefont
  {Stillinger}},\ }\href {\doibase 10.1063/1.4775741} {\bibfield  {journal}
  {\bibinfo  {journal} {The Journal of Chemical Physics}\ }\textbf {\bibinfo
  {volume} {138}},\ \bibinfo {pages} {12A526} (\bibinfo {year}
  {2013})}\BibitemShut {NoStop}%
\bibitem [{\citenamefont {Mei}\ \emph {et~al.}(2019)\citenamefont {Mei},
  \citenamefont {Lu}, \citenamefont {An},\ and\ \citenamefont
  {Wang}}]{Mei2019}%
  \BibitemOpen
  \bibfield  {author} {\bibinfo {author} {\bibfnamefont {B.}~\bibnamefont
  {Mei}}, \bibinfo {author} {\bibfnamefont {Y.}~\bibnamefont {Lu}}, \bibinfo
  {author} {\bibfnamefont {L.}~\bibnamefont {An}}, \ and\ \bibinfo {author}
  {\bibfnamefont {Z.~G.}\ \bibnamefont {Wang}},\ }\href {\doibase
  10.1103/PhysRevE.100.052607} {\bibfield  {journal} {\bibinfo  {journal}
  {Physical Review E}\ }\textbf {\bibinfo {volume} {100}},\ \bibinfo {pages}
  {052607} (\bibinfo {year} {2019})}\BibitemShut {NoStop}%
\bibitem [{\citenamefont {Flenner}\ and\ \citenamefont
  {Szamel}(2005{\natexlab{a}})}]{Flenner2005a}%
  \BibitemOpen
  \bibfield  {author} {\bibinfo {author} {\bibfnamefont {E.}~\bibnamefont
  {Flenner}}\ and\ \bibinfo {author} {\bibfnamefont {G.}~\bibnamefont
  {Szamel}},\ }\href {\doibase 10.1103/PhysRevE.72.011205} {\bibfield
  {journal} {\bibinfo  {journal} {Physical Review E}\ }\textbf {\bibinfo
  {volume} {72}},\ \bibinfo {pages} {011205} (\bibinfo {year}
  {2005}{\natexlab{a}})},\ \Eprint {http://arxiv.org/abs/0505173}
  {arXiv:0505173 [cond-mat]} \BibitemShut {NoStop}%
\bibitem [{\citenamefont {Flenner}\ \emph {et~al.}(2014)\citenamefont
  {Flenner}, \citenamefont {Staley},\ and\ \citenamefont
  {Szamel}}]{Flenner2014}%
  \BibitemOpen
  \bibfield  {author} {\bibinfo {author} {\bibfnamefont {E.}~\bibnamefont
  {Flenner}}, \bibinfo {author} {\bibfnamefont {H.}~\bibnamefont {Staley}}, \
  and\ \bibinfo {author} {\bibfnamefont {G.}~\bibnamefont {Szamel}},\ }\href
  {\doibase 10.1103/PhysRevLett.112.097801} {\bibfield  {journal} {\bibinfo
  {journal} {Physical Review Letters}\ }\textbf {\bibinfo {volume} {112}},\
  \bibinfo {pages} {097801} (\bibinfo {year} {2014})},\ \Eprint
  {http://arxiv.org/abs/1310.1029} {arXiv:1310.1029} \BibitemShut {NoStop}%
\bibitem [{\citenamefont {Sengupta}\ \emph {et~al.}(2013)\citenamefont
  {Sengupta}, \citenamefont {Karmakar}, \citenamefont {Dasgupta},\ and\
  \citenamefont {Sastry}}]{Sengupta2013}%
  \BibitemOpen
  \bibfield  {author} {\bibinfo {author} {\bibfnamefont {S.}~\bibnamefont
  {Sengupta}}, \bibinfo {author} {\bibfnamefont {S.}~\bibnamefont {Karmakar}},
  \bibinfo {author} {\bibfnamefont {C.}~\bibnamefont {Dasgupta}}, \ and\
  \bibinfo {author} {\bibfnamefont {S.}~\bibnamefont {Sastry}},\ }\href
  {\doibase 10.1063/1.4792356} {\bibfield  {journal} {\bibinfo  {journal} {The
  Journal of Chemical Physics}\ }\textbf {\bibinfo {volume} {138}},\ \bibinfo
  {pages} {12A548} (\bibinfo {year} {2013})},\ \Eprint
  {http://arxiv.org/abs/1211.0686} {arXiv:1211.0686} \BibitemShut {NoStop}%
\bibitem [{\citenamefont {Risken}(1989)}]{Risken_FP_book}%
  \BibitemOpen
  \bibfield  {author} {\bibinfo {author} {\bibfnamefont {H.}~\bibnamefont
  {Risken}},\ }\href@noop {} {\emph {\bibinfo {title} {{The Fokker-Planck
  Equation: Methods of Solution and Applications}}}},\ \bibinfo {edition}
  {2nd}\ ed.\ (\bibinfo  {publisher} {Springer},\ \bibinfo {year}
  {1989})\BibitemShut {NoStop}%
\bibitem [{\citenamefont {Weeks}\ \emph {et~al.}(2000)\citenamefont {Weeks},
  \citenamefont {Crocker}, \citenamefont {Levitt}, \citenamefont {Schofield},\
  and\ \citenamefont {Weitz}}]{Weeks2000}%
  \BibitemOpen
  \bibfield  {author} {\bibinfo {author} {\bibfnamefont {E.~R.}\ \bibnamefont
  {Weeks}}, \bibinfo {author} {\bibfnamefont {J.~C.}\ \bibnamefont {Crocker}},
  \bibinfo {author} {\bibfnamefont {A.~C.}\ \bibnamefont {Levitt}}, \bibinfo
  {author} {\bibfnamefont {A.}~\bibnamefont {Schofield}}, \ and\ \bibinfo
  {author} {\bibfnamefont {D.~A.}\ \bibnamefont {Weitz}},\ }\href {\doibase
  10.1126/science.287.5453.627} {\bibfield  {journal} {\bibinfo  {journal}
  {Science}\ }\textbf {\bibinfo {volume} {287}},\ \bibinfo {pages} {627}
  (\bibinfo {year} {2000})}\BibitemShut {NoStop}%
\bibitem [{\citenamefont {Flenner}\ and\ \citenamefont
  {Szamel}(2005{\natexlab{b}})}]{Flenner2005b}%
  \BibitemOpen
  \bibfield  {author} {\bibinfo {author} {\bibfnamefont {E.}~\bibnamefont
  {Flenner}}\ and\ \bibinfo {author} {\bibfnamefont {G.}~\bibnamefont
  {Szamel}},\ }\href {\doibase 10.1103/PhysRevE.72.031508} {\bibfield
  {journal} {\bibinfo  {journal} {Physical Review E}\ }\textbf {\bibinfo
  {volume} {72}},\ \bibinfo {pages} {031508} (\bibinfo {year}
  {2005}{\natexlab{b}})},\ \Eprint {http://arxiv.org/abs/0508102}
  {arXiv:0508102 [cond-mat]} \BibitemShut {NoStop}%
\bibitem [{\citenamefont {Scalliet}\ \emph {et~al.}(2022)\citenamefont
  {Scalliet}, \citenamefont {Guiselin},\ and\ \citenamefont
  {Berthier}}]{Scalliet2022}%
  \BibitemOpen
  \bibfield  {author} {\bibinfo {author} {\bibfnamefont {C.}~\bibnamefont
  {Scalliet}}, \bibinfo {author} {\bibfnamefont {B.}~\bibnamefont {Guiselin}},
  \ and\ \bibinfo {author} {\bibfnamefont {L.}~\bibnamefont {Berthier}},\
  }\href {\doibase 10.1103/PhysRevX.12.041028} {\bibfield  {journal} {\bibinfo
  {journal} {Physical Review X}\ }\textbf {\bibinfo {volume} {12}},\ \bibinfo
  {pages} {041028} (\bibinfo {year} {2022})},\ \Eprint
  {http://arxiv.org/abs/2207.00491} {arXiv:2207.00491} \BibitemShut {NoStop}%
\bibitem [{\citenamefont {Flenner}\ and\ \citenamefont
  {Szamel}(2010)}]{Flenner2010}%
  \BibitemOpen
  \bibfield  {author} {\bibinfo {author} {\bibfnamefont {E.}~\bibnamefont
  {Flenner}}\ and\ \bibinfo {author} {\bibfnamefont {G.}~\bibnamefont
  {Szamel}},\ }\href {\doibase 10.1103/PhysRevLett.105.217801} {\bibfield
  {journal} {\bibinfo  {journal} {Physical Review Letters}\ }\textbf {\bibinfo
  {volume} {105}},\ \bibinfo {pages} {217801} (\bibinfo {year} {2010})},\
  \Eprint {http://arxiv.org/abs/1005.3794} {arXiv:1005.3794} \BibitemShut
  {NoStop}%
\bibitem [{\citenamefont {Kob}\ and\ \citenamefont {Barrat}(1997)}]{Kob1997}%
  \BibitemOpen
  \bibfield  {author} {\bibinfo {author} {\bibfnamefont {W.}~\bibnamefont
  {Kob}}\ and\ \bibinfo {author} {\bibfnamefont {J.~L.}\ \bibnamefont
  {Barrat}},\ }\href {\doibase 10.1103/PhysRevLett.78.4581} {\bibfield
  {journal} {\bibinfo  {journal} {Physical Review Letters}\ }\textbf {\bibinfo
  {volume} {78}},\ \bibinfo {pages} {4581} (\bibinfo {year} {1997})},\ \Eprint
  {http://arxiv.org/abs/9704006} {arXiv:9704006 [cond-mat]} \BibitemShut
  {NoStop}%
\bibitem [{\citenamefont {Hodge}(1995)}]{Hodge1995}%
  \BibitemOpen
  \bibfield  {author} {\bibinfo {author} {\bibfnamefont {I.~M.}\ \bibnamefont
  {Hodge}},\ }\href@noop {} {\bibfield  {journal} {\bibinfo  {journal}
  {Science}\ }\textbf {\bibinfo {volume} {267}},\ \bibinfo {pages} {1945}
  (\bibinfo {year} {1995})}\BibitemShut {NoStop}%
\bibitem [{\citenamefont {Angelini}\ \emph {et~al.}(2013)\citenamefont
  {Angelini}, \citenamefont {Zulian}, \citenamefont {Fluerasu}, \citenamefont
  {Madsen}, \citenamefont {Ruocco},\ and\ \citenamefont
  {Ruzicka}}]{Angelini2013}%
  \BibitemOpen
  \bibfield  {author} {\bibinfo {author} {\bibfnamefont {R.}~\bibnamefont
  {Angelini}}, \bibinfo {author} {\bibfnamefont {L.}~\bibnamefont {Zulian}},
  \bibinfo {author} {\bibfnamefont {A.}~\bibnamefont {Fluerasu}}, \bibinfo
  {author} {\bibfnamefont {A.}~\bibnamefont {Madsen}}, \bibinfo {author}
  {\bibfnamefont {G.}~\bibnamefont {Ruocco}}, \ and\ \bibinfo {author}
  {\bibfnamefont {B.}~\bibnamefont {Ruzicka}},\ }\href {\doibase
  10.1039/c3sm52173g} {\bibfield  {journal} {\bibinfo  {journal} {Soft Matter}\
  }\textbf {\bibinfo {volume} {9}},\ \bibinfo {pages} {10955} (\bibinfo {year}
  {2013})}\BibitemShut {NoStop}%
\bibitem [{\citenamefont {Segura}\ \emph {et~al.}(2022)\citenamefont {Segura},
  \citenamefont {Montoya},\ and\ \citenamefont {Tapias}}]{Segura2022a}%
  \BibitemOpen
  \bibfield  {author} {\bibinfo {author} {\bibfnamefont {C.~H.}\ \bibnamefont
  {Segura}}, \bibinfo {author} {\bibfnamefont {E.}~\bibnamefont {Montoya}}, \
  and\ \bibinfo {author} {\bibfnamefont {D.}~\bibnamefont {Tapias}},\ }\href
  {\doibase 10.1088/2632-2153/ac8f1b} {\bibfield  {journal} {\bibinfo
  {journal} {Machine Learning: Science and Technology}\ }\textbf {\bibinfo
  {volume} {3}},\ \bibinfo {pages} {035013} (\bibinfo {year}
  {2022})}\BibitemShut {NoStop}%
\bibitem [{\citenamefont {LeCun}\ \emph {et~al.}(1989)\citenamefont {LeCun},
  \citenamefont {Denker},\ and\ \citenamefont {Solla}}]{LeCun1990}%
  \BibitemOpen
  \bibfield  {author} {\bibinfo {author} {\bibfnamefont {Y.}~\bibnamefont
  {LeCun}}, \bibinfo {author} {\bibfnamefont {J.~S.}\ \bibnamefont {Denker}}, \
  and\ \bibinfo {author} {\bibfnamefont {S.~A.}\ \bibnamefont {Solla}},\
  }\href@noop {} {\bibfield  {journal} {\bibinfo  {journal} {Advances in neural
  information processing systems}\ }\textbf {\bibinfo {volume} {2}},\ \bibinfo
  {pages} {598} (\bibinfo {year} {1989})}\BibitemShut {NoStop}%
\bibitem [{\citenamefont {Han}\ \emph {et~al.}(2015)\citenamefont {Han},
  \citenamefont {Pool}, \citenamefont {Tran},\ and\ \citenamefont
  {Dally}}]{Han2015}%
  \BibitemOpen
  \bibfield  {author} {\bibinfo {author} {\bibfnamefont {S.}~\bibnamefont
  {Han}}, \bibinfo {author} {\bibfnamefont {J.}~\bibnamefont {Pool}}, \bibinfo
  {author} {\bibfnamefont {J.}~\bibnamefont {Tran}}, \ and\ \bibinfo {author}
  {\bibfnamefont {W.~J.}\ \bibnamefont {Dally}},\ }in\ \href@noop {} {\emph
  {\bibinfo {booktitle} {Advances in Neural Information Processing Systems}}},\
  Vol.~\bibinfo {volume} {28}\ (\bibinfo {year} {2015})\ \Eprint
  {http://arxiv.org/abs/1506.02626} {arXiv:1506.02626} \BibitemShut {NoStop}%
\bibitem [{\citenamefont {Blalock}\ \emph {et~al.}(2020)\citenamefont
  {Blalock}, \citenamefont {{Javier Gonzalez Ortiz}}, \citenamefont {Frankle},\
  and\ \citenamefont {Guttag}}]{Blalock2020}%
  \BibitemOpen
  \bibfield  {author} {\bibinfo {author} {\bibfnamefont {D.}~\bibnamefont
  {Blalock}}, \bibinfo {author} {\bibfnamefont {J.}~\bibnamefont {{Javier
  Gonzalez Ortiz}}}, \bibinfo {author} {\bibfnamefont {J.}~\bibnamefont
  {Frankle}}, \ and\ \bibinfo {author} {\bibfnamefont {J.}~\bibnamefont
  {Guttag}},\ }\href@noop {} {\bibfield  {journal} {\bibinfo  {journal}
  {Proceedings of the 3rd MLSys Conference}\ } (\bibinfo {year} {2020})},\
  \Eprint {http://arxiv.org/abs/2003.02389} {arXiv:2003.02389} \BibitemShut
  {NoStop}%
\bibitem [{\citenamefont {Elam}\ \emph {et~al.}(1984)\citenamefont {Elam},
  \citenamefont {Kerstein},\ and\ \citenamefont {Rehr}}]{Elam1984}%
  \BibitemOpen
  \bibfield  {author} {\bibinfo {author} {\bibfnamefont {W.~T.}\ \bibnamefont
  {Elam}}, \bibinfo {author} {\bibfnamefont {A.~R.}\ \bibnamefont {Kerstein}},
  \ and\ \bibinfo {author} {\bibfnamefont {J.~J.}\ \bibnamefont {Rehr}},\
  }\href {\doibase 10.1103/PhysRevLett.52.1516} {\bibfield  {journal} {\bibinfo
   {journal} {Physical Review Letters}\ }\textbf {\bibinfo {volume} {52}},\
  \bibinfo {pages} {1516} (\bibinfo {year} {1984})}\BibitemShut {NoStop}%
\bibitem [{\citenamefont {Hofling}\ \emph {et~al.}(2006)\citenamefont
  {Hofling}, \citenamefont {Franosch},\ and\ \citenamefont
  {Frey}}]{Hofling2006}%
  \BibitemOpen
  \bibfield  {author} {\bibinfo {author} {\bibfnamefont {F.}~\bibnamefont
  {Hofling}}, \bibinfo {author} {\bibfnamefont {T.}~\bibnamefont {Franosch}}, \
  and\ \bibinfo {author} {\bibfnamefont {E.}~\bibnamefont {Frey}},\ }\href
  {\doibase 10.1103/PhysRevLett.96.165901} {\bibfield  {journal} {\bibinfo
  {journal} {Physical Review Letters}\ }\textbf {\bibinfo {volume} {96}},\
  \bibinfo {pages} {165901} (\bibinfo {year} {2006})}\BibitemShut {NoStop}%
\bibitem [{\citenamefont {Spanner}\ \emph {et~al.}(2016)\citenamefont
  {Spanner}, \citenamefont {H{\"{o}}fling}, \citenamefont {Kapfer},
  \citenamefont {Mecke}, \citenamefont {Schr{\"{o}}der-Turk},\ and\
  \citenamefont {Franosch}}]{Spanner2016}%
  \BibitemOpen
  \bibfield  {author} {\bibinfo {author} {\bibfnamefont {M.}~\bibnamefont
  {Spanner}}, \bibinfo {author} {\bibfnamefont {F.}~\bibnamefont
  {H{\"{o}}fling}}, \bibinfo {author} {\bibfnamefont {S.~C.}\ \bibnamefont
  {Kapfer}}, \bibinfo {author} {\bibfnamefont {K.~R.}\ \bibnamefont {Mecke}},
  \bibinfo {author} {\bibfnamefont {G.~E.}\ \bibnamefont
  {Schr{\"{o}}der-Turk}}, \ and\ \bibinfo {author} {\bibfnamefont
  {T.}~\bibnamefont {Franosch}},\ }\href {\doibase
  10.1103/PhysRevLett.116.060601} {\bibfield  {journal} {\bibinfo  {journal}
  {Physical Review Letters}\ }\textbf {\bibinfo {volume} {116}},\ \bibinfo
  {pages} {060601} (\bibinfo {year} {2016})}\BibitemShut {NoStop}%
\bibitem [{\citenamefont {H{\"{o}}fling}\ \emph {et~al.}(2008)\citenamefont
  {H{\"{o}}fling}, \citenamefont {Munk}, \citenamefont {Frey},\ and\
  \citenamefont {Franosch}}]{Hofling2008}%
  \BibitemOpen
  \bibfield  {author} {\bibinfo {author} {\bibfnamefont {F.}~\bibnamefont
  {H{\"{o}}fling}}, \bibinfo {author} {\bibfnamefont {T.}~\bibnamefont {Munk}},
  \bibinfo {author} {\bibfnamefont {E.}~\bibnamefont {Frey}}, \ and\ \bibinfo
  {author} {\bibfnamefont {T.}~\bibnamefont {Franosch}},\ }\href {\doibase
  10.1063/1.2901170} {\bibfield  {journal} {\bibinfo  {journal} {The Journal of
  Chemical Physics}\ }\textbf {\bibinfo {volume} {128}},\ \bibinfo {pages}
  {164517} (\bibinfo {year} {2008})},\ \Eprint {http://arxiv.org/abs/0712.2313}
  {arXiv:0712.2313} \BibitemShut {NoStop}%
\bibitem [{\citenamefont {Bauer}\ \emph {et~al.}(2010)\citenamefont {Bauer},
  \citenamefont {H{\"{o}}fling}, \citenamefont {Munk}, \citenamefont {Frey},\
  and\ \citenamefont {Franosch}}]{Bauer2010}%
  \BibitemOpen
  \bibfield  {author} {\bibinfo {author} {\bibfnamefont {T.}~\bibnamefont
  {Bauer}}, \bibinfo {author} {\bibfnamefont {F.}~\bibnamefont
  {H{\"{o}}fling}}, \bibinfo {author} {\bibfnamefont {T.}~\bibnamefont {Munk}},
  \bibinfo {author} {\bibfnamefont {E.}~\bibnamefont {Frey}}, \ and\ \bibinfo
  {author} {\bibfnamefont {T.}~\bibnamefont {Franosch}},\ }\href {\doibase
  10.1140/epjst/e2010-01313-1} {\bibfield  {journal} {\bibinfo  {journal}
  {European Physical Journal: Special Topics}\ }\textbf {\bibinfo {volume}
  {189}},\ \bibinfo {pages} {103} (\bibinfo {year} {2010})},\ \Eprint
  {http://arxiv.org/abs/1003.2918} {arXiv:1003.2918} \BibitemShut {NoStop}%
\bibitem [{CIF()}]{CIFAR10_dataset}%
  \BibitemOpen
  \href {https://www.cs.toronto.edu/$\sim$kriz/cifar.html} {\enquote {\bibinfo
  {title} {{CIFAR10 dataset,
  https://www.cs.toronto.edu/$\sim$kriz/cifar.html}},}\ }\BibitemShut {NoStop}%
\end{thebibliography}%

\end{document}